\documentclass[a4paper,12pt,times,numbered,print,index]{Classes/PhDThesisPSnPDF}

\title{Architecture Models Refinements for Critical Real-time Embedded Systems}

\author{Etienne Borde}

\dept{École doctorale Informatique, Télécommunications et Électronique}

\university{Sorbonnes University}
\crest{\includegraphics[width=0.8\textwidth]{Titlepage/logo_su}}

\degreetitle{Habilitation à Diriger des Recherches}

\college{Sorbonnes University}

\subject{HDR report} \keywords{{HDR} {Engineering} {Sorbonnes University}}

\ifdefineAbstract
\fi

\ifdefineChapter
\fi

\usepackage{amssymb}
\usepackage{amsmath}
\usepackage{amsthm}
\usepackage{minitoc}

\newcommand{\sem}[1]{\llbracket #1 \rrbracket}
\newcommand{\tightdownarrow}{\!\downarrow}

\newtheoremstyle{mydefstyle}%
{15pt}
{15pt}
{}
{}
{\bfseries}
{.}
{ }
{\thmname{#1}\thmnumber{ #2}\thmnote{ (#3)}}

\theoremstyle{mydefstyle}
\newtheorem{defthm}{Definition}[chapter]

\usepackage{xparse}

\NewDocumentEnvironment{deft}{o}
{\IfNoValueTF{#1}{\begin{defthm}}{\begin{defthm}[#1]}}
{\ignorespaces \qed \end{defthm}}

\newtheorem{mydef}{Definition}
\newenvironment{pushright}{
  \begin{itemize}\item[\hspace{12pt}]}{\end{itemize}
}

\usepackage{listings}

\lstdefinelanguage{aadl}
{morekeywords={in,out,package,end,bus,data,thread,port,group,process,processor,
system,memory,device,subprogram,private,event,property,set,applies,to,
units,type,implementation,parameter,reference,mode,final,initial,
transitions},
morekeywords={extends,properties,features,annex,modes,connections,flows,
subcomponents,calls,binding,states,initial,requires,access,prototypes,state,
in, binding, computation, !<, !>, {**, **}},
morekeywords={aadlinteger,aadlboolean,aadlstring,aadlfloat, behavior_specification},
morecomment=[l]{--}
}

\lstdefinelanguage{ATL}
{
  morekeywords={module,create,rule,using,from,to,not},
  otherkeywords={:,<-,!,\{,\},\,,->},
  morecomment=[l]{--}
}

\usepackage{pdfpages}
\usepackage{graphics}
\usepackage{subfig}
\usepackage{graphicx}
\usepackage{pgfplots}
\usepackage{pgfplotstable}

\newcommand{\galap}{\textsc{G-alap-LLF}}
\newcommand{\condmc}{\textbf{Safe Trans. Prop.}}
\newcommand{\condhi}{\textbf{Condition HI-Mode}}
\newcommand{\condlo}{\textbf{Condition LO-Mode}}

\usepackage{enumitem}

\usepackage[htt]{hyphenat}

\usepackage{xcolor}

\dominitoc

\begin{document}


\includepdf{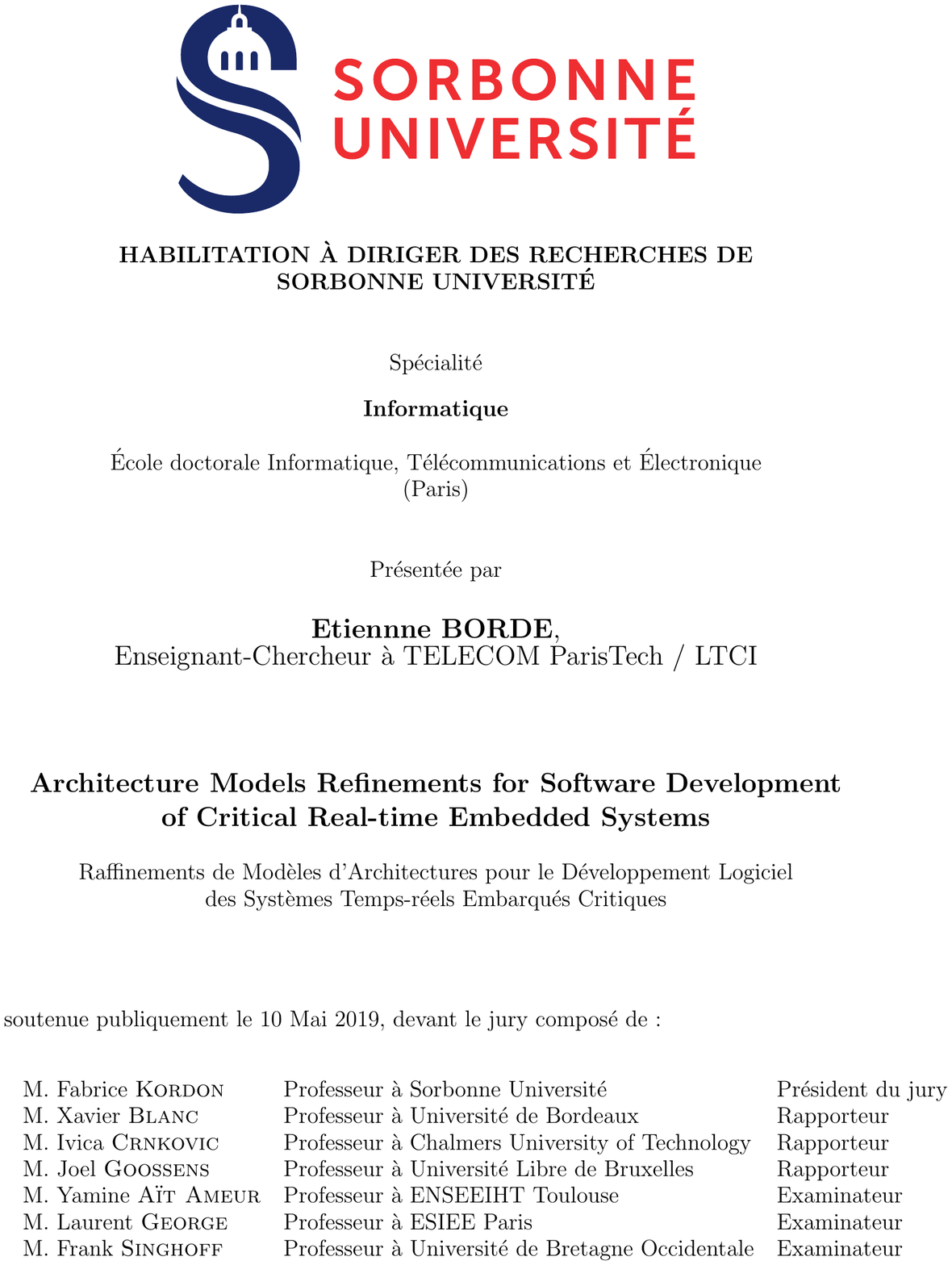}


\begin{acknowledgements}      

I would like to sincerely thank Professors Xavier Blanc, Ivica
Crnkovic, and Joël Goossens for giving me the great honor of reviewing
my habilitation thesis. I greatly appreciated their feedback on my
work and the accuracy of their assessment.

I would also like to thank Professors Yamine Ait Ameur, Laurent
George, and Frank Singhoff for having accepted to serve on the jury, as
well as Professor Fabrice Kordon for having accepted to be the president of
the jury. I would especially like to thank Professor Fabrice Kordon for his
help in preparing for this habilitation at Sorbonne Université.

The work presented in this document is the result of many and very
rich interactions with colleagues from academia and industry.

First of all, I want to warmly thank Laurent Pautet who has been
working with me since the very beginning of my career: from the PhD to
this habilitation. It has been a pleasure to work with him, and learn
from him. I also want to thank my colleagues from TELECOM ParisTech
for the quality of their feedback on my work, and all the good times
spent together: Thomas Robert, Florian Brandner, Rémi Sharrock, James
Eagan, Ada Diaconescu, Elie Najm, Sylvie Vignes, and Petr
Kuznetsov. Not to mention Sébastien Gardoll from CNRS.

This work is also the result of the work of brilliant PhD students that
I would like to congratulate again for their achievements, and thank
for their collaborations: Fabien Cadoret, Cuauhtémoc Castellanos, Elie
Richa, Smail Rahmoun, and Roberto Medina.

Finally, this work is the result of numerous interactions with
researchers and engineers from both academic and industrial
insitutes: I would like to thank the Chaire ISC (in particular Eric
Goubault from Polytechnique, and Alexandre Chapoutot from ENSTA), the
IRT SystemX (along with Alstom, Renault, Thales, Safran), the AADL
standardization committee (in particular Peter Feiler from the
Software Engineering Institute, Jean-Pierre Talpin from INRIA, Pierre
Dissaux from Ellidiss, and Brian Larson from Kansas State University)
for their support, their contributions, and feedback.

Last but not least, I want to thank my wife Sara for her support in
both my professional and personal achievements.

\end{acknowledgements}

\begin{abstract}
  Cyber Physical Systems are systems controlled or monitored by
  computer-based programs, tightly integrated networks, sensors, and
  actuators. Trains, aircrafts, cars, and some medical equipments are
  examples of complex CPS. Software development of complex CPS has
  become so difficult that it represents most of the cost of CPS
  production. According to domain experts, this trend is going to
  reach a point where software development would represent the main
  source of cost of a CPS production.

  \noindent
  In addition, it is interesting to note that the integration,
  verification and validation of software in CPS require more efforts
  than the analysis, design, and implementation activities. The main
  reason is that these activities are conducted late in the
  development process and issues discovered at this stage of the
  process will require to rework artifacts produced in the previous
  activities (\emph{i.e.} analysis, design and/or implementation).

  \noindent
  In this document, we present our work aiming to improve the
  reliability of software development in the domain of CPS. In this
  context, we define the reliability of the development process as its
  capacity to deliver intermediate artifacts for which the rework
  effort would be as small as possible. 

  \noindent
  This problem is very difficult for general purpose software
  (\emph{i.e.} used on desktop computers or servers), and even more
  difficult for software in CPS. The main reason is that software in
  CPS is often \emph{critical, real-time and embedded on domain
    specific execution platforms}. As a consequence, non-functional
  properties (also called quality attributes) of software applications
  in CPS are often as important and difficult to satisfy as the
  logical correctness of these applications.

  \noindent
  In order to the improve the reliability of software development in
  the domain of CPS, we propose a Model Driven Engineering (MDE)
  method based on step-wise refinements of software architecture
  descriptions (also called architectural models). An architecture
  description being an abstraction of the software being developed,
  the implementation of this software (\emph{i.e.}  source or binary
  code) is an \textbf{interpretation} of the architecture model. In
  the framework we propose, such interpretations are automated using
  \textbf{model refinements}, \emph{i.e.} model to model
  transformations lowering the abstraction level of the architecture
  description.

  \noindent
  However, models interpretation may introduce faults such as bugs or
  invalidation of non-functional requirements. It is hence necessary
  to control as much as possible the correctness, consistency, and
  optimality of artifacts produced along the model refinement steps.

  To reach this objective, we propose to
  \begin{enumerate}
    \item define model transformations so as to interleave refinement
      steps with analysis of the resulting artifacts. We thus improve
      the consistency between the analysis results and the software
      implementation by analyzing models as close as possible to the
      implementation.
    \item define timing analysis and real-time scheduling techniques
      to ensure the correctness of software architectures from a
      timing perspective.
    \item formalize model transformations in order to ensure their
      correctness using formal verification techniques.
    \item compose model transformations in order to automate the
      search for optimal (or near-optimal) architectures.
  \end{enumerate}

  The work presented in this document is thus at the frontier among
  different research domains: MDE, real-time systems scheduling,
  formal verification, and operational research.

  \noindent
  In this work, we chose to rely and extend the Architecture Analysis
  and Design Language (AADL) to model the cyber part of CPS. The
  reasons for this choice are simple: Firstly, AADL is a standard and
  a domain specific language for real-time embedded systems. Secondly,
  It allows to represent software architectures with different
  abstraction levels. Last but not least, AADL supports different
  types of models of computations communications, some of which being
  deterministic.

  \noindent
  As a guideline for our work, we developed the methodology we propose
  in a MDE framework called RAMSES (Refinement of AADL Models for the
  Synthesis of Embedded Systems). This document presents both the
  methodology and some illustrations of its implementation in RAMSES.
  
\end{abstract}


\tableofcontents

\listoffigures
\adjustmtc



\printnomenclature

\mainmatter


\chapter{Introduction}  

\minitoc

\section{Industrial Context and Scientific Challenges}

A Cyber Physical System (CPS) is a system that is controlled or
monitored by computer-based programs, tightly integrated networks,
sensors, and actuators. ``In cyber-physical systems, physical and
software components are deeply intertwined, each operating on
different spatial and temporal scales, exhibiting multiple and
distinct behavioral modalities, and interacting with each other in a
lot of ways that change with
context.''\footnote{https://www.nsf.gov/pubs/2010/nsf10515/nsf10515.htm}
Robotic systems of course, but also transportation systems, medical
devices or power plants are example of CPSs.

\noindent
The evolution of software applications deployed in CPS shows a
significant increase in their complexity. Measured in terms of lines
of code (SLOC) embedded in different generations of aircraft systems,
this complexity indicator shows the significance of their software
evolution: Airbus A310: \~{}400 KSLOC, A320: \~{}800KSLOC, A330/340:
\~{}2MSLOC, Boeing 777: \~{}4MSLOC, Airbus A380: \~{}8MSLOC, Boeing
787: \~{}10MSLOC. These impressive numbers are still below the numbers
of estimated lines of code in military aircrafts or luxurious modern
cars.

\noindent
As a consequence, software development, takes an important role in the
production of such systems. Our industrial partners in the avionics
domain estimate that 70\% of systems production cost is due to
software development, mainly because of very demanding validation,
documentation, and integration activities. They estimate this ratio
would grow up to about 90\% in 2024.

\noindent
The cost presented here above measures how difficult software
development has become in the domain of CPS. This cost actually
aggregates several sources of difficulties over the development
life-cycle. In particular, our industrial partners (still in the
avionics domain) estimate that 70\% of software development cost is
due to rework, validation, and verification activities, mostly because
of faults introduced during the early phases of the development
process. Counterintuitively, this estimation highlights that
integration, verification, and validation activities represent more
efforts than analysis, design, and implementation activities.

\noindent
Verification and validation of software in CPS is particularly
difficult because CPS are often mission or safety critical: failures
of such systems could have catastrophic consequences. As a result,
developers of CPS are often required to conform to certification
processes aiming at ensuring these systems meet safety requirements.

\noindent
On the other hand, software integration in CPS is also challenging
because these systems have meet various requirements such as timing
performance, energy consumption, weight, availability,
maintainability, robustness, etc.

\begin{figure}[h]
\centering
\includegraphics[width=\textwidth, trim={0cm 17cm 0cm 1cm}, clip]{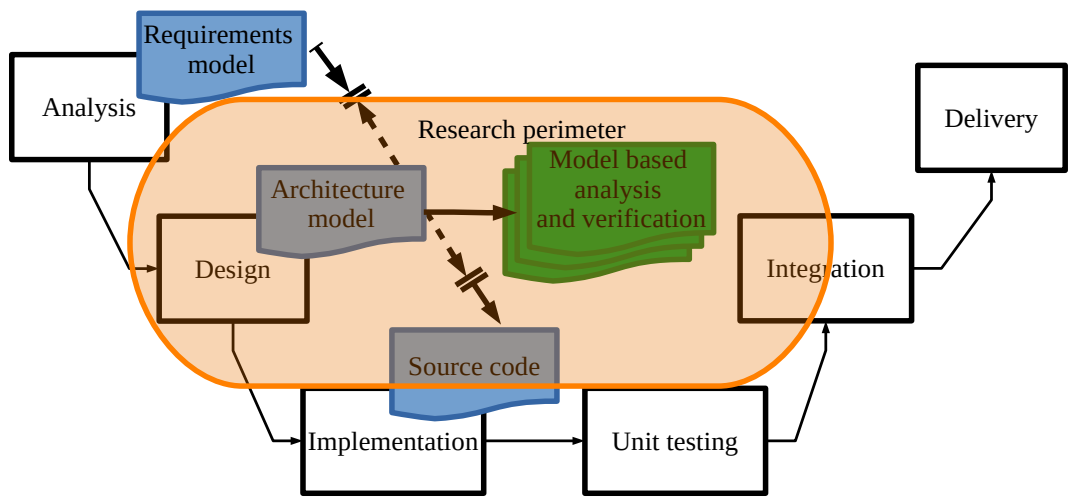}
\caption{Context: Critical CPSs Development Process}
\label{fig:general_context}
\end{figure}

\noindent
Figure~\ref{fig:general_context} gives an overview of a traditional
development process, called V-cycle, generally used in the development
of CPSs. The objective of the V-cycle development process is to
anticipate validation activity by preparing verification and
validation artifacts along with requirements, design, and
implementation activities.

\noindent
In addition, Model Driven Engineering (MDE) advocates for the use of
models in order to improve the development process of software
applications and in order to increase products quality. For example,
models can be used to improve the development process by enabling
early estimation of software applications performance. In addition,
such estimation will help designers of software applications to
compare different solutions and select the most appropriate one(s). As
shown in figure~\ref{fig:general_context}, models can be the result of
the requirements definition and design activities of the development
process, while source code is produced during the implementation phase
of the process.

\noindent
In the work we present in this document, architecture models produced
during the design activities play a major role: firstly, we combine
them with verification and analysis techniques in order to detect
design flaws as early as possible in the process. Secondly, as
illustrated on the figure, we aim at using architecture models to
bridge the gap between requirements models and source code. Therefore,
the work presented in this document aims to improve methods dedicated
to the design of software architecture(s) for embedded systems which
are also critical, real-time, and distributed systems. Because the
type of systems we consider are mission or safety critical, the
methods we aim for must rely on rigorous models so as to guarantee
safety related properties.

\noindent
The targeted application domain is the domain of CPS, in
which physical systems are controlled by a set of interconnected
computation units (CUs) executing control and/or monitoring
algorithms. More specifically, industrial partners involved in the
definition of research problems presented in this document work on
transportation systems (cars, planes, and/or trains). In next
subsection, we present the general problems for which we proposed the
scientific contributions presented in this document.

\newpage

\section{Problem statement}

Given the increasing complexity and cost of software development in
CPSs, a zoom on the research perimeter presented in previous section
(see the orange part of figure~\ref{fig:general_context}) led us to
raise the following research question: how to improve the reliability
of CPS design activities?

\noindent
In the context of software development for CPS, we define the
reliability of design activities as their capacity to deliver
intermediate artifacts for which the rework effort would be as small
as possible.
  
\begin{figure}[h!]
\centering
\includegraphics[width=\textwidth, trim={0cm 15cm 0cm 1cm}, clip]{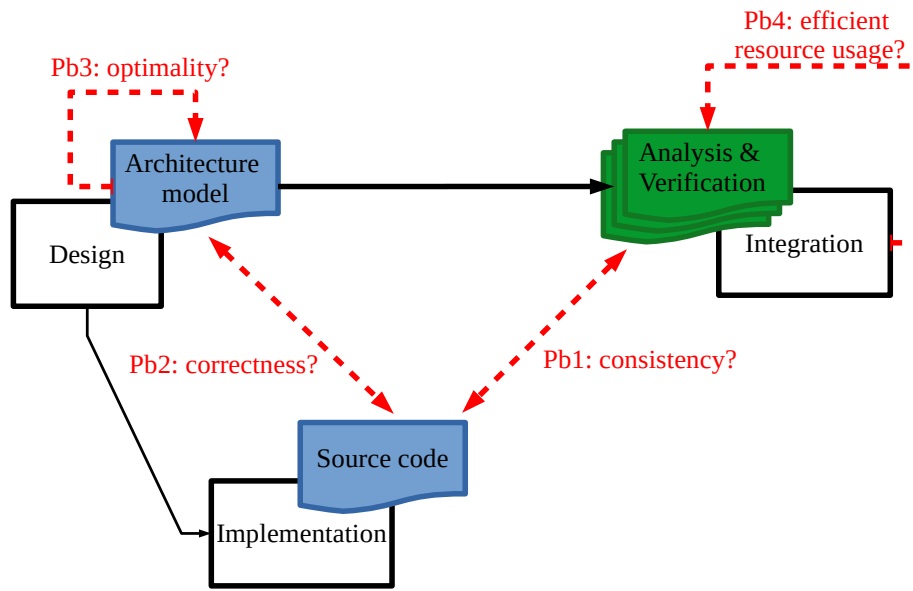}
\caption{Problems in Critical CPSs Development Process}
\label{fig:pb_overview}
\end{figure}

\noindent
On figure~\ref{fig:pb_overview}, we provide a decomposition of this
research question into different but connected research problems:

\begin{itemize}[leftmargin=3cm, rightmargin=1cm]
\item[Problem 1:] Are analysis results, obtained on an architecture
  model, consistent with the implementation of this architecture with
  source code?

  By essence, architecture models are abstractions of the reality
  aiming to enable the analysis of a system under design. This
  analysis can evaluate quality of services or even safety properties
  of the system. On the other hand, the implementation of these models
  will require to interpret this abstract view and translate it into
  source code by introducing missing details. If these details had an
  impact on the analysis results, these results obtained with the
  model may no longer be consistent with the properties exhibited by
  its implementation. The corresponding design would become obsolete,
  early design choices would be invalidated, leading to extra
  re-engineering efforts.

\item[Problem 2:] How to ensure the correctness of an architecture
  model implementation?

  The activity transforming an architecture model into the
  implementation of a software application boils down to translate an
  abstract view of the system to provide an equivalent executable
  artifact. By essence, there exist several variants of such
  translation: an abstract model may have several possible
  implementations. In addition, this activity my rapidly become
  repetitive, thus error prone. It is therefore very important to
  ensure the correctness of the implementation by making sure the
  translation effort did not introduce flaws in the resulting source
  code. Such flaws may, again, invalidate analysis results, or even
  worse: introduce bugs in the source code of the application.
  
\item[Problem 3:] Is the input architecture optimal?

  As explained in the presentation of Problem~1, the translation of
  abstract models into source code may invalidate early analysis
  results. One may conclude it would be sufficient to model systems
  with sufficient margins to ensure the preservation of analysis
  results along the development life-cycle. However, the values of
  such margins are difficult to anticipate, and it would be necessary
  to take very pessimistic estimations to make sure there would not be
  extra re-engineering efforts eventually. In practice, this is
  infeasible since big margins also means poor quality: for instance,
  a computer loaded at 20\% exhibits a big margin but is poorly
  exploited (which means more functions of more complex functions
  could have been deployed on it). In addition, quality attributes are
  often in conflict as improving one quality attribute requires to
  degrade another one. For instance, the deployment of replicated
  functions improves the availability of these functions to the price
  of extra weight, energy consumption, and data flow latency. It is
  thus important to deal with the problem of providing optimal (or
  near optimal) architecture models, otherwise the chance to face
  integration issues grows rapidly for complex systems.

\item[Problem 4:] Are computation resources efficiently allocated to
  software applications?

  In CPS, software architects have to pay extra attention to the
  allocation of computation resources to software applications. The
  main reason is that some of these applications will have a direct
  impact on the safety of the system. Such applications would be
  classified with a high level of criticality whereas others would be
  classified with a lower level of criticality. In most cyber physical
  systems, the provision of enough computation resource to high
  criticality functions is thus a safety requirement, whereas the
  provision of computation resources to lower criticality functions is
  a quality of service requirement. Note that quality of service,
  though not critical, is of prime importance as it has a direct
  impact on consumers satisfaction. In a CPS, an inefficient resource
  allocation may lead to a poor quality of services, and even worse,
  to safety requirements violation.

\end{itemize}

\noindent
Problems 1 and 2 are obviously connected to our research question
(\emph{i.e.} how to improve the reliability of CPS design activities?)
whereas the link with problems 3 and 4 may seems less direct. Yet, it
is important to consider that software applications in CPS have to
meet stringent requirements in terms of timing performance, memory
footprint, safety, security, and/or energy consumption. These
requirements, usually called Non-Functional Requirements (NFR), are
often as important as functional requirements in CPSs. Thus, if errors
are discovered late in the development process because of poor Non
Functional Properties (NFPs) due to design flaws, a design rework is
necessary and its cost will raise fast. It is thus important to
ensure, as soon as possible in the design process, that considered
architectures respect NFRs but also provide the best possible margin
with respect to the limit imposed by these NFRs. This boils down to
optimize these architectures, either during the modeling phase
(problem 3) or during the deployment phase (problem 4).

\noindent
Even though our work focuses on software architectures, specificities
of CPSs require to take into consideration hardware platform
characteristics as well. Indeed, the adequacy of a software
architecture with respect to these requirements, called Non Functional
Requirements (NFRs), cannot be assessed without a knowledge of the
underlying hardware and/or network architecture.

\noindent
This is the reason why we used in our work the Architecture Analysis
and Design Language (AADL), an architecture description language
offering the capabilities to model both the software and hardware
architecture of a CPS, as well as the binding of software components
onto hardware components.

\noindent
In addition, Non Functional Properties (NFPs) evaluation requires
dedicated models in which implementation details are abstracted away
in order to focus on most relevant characteristics of the architecture
for a given property. To extract such characteristics from a model,
model transformations are often used: a model transformation is a
software application that takes as input a model, and/or produce a
model as output. Model transformations may be used to translate a
given model from one formalism to another one with the same
abstraction level. Such transformations are called horizontal
transformations. Model transformations may also be used to change the
level of abstraction of a model by adding or abstracting away modeling
details. Such transformations are called vertical transformations. In
this document, we call \emph{refinement} a vertical model
transformation adding modeling details.

\noindent
Last but not least, CPS architects often consider design alternatives
as decision variables in an optimization problem aiming at minimizing
or maximizing NFPs. However, design alternatives often come into
conflict with respect to their impacts on NFP: most of the time, a
design alternative improves a NFP at the cost of degrading another NFP
of a CPS. As a consequence, designers aim at providing the best
possible trade-off among NFPs of a CPS.

\section{Organization of the document}

This document is organized as follows.

\noindent
Chapter~\ref{chap:chapter2} contains an overview of the approach we
propose to contribute to the resolution of the problems presented
above. A brief presentation of related works helps to understand the
originality of our approach.

\noindent
In chapter~\ref{chap:chapter3}, we present the core ideas our work
relies on: model transformations of architecture models for the
analysis, design, and optimisation of critical
CPSs. Chapter~\ref{chap:chapter4} gives more details on the work we
undertook on the composition and formalization of model
transformations.

\noindent
This work is integrated in RAMSES, which is to the best of our
knowledge the only AADL to code generation framework implementing
deterministic subsets of AADL. It is also the only AADL to code
generation framework allowing fine-grained timing analysis of the
non-deterministic subsets of AADL. Last but not least, model
transformation compositions for design space exploration have been
experimented in RAMSES on complex optimization problems with very
satisfying results.

\noindent
Finally, chapter~\ref{chap:chapter5} concludes this document and
provides research perspectives for the work presented in this
document.


\chapter{Overview of contributions}
\label{chap:chapter2}

\minitoc

\vspace{1.5cm}

In the previous chapter, we have defined ambitious and difficult
research problems. We contributed to their resolution through the
definition of an architecture models refinement framework we present
in section~\ref{sec:2.1}. We then present in section~\ref{sec:2.2} a
brief overview of the state of the art in this domain, before to
summarize this activity in terms of PhD students supervisions
(section~\ref{sec:2.3}).

\section{Approach: Architecture refinement framework}
\label{sec:2.1}
In order to answer the research problems introduced in previous
chapter, we proposed a method based on model refinements, analysis,
and optimization. The general idea behind this approach is to bridge
the gap between requirements model and source code by defining model
transformations that progressively lower the abstraction level of
design models. Thus, from an abstract model provided by a CPS
architect, we propose to define and compose model transformations
which produce refined and optimal (or near optimal) architecture
models. Such transformations would, for instance, integrate design
patterns in the initial architecture model.

\begin{figure}[h!]
\centering
\includegraphics[width=0.9\textwidth, trim={0cm 8cm 0cm 0cm}, clip]{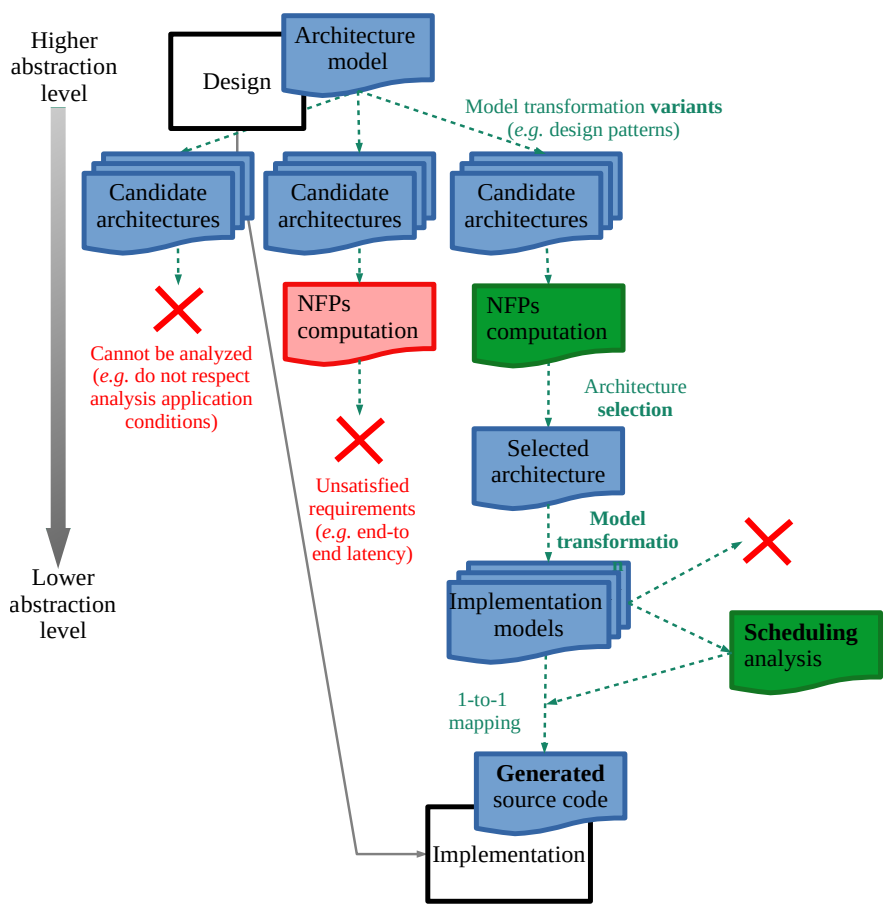}
\caption{Proposed refinement framework for CPS design models}
\label{fig:approach_overview}
\end{figure}

\noindent
We then analyze the resulting models in order to evaluate the impact
of the refinement on the system's NFPs and select architectures
answering at best the trade-off among NFRs. We continue this process
until we define implementation models, i.e. architecture models with a
straightforward correspondence between model elements and source code
(\emph{e.g.} a one-to-one mapping between each modeling element and a
construction in the underlying programming language and/or operating
system configuration). Figure~\ref{fig:approach_overview} illustrates
this approach in a two stages refinement process: the input
architecture model is refined into a set of architecture
candidates. These candidates, are, when possible, analyzed to compute
their NFPs and verified to check the design meets predefined NFR and
structural constraints. Architecture candidates satisfying predefined
NFRs and structural constraints are then selected and further refined
into a set of implementation models. Again, these models are analyzed
and the most appropriate model(s) is (are) used to automatically
generate the corresponding source code.

\noindent
By lowering the abstraction level of models used for both code
generation and analysis, we improve consistency between analysis
results and software implementation. More generally, we aim at
applying model transformations and analyze output models to control
the impact of these transformations on NFPs. This work contributes to
the resolution of problem 1.

\noindent
Because timing properties and in particular real-time tasks scheduling
plays an important role in CPSs design, we first experimented these
ideas so as to improve precision and/or reduce the pessimism of timing
analysis based on architecture models. This work contributes to the
resolution of problem 4.

\noindent
We also worked on model transformations formalization as a mean to
verify their correctness: we defined structural constraints on output
models and proposed a method to define model transformation chains
enforcing the respect of these constraints; we also defined a method to
validate model transformation chains by producing integration test
cases covering unit test requirements. This work contributes to the
resolution of problem 2.

\noindent
Last but not least, we proposed a design space exploration framework
combining multi-objectives optimization techniques and model
transformations composition in order to produce architecture models
answering at best a trade-off among NFPs. This work contributes to the
resolution of problem 3.

\noindent
Pursuing the objective to provide \emph{Refinement Techniques for
  Real-time Embedded Systems Architectures}, the work we present in
this document covers different research domains:
\begin{enumerate}
\item Real-time scheduling analysis,
\item Model Driven Engineering,
\item Design Space exploration,
\item Safety and Security.
\end{enumerate}

\noindent
We give a rapid overview of the state of art of these domains in next
section.

\section{Overview of the State-of-Art}
\label{sec:2.2}

Research areas mentioned at the end of the previous section cover a
huge number of relevant research works. In this section, we do not
seek for an exhaustive presentation of these areas. Instead, we aim at
providing a minimal background information to have our readers
understand the scientific context in which our research efforts were
undertaken.\\ \textbf{Architectures description Languages.}  Works in
the domain of architecture description languages aim at defining
modeling languages for software architectures, hardware architectures,
or a combination of both. Historically, research works in this area
have produced formalisms, which may be Modeling Languages,
Architecture Description Languages and/or Component-Based Modeling
languages. In the domain of CPS, different languages have been
proposed, such as
SCADE~\footnote{https://www.ansys.com/products/embedded-software/ansys-scade-suite},
MATLAB/SIMULINK~\footnote{https://www.mathworks.com/products/simulink.html},
LUSTRE~\cite{Halbwachs91thesynchronous},
Giotto~\cite{Henzinger:2001:GTL:646787.703890},
Polychrony~\cite{DBLP:journals/jcsc/GuernicTL03}, the Synchronous Data
Flows Graphs (SDFGs)~\cite{Lee1987SynchronousDF},
Wright~\cite{Allen97Thesis},
BIP~\cite{Bozga:2009:MSS:1629335.1629347}, ProCom~\cite{Bures1279},
the Palladio Component Model (PCM)~\cite{reussner2011a},
Fractal~\cite{Bruneton:2006:FCM:1152333.1152345},
UML/MARTE~\cite{OMG2009}, and the Architecture Analysis and Design
Language (AADL)~\cite{aadl04}. These languages differ in many aspects,
including the Model of Computation and Communication (MoCC) they
define. For instance, SCADE relies on the synchronous model of
computation in which computations and communications are assumed to
take zero time. This hypothesis is satisfied if the underlying
Computation Unit (CU) is fast enough to process input and produce
results before the acquisition of the next input. Relaxing the
synchronous hypothesis, Giotto defines a Logical Execution Time (LET)
model of computation where components, modeled as tasks, take a
predefined amount of time (the LET) to execute. Communications with a
component (\emph{i.e.}  inputs reading, outputs writing) can only
occur outside its LET interval (i.e. when the component is not
executed). In SDFGs, the focus is more on communications among
components than on their independent execution: applications are
described as a set of communication channels connecting applications
(also called actors, processes or tasks). These channels model
communication FIFO queues, and the execution of an application is
triggered by the content of its input queues. Last but not least, AADL
is a standardized modeling language aiming at gathering both a
representation of the software architecture, the hardware
architecture, and the binding of software components onto hardware
components.

\noindent
In our work, we consider source code generation as the final objective
of an efficient MDE process. In some industry, and in particular in
the transportation domain, several success stories show the added
value of automated source code generation techniques. MATLAB/SIMULINK
and SCADE SUITE provide source code generator widely used today. Note
that these code generators produce the so-called functional code,
\emph{i.e.} software responsible to answer functional
requirements. Another part of software applications for CPSs is called
technical code, \emph{i.e.} software responsible for interfacing
functional code with the hardware platform of the CPS.

\noindent
In this context, AADL is an interesting architecture description
language since it allows to represent both the software architecture,
the hardware architecture, and the allocation of software components
on hardware components. In addition, as illustrated on
figure~\ref{fig:approach_overview}, we aim at representing CPSs
architecture at different abstraction level. This is also a facility
offered by AADL, which was experimented during the PhD of Fabien
Cadoret (2010 - 2014). With respect to its usability in industrial
applications, AADL is a standard with a high visibility. This is an
important asset when it comes to experiment our work on industrial
case studies. Last but not least, AADL defines a MoCC which is
configurable thanks to standardized properties. For some subsets of
AADL configurations, the MoCC is deterministic and matches existing
formally defined languages (\emph{e.g.}  LET or SDFGs). In the PhD of
Fabien Cadoret and Roberto Medina, we identified subsets of AADL with
a deterministic and formally defined MoCC.

\noindent
Being advanced users of the language, we have also contributed to its
evolution by (i) providing regular feedback to the AADL
standardization committee, and (ii) by leading the revision of its
Behavior Annex: a sub-language of AADL dedicated to modeling components
behavior with state machines.
\newline
\textbf{Scheduling and analysis of real-time systems.} Among NFRs of
CPSs, timing requirements plays an important role. Indeed, one of the
specificities of CPSs is that they control physical systems. This
means CPSs implement control loops executed repeatedly with a
frequency that is derived from an analysis of the system's physics. If
the results of this control law are not produced in time, the physical
system does not wait. This is why, in real-time systems, the outputs
produced by a software function are valid if their computation is
correct, \emph{and} they are produced before a predefined
\emph{deadline}.
\newline
In order to ensure timing requirements are always satisfied, Liu and
Layland~\cite{journals/jacm/LiuL73} proposed to model software
applications with a set of tasks $\tau = \{\tau_i\}_{i=1..N}$
characterized by:
\begin{itemize}
\item a period $T_i$: the minimum delay between two consecutive
  executions of a task $\tau_i$.
\item a capacity $C_i$: the time required for the CU to execute task
  $\tau_i$. $C_i$ is usually set to the worst-case execution time
  (WCET) of $\tau_i$.
\item a deadline $D_i$: time interval between the release of $\tau_i$
  and the date at which $\tau_i$ must have finished its execution.
\end{itemize}
Since the definition of this very first task model, real-time systems
have been intensively studied and this research field has significantly
matured.
\newline
In particular, one important issue with the initial model introduced
Liu and Layland is the induced pessimism on tasks response time which
leads to a poor resource usage. Different sources of pessimism are
indeed cumulated when verifying tasks always meet their deadlines,
since the verification methods assume:
\begin{itemize}
\item tasks always execute all together for their worst case execution
  time;
\item if tasks share data is a protected access, they execute all
  together spending the longest possible time in all their critical
  sections, causing significant blocking times.
\end{itemize}
In practice, the probability that one task executes for its WCET is
low, so the probability they all together execute for their WCET is
very low. To overcome this limitation, methods based on
allowance~\cite{DBLP:conf/pdcn/BouguerouaGM04} or mixed-criticality
scheduling~\cite{Burns:2017:SRM:3161158.3131347} have been proposed.
\newline
In the PhD of Fabien Cadoret, we first proposed to use AADL in order
to implement fine-grain and less pessimistic response time
analysis~\cite{DBLP:conf/rsp/BordeRCPSD14}. The objective was to
reduce pessimism due to the presence of critical sections. We studied
this problem in the context of avionics systems using the ARINC653
standard: in this context, blocking time induced by inter-partition
communications is particularly significant. We also proposed a lock
free implementation for a deterministic
MoCC~\cite{DBLP:conf/isorc/CadoretRBPS13,DBLP:conf/adaEurope/JaouenBPR14}. More
recently, in the PhD of Roberto Medina, we have considered a MoCC for
which, by construction, data access protection is not needed: Directed
Acyclic Graphs (DAGs) of tasks. Using this well-known MoCC, we have
proposed new methods to schedule DAGs of mixed-criticality tasks.

\noindent
In this research work, we showed the capacity of AADL to model such
MoCCs. Even more significant, we used AADL to implement step-wise
architecture refinements and proceed to timing analysis at different
abstraction levels. This work has been integrated in the RAMSES
platform\footnote{https://mem4csd.telecom-paristech.fr/blog/}, an open
source AADL to AADL model transformation and code generation platform.
\newline
Building on the knowledge gained designing model transformations in
RAMSES, we developed new research activities dedicated to (i) compose
and validate model transformations, and (ii) implement model-based
design exploration techniques. We briefly introduce these methods in
next subsection.
\newline
\textbf{MDE and Model transformations for CPS.} Model transformations
are software applications taking model(s) as input and producing
model(s) as output. Even though model transformations can be written
in any programming language, they are by essence difficult to
write. Indeed, a model transformation consists in transforming a typed
graph into another typed graph. Writing and maintaining such
applications rapidly becomes
difficult~\cite{DBLP:conf/models/EtienABP12}. For these reasons,
dedicated model transformation
methods~\cite{Czarnecki:2006:FSM:1165093.1165106,
  Mens:2006:TMT:1706639.1706924} and languages~\cite{MTIP05,
  QVT-specification, Arendt:2010:HAC:1926458.1926471} and have been
proposed. In RAMSES, we have initially decided to use ATL as a
trade-off between the rigor of its semantics, and its simplicity of
use. In addition, we decided to decompose model transformations as a
chain of AADL to AADL transformations in order to enable verifications
on intermediate AADL models.
\newline
However, we use model transformations in the context of critical CPSs,
hence we have to pay extra attention to the validation of these
transformations. This is why we proposed to formalize model
transformations and more specifically model transformation chains.
\newline
In the PhD of Cuauhtémoc Castellanos, we proposed a formalization of
model transformations in Alloy~\cite{Jackson02}. From this
formalization, and the specification of constraints on the output model
(expressed with first order logic) we were able to automate the
production of model transformation chains to produce output models
satisfying the
constraints~\cite{DBLP:conf/euromicro/CastellanosBPVD14,
  DBLP:conf/euromicro/CastellanosBPSV15}.
\newline
In the PhD of Elie Richa, we proposed a formalization of ATL as
Algebraic Graph Transformations
(AGT)~\cite{boo:ehrigfundamentals-algebraic-graph-transformation}. An
automated mapping to Henshin, as well as the automated construction of
the weakest liberal precondition paved the way towards different kinds
of ATL transformations
verification~\cite{phd:poskitt2013verification-graph-programs,
  inc:pennemann2008development-correct-graph-transformation}. This
work was defined as part of an integration test case generation, but
it could also be used for proving model transformations.
\newline
In addition to our contributions on model transformations
formalization, we proposed to use model transformations as a medium
for design space exploration. We describe this work in next
subsection.
\newline
\textbf{Design space exploration.} As CPS have to meet conflicting
objectives with respect to their NFPs, lots of work have been
conducted to automate design space exploration for CPSs. In
particular, frameworks such as
ArcheOpterix~\cite{DBLP:conf/mompes/AletiBGM09},
PerOpteryx~\cite{PerOpteryx}, and AQOSA~\cite{AQOSA} are model-based
DSE frameworks in the sense that they rely on an input modeling
language (\emph{i.e.}  AADL, PCM) and they provide interfaces for
models analysis, optimization problems definition, and constraints
validation.
\newline
As an extension to these principles, we proposed in the PhD of Smail
Rahmoun to define design space exploration problems by composition of
model transformation variants~\cite{Rahmoun2015,Rahmoun2017}. The DSE
framework we developed relies on genetic algorithms, and our method
based on model transformations composition had the following
advantages: first, by transferring structural validity constraints from
the output model to the composition process, only valid architectures
are considered during the exploration process. We expressed validity
constraints with boolean formula and used SAT solving techniques to
ensure explored architecture satisfy the validity constraints. Second,
by defining model transformation composition techniques for DSE, we
keep the optimization framework (based on genetic algorithm)
completely generic: it only requires the definition of alternative
model transformations as the definition of a new optimization
problem. Following very similar ideas, the MOMoT framework was
designed in parallel~\cite{10.1007/978-3-319-42064-6_6}. This work,
as well as ours, fulfill the objective to make DSE generic. However,
as opposed to our work, this work does not proceed to the validation
of structural constraints during the composition of model
transformation rules but after their application. As shown
in~\cite{Rahmoun2017}, this would lead to a prohibitive loss of
performance on complex optimization problems.
\newline
\textbf{Safety and Security.} When defining model driven methods for
critical software applications, safety and security issues have to be
considered. Safety related concerns have always been considered in our
work, either by considering platforms (\emph{e.g.} operating
systems), standards (\emph{e.g.} for certification), or design
patterns (\emph{e.g.} triple modular redundancy) dedicated to improve
systems safety. More recently, we have started a PhD with Jean Oudot
aiming at defining quantitative evaluation methods for CPS security,
as well as architecture optimization methods for CPS security. We also
started another PhD with Maxime Ayrault aiming at defining design
methods and runtime mechanisms to improving the resilience of CPSs to
cyber attacks.

\section{Supervised PhD students}
\label{sec:2.3}

\begin{figure}[h]
\centering
\includegraphics[width=0.7\textwidth, trim={0cm 9cm 8cm 0cm}, clip]{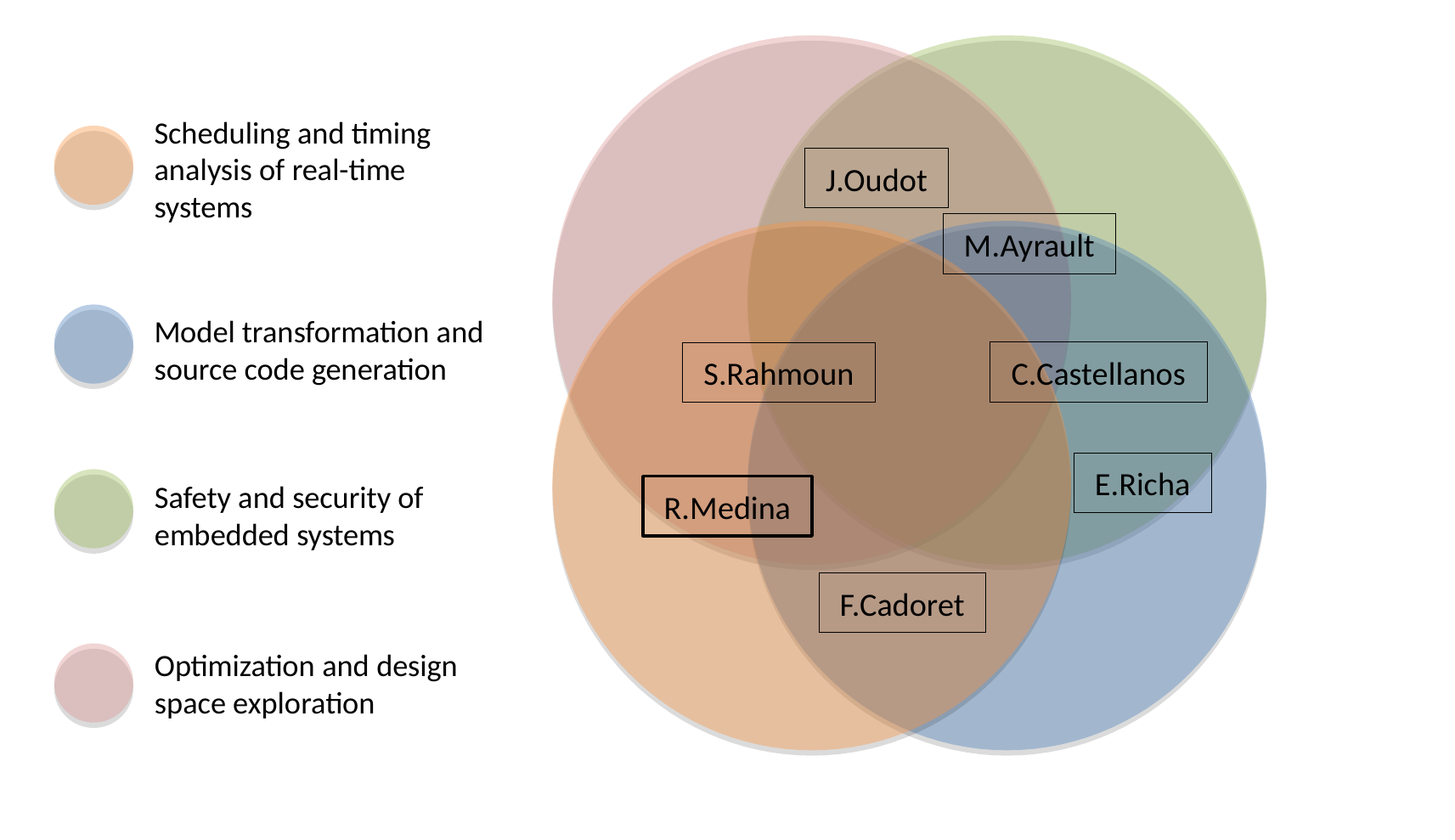}
\caption{Thematic distribution of co-supervised PhD students}
\label{fig:thematic_distribution}
\end{figure}

Fig.~\ref{fig:thematic_distribution} provides a rapid overview of the
PhDs undertaken with the aforementioned general objectives. In this
figure, PhDs have been placed with respect to the research areas they
contribute to.

These students have studied, with different viewpoints and objectives,
the problems presented in section~\ref{sec:2.3}:

\begin{itemize}
\item Fabien Cadoret (02/2010 - 05/2014): initial proposal of the
  approach presented on figure~\ref{fig:approach_overview}, with a
  focus on consistency between models used for schedulability analysis
  and models used for code generation (thus contributing to solve
  problem 1). Fabien Cadoret also studied a deterministic MoCC of
  AADL, for which he proposed lock-free implementation variants (thus
  contributing to solve problem 4).
\item Cuauhtémoc Castellanos (09/2011 - 12/2015) proposed to formalize
  model transformations into Alloy in order to ensure their
  composition always leads to analyzable models. With respect to the
  approach presented in figure~\ref{fig:approach_overview}, this work
  aimed at chaining model transformations in order to reduce the
  abstraction level of models used for analysis (contribution to solve
  problem 1) while making sure the model transformation is correct in
  the following sense: the model it produces respects the application
  condition of predefined analysis techniques (contribution to solve
  problem 2).
\item Elie Richa (01/2012 - 12/2015) also studied the correctness of
  model transformations (problem 2) with another viewpoint: how to
  ease the verification and validation of their implementation? In
  this PhD, we defined a test framework for model transformation
  chains, allowing to enforce unit test coverage (where a unit is a
  single model transformation) using integration tests (\emph{i.e.}
  tests on model transformation chains).
\item Smail Rahmoun (11/2013 - 02/2017) extended the work of Fabien
  Cadoret and Cuauhtémoc Castellanos by considering design space
  exploration based on model transformations as a multi-objective
  optimization problem aiming to improve NFPs. This extension is
  presented in the upper part of figure~\ref{fig:approach_overview},
  were candidate architecture are selected according to their
  NFPs. This work helped to automate the definition of near-optimal
  architectures, which is an important part of our initial goal: make
  the design process of CPSs more reliable by starting with the best
  possible design. In addition, the proposed approach was both applied
  on design patterns for safety (thus reusing and extending model
  transformations proposed by Cuauhtémoc Castellanos) and code
  generation (thus reusing and extending model transformations
  proposed by Fabien Cadoret).
\item Roberto Medina (11/2015 - 01/2019) studied more specifically the
  problem of resource usage efficiency in real-time embedded systems
  (problem 4). This is an important topic in our context as improving
  the reliability of software development in real-time CPS often leads
  to consider margins on tasks execution time. As a consequence,
  computation resources are poorly used whereas system designers aim
  at deploying more and more complex applications in CPSs. As a
  contribution to solve this problem, Roberto Medina proposed new
  scheduling techniques on multi-core architectures based on the
  concepts of Mixed-Criticality, applied to directed acyclic graphs
  of tasks. In particular, we defined in this work new methods to (i)
  ensure schedulability of high criticality functions and (ii) evaluate
  the impact of sharing computation resource among functions of
  different criticality levels on the quality of service of low
  criticality functions.
\item Jean Oudot (09/2017 - ) is working on the definition of
  quantification methods for cyber-security of CPSs. Indeed, cyber
  security is becoming an important problem in the design of CPS since
  these systems are becoming more and more connected to their
  environment. These interactions with their environment make CPSs
  subjects to cyber attacks with a more and more important surface
  attack. As opposed to traditional information systems or desktop
  computers, CPSs also have to meet stringent safety, performance,
  and/or energy consumption properties. For this reason, integrating
  security counter measures in CPS architectures raises important
  challenges in terms of multi-objective optimization problems. In
  particular, we aim at defining a methodology to select the set of
  sufficient counter measures to reach a level of acceptable risk
  while minimizing the impact of these counter measures on safety and
  performance properties. This work will contribute to the resolution
  of problem 3, with an emphasis on security counter measure selection
  and configuration.
\item Maxime Ayrault (10/2018 - ) is also studying cyber security of
  CPS but with a different viewpoint: how to improve their resilience
  to attacks? Indeed, it is impossible to anticipate all the potential
  vulnerabilities of a complex CPS. In addition, once a CPS is
  deployed and used, attackers have time to study the system and
  discover new vulnerabilities. For this reason, it is important to
  deploy resilience mechanisms in CPSs to delay as much as possible
  the effectiveness of an attack and/or its propagation. This problem
  is obviously connected to the design of CPS architectures since
  resilience mechanisms have to be defined at design time. This work
  is going to contribute to the resolution of problem 3, with an
  emphasis on resilience to cyber attacks.
\end{itemize}

I have been the advisor of the first 6 PhD students listed above, and
I am the supervisor of Maxime Ayrault's PhD.

In next chapters~(\ref{chap:chapter3} and~\ref{chap:chapter4}), we
describe more precisely these research contributions and their link
with the general approach presented on
figure~\ref{fig:approach_overview}.


\chapter{Architecture Models Timing Analysis}
\label{chap:chapter3}

\minitoc

\vspace{1.5cm}


CPSs are subject to various non-functional requirements, and timing
performance is an important class of such requirements when
architecting software applications of a CPS. In addition, Model Driven
Engineering (MDE) advocates for the use of models in order to improve
the development process, as well as the quality, of these
applications. In this chapter, we present a MDE framework aiming at
automating the production of software applications of CPSs.

\noindent
In particular, we consider source code generation as the final
objective of an efficient MDE process. In some industry, and
in particular in the transportation domain, several success stories
show the added value of automated source code generation
techniques. MATLAB/SIMULINK and SCADE SUITE provide source code
generator widely used today. Note that these code generators produce
the so-called functional code, \emph{i.e.} software responsible to
answer functional requirements. Another part of software applications
for CPSs is called technical code, \emph{i.e.} software responsible for
interfacing functional code with the hardware platform of the CPS.

\noindent
When it comes to the evaluation of timing performance in software
architectures, both functional and technical concerns can have a great
impact on the result. For instance, functional code is the main
software artifact used to compute the execution time of tasks,
\emph{i.e.} their capacity $C_i$ as defined in the previous
chapter. On the other hand, technical code may implement complex
communication mechanisms with a significant impact on timing
performance. This is particularly true in avionics ARINC653 systems
with inter partition communications. Nowadays, very few MDE frameworks
are able to consider both the functional and technical code of a
software application when evaluating their timing performances.

\noindent
In this chapter, we present our contributions on the timing analysis
of software architectures for CPSs.
The remainder of this chapter is organized as
follows. Section~\ref{sec:section_3.1} gives an overview of the MDE
framework we have designed to experiment our research activities. In
the following sections, we present subsets of the AADL modeling
language for which this framework was experimented: partitioned
ARINC653 systems in section~\ref{sec:section_3.2}, periodic delayed
queued communications in section~\ref{sec:section_3.3}, and DAGs of
mixed-criticality tasks in section~\ref{sec:section_3.4}.

\begin{figure}[ht!]
\centering
\includegraphics[width=0.9\textwidth, trim={0cm 7cm 0cm 0cm}, clip]{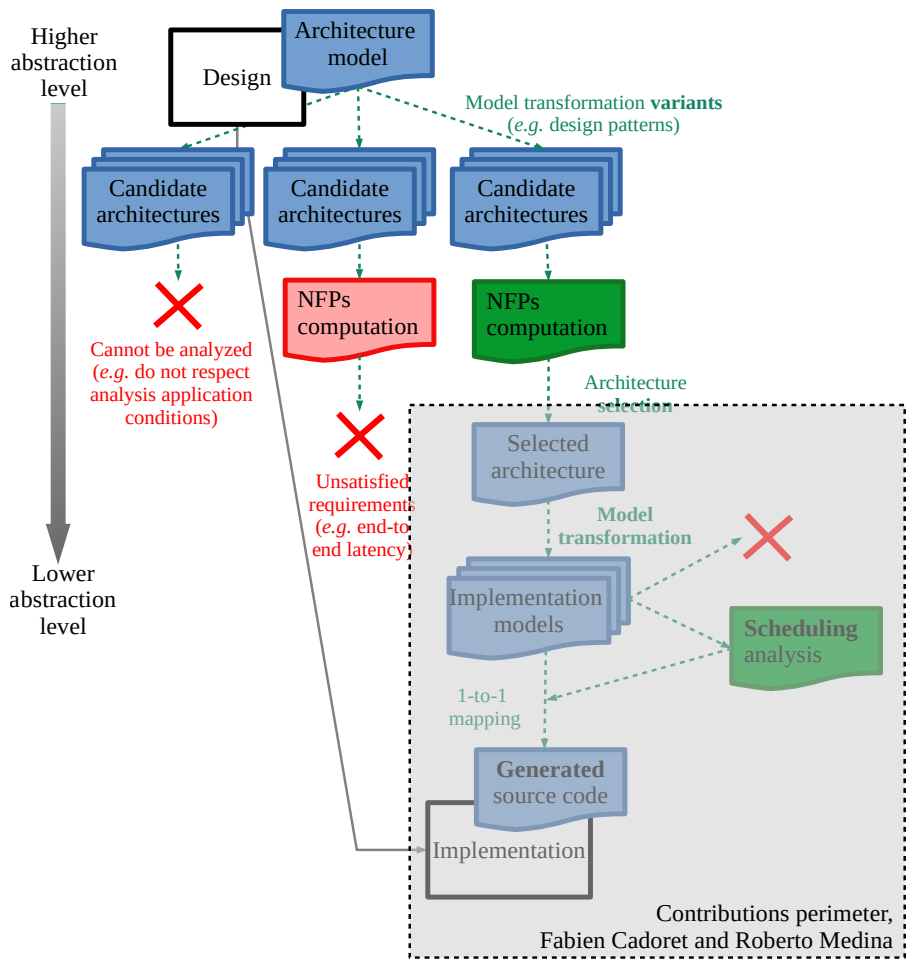}
\caption{Perimeter of research activities presented in this chapter}
\label{fig:approach_overview_part1}
\end{figure}

\noindent
Results presented in sections~\ref{sec:section_3.2}
and~\ref{sec:section_3.3} are results of Fabien Cadoret's PhD. Results
presented in section~\ref{sec:section_3.4} were obtained during
Roberto Medina's
PhD~\cite{MedinaBP17,MedinaBP18,medina2018scheduling}. Figure~\ref{fig:approach_overview_part1}
show how these contributions are positioned with respect to the
approach we described in chapter~\ref{chap:chapter2}.

\newpage

\section{AADL Refinement Framework}
\label{sec:section_3.1}

\begin{figure}[h]
  \centering
  \includegraphics[width=\textwidth, trim={0cm 0cm 0cm 0cm},
  clip]{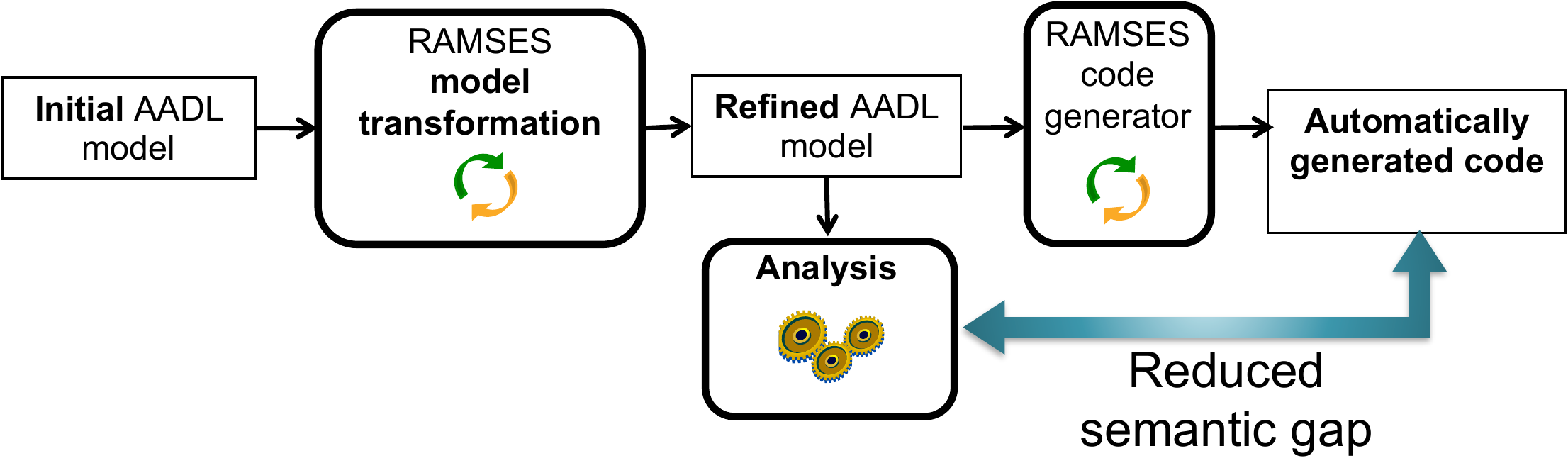}
  \caption{Overview of RAMSES functionalities}
  \label{fig:ramses_overview}
\end{figure}

In order to automate code generation for CPS while mastering the
impact of generated code on timing performance, we proposed in 2011 a
model transformation and code generation framework based on AADL,
called RAMSES~\cite{DBLP:conf/rsp/BordeRCPSD14}. The basic principles
of this framework are depicted on
figure~\ref{fig:ramses_overview}. The idea is to proceed to code
generation in a step-wise model transformation process which would (i)
exhibit the generated code into intermediate AADL models (\emph{e.g.}
refined AADL model on figure~\ref{fig:ramses_overview}), (ii) analyze
these intermediate models, until (iii) the AADL model reaches an
abstraction level leading to a very simple mapping from AADL to source
code constructions. One of the major benefit of this approach is to
reduce the semantic gap between models used for analysis purpose, and
models used for code generation per se.

\noindent
This framework has been the playground of several ideas and
experiments presented throughout this document. We describe those
related to timing analysis in the remainder of this chapter.

\section{Code generation and fine-grain analysis of partitioned systems}
\label{sec:section_3.2}

As mentioned in the introduction of this document, CPSs are often
critical systems. However, among software components of a CPS, only a
few are expected to have a high criticality level. As a consequence,
CPS designers have to provide safe methods to share computation and
storage resources among software components of different levels of
criticality.

\noindent
In the avionics domain, this problem has been solved by developing
dedicated fault containment mechanisms in operating systems. These
mechanisms are called time and space partitioning: applications are
statically provisioned with dedicated memory and execution time slots
and the operating system is in charge of enforcing the applications to
remain within these predefined slots. Partitioned operating systems
are known to ensure a good time and space isolation among software
applications they execute, to the price of a timing overhead in
communication mechanisms. This timing overhead is even more
significant when considering communications among different
partitions.

\noindent
\textbf{Input models description.} Following the general principles
described in section~\ref{sec:section_3.1}, we have proposed a method
to precisely take into account this overhead when verifying timing
requirements of a CPS. This method was first published at the
International Conference on Complex Computer
Systems~\cite{DBLP:conf/iceccs/CadoretBGP12} and specialized to a case
study from the real-time systems domain in a publication at the
international symposium on Rapid Systems Prototyping in
2014~\cite{DBLP:conf/rsp/BordeRCPSD14}. The proposed method takes as
input (i) the AADL model of applications deployed on a partitioned
system using the ARINC653 annex of AADL, and (ii) a behavioral
description of the runtime services of an ARINC653 system provider,
using the behavior annexe of AADL.

\noindent
The AADL model of applications is composed of a set of interconnected
processes, themeselves composed of interconnected tasks. In addition
to these structural characteristics, models may come with a
description of tasks internal behavior. The task set is expected to be
described with the following information:

\begin{itemize}
\item timing consumption of each subprogram or thread component:
  either as a timing interval (bounded by best and worst case
  execution time) for subprograms or threads, or timed behavior
  actions (in the behavior annex), or a set of properties that enable
  to compute such timing consumptions from the control flow graph of
  the components (\emph{e.g.} time of assignment actions, subprogram
  calls, expressions, etc.);

\item accesses to shared data, in order to describe which component
  has access to a shared data, when does it access it, and what is the
  access policy to be considered for schedulability analysis;

\item scheduling properties of the task set: scheduling protocol,
  periods, deadlines, and priorities (if needed, depending on the
  scheduling protocol).
\end{itemize}

\noindent
AADL models of the runtime services are provided by RAMSES, with the
support of operating systems provider. These models takes the form of
a library of AADL subprograms and data components definition. Their
behavior is described with the same elements as those described in
previous paragraph for threads description. Their timing
characteristics are supposed to be provided by operating systems
vendors.

\begin{figure}[h!]
  \centering
  \includegraphics[trim = 0mm 0mm 0mm 0mm, clip, width=0.7\linewidth]{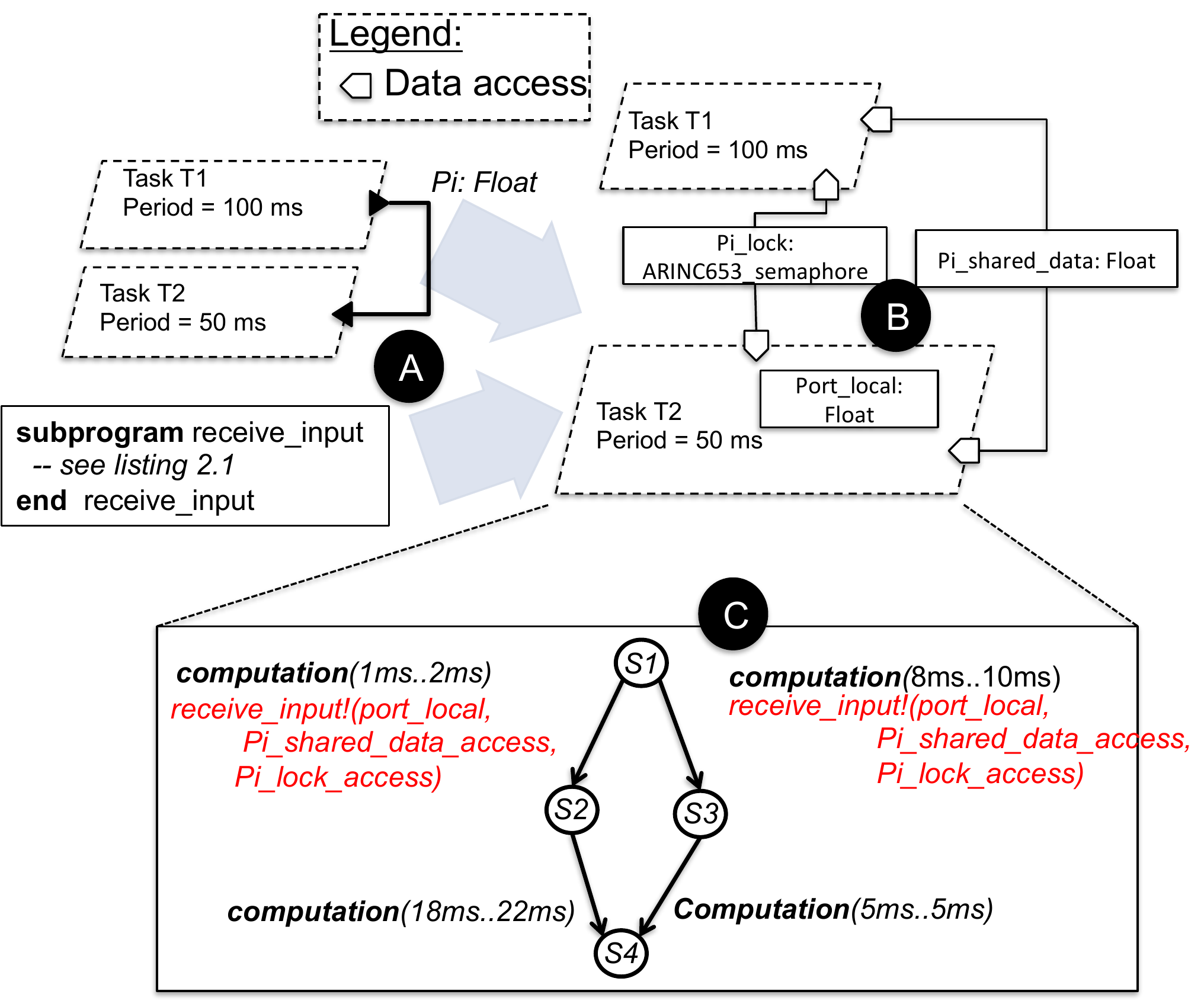}
  \caption{Model Refinement for Timing Analysis}
  \label{fig:refinement_principles}
\end{figure}

\noindent
\textbf{Model refinements in RAMSES.}
Figure~\ref{fig:refinement_principles} illustrates the principles of
the architecture refinement implemented in RAMSES in order to provide
fine-grain schedulability analysis.

On the part \textbf{A} of the figure, we represented a summary of the
input models : the architecture model on top, provided by an end-user
of the framework, and the runtime services on the lower part, provided
by an operating system vendor. As specified on the figure, the AADL
code of the receive\_input runtime service is provided in
listing~\ref{lst:middleware}. Part \textbf{B} of the figure represents
the result of the refinement implemented as an automatic model
transformation in RAMSES. This transformation expands abstract
communication interfaces (\textit{i.e.}  AADL ports) into data
accesses and subprogram calls. Data accesses enable threads to write
or read the value of shared variable $Pi\_shared\_data$, which
contains data exchanged between threads through these ports. Global
variable $Pi\_lock$ enables to protect accesses to
$Pi\_shared\_data$. Subprogram calls are represented on part
\textbf{C} of the figure (see $receive\_input!(...)$).



Listing~\ref{lst:middleware} shows the AADL model of the
$receive\_input$ subprogram of the runtime services we use to
implement communications between AADL threads.
In addition, this listing provides the definition of a processor
component which according to the AADL standard, is an abstract
execution platform that represents both the hardware execution unit,
and the operating system running on it. The reason for modeling the
processor at this stage is that execution times of threads and
subprograms obviously depend on the processor they are executed on.

\begin{figure}[h!]
\begin{lstlisting}[basicstyle=\scriptsize,captionpos=b,caption=AADL
    Runtime Services
    Component,label=lst:middleware,frame=single,xleftmargin=\parindent,language=aadl,numbers=left,
  escapeinside={/*}{*/}]
subprogram receive_input
features
  value_out: out parameter Float; /*\label{feature2}*/
  data_storage: requires data access Float; /*\label{feature3}*/
  lock_access: requires data access ARINC653_semaphore; /*\label{feature1}*/
annex behavior_specification {** /*\label{ba_start}*/
  states
    s1: initial final state;
  transitions
    t1: s1 -[]-> s1
    {
      computation(1 ms .. 2 ms) in binding (x86); /*\label{abst_compute}*/
      lock_access!<; /*\label{lock}*/
      value_out := data_storage; /*\label{assignment}*/
      lock_access!>; /*\label{unlock}*/
    }
**};
end send_output;

processor x86
properties
  Assign_Time => [Fixed => 0us; Per_Byte => 50 us]; /*\label{assignment_time_property}*/
end x86;
\end{lstlisting}
\end{figure}

\noindent
From a timing analysis viewpoint, this model contains the following
information:

\begin{itemize}
\item A computation statement describes that an execution time
  interval of one to two milliseconds is necessary at the beginning of
  the execution of this subprogram (see line \ref{abst_compute} of
  listing \ref{lst:middleware}). Note that these timing
  characteristics are only valid when the subprogram is executed on
  the \textit{x86} processor (modeled in the same listing) as
  specified by the \textit{``in binding''} statement line
  \ref{abst_compute}.

\item Execution time of subprogram $receive\_input$ can also be
  deduced from the assignment action line \ref{assignment}, combined
  with (i) the data size of operands of the assignment and (ii) the
  assignment time property given in line
  \ref{assignment_time_property} of the listing.

\item Data accesses are represented in lines \ref{lock} and
  \ref{unlock}, with respectively a locking and unlocking access to
  shared data that will be connected to interface
  \textit{lock\_access} (line \ref{feature1}).
\end{itemize}

In order to analyze the refined model illustrated on parts \textbf{B}
and \textbf{C} of figure~\ref{fig:refinement_principles}, at least
three alternatives exist:
\begin{enumerate}
\item transform the intermediate model into a formal model to apply
  model checking techniques. Given its features,
  TIMES~\cite{10.1007/978-3-540-40903-8_6} would be a good candidate
  but to the best of our knowledge, it does not cover hierarchical
  scheduling (which is an important feature of ARINC653
  systems). Another possibility could be to use more generic formal
  models such as timed automata or timed Petri nets. However, the
  translation of AADL to such models is a difficult task for which
  different research works were already
  undertaken~\cite{DBLP:conf/adaEurope/BerthomieuBCDFV09,Renault:2009:AMM:1590961.1591431}. These
  works only cover a subset of AADL which is not he one considered in
  our work.
\item transform the intermediate model into a single task set with the
  following characteristics: each task is given for its capacity its
  WCET, and each critical section of each task is characterized by its
  WCET as well. The resulting model is simple to analyze with tools
  such as Cheddar~\cite{Singhoff:2004:CFR:1032297.1032298} but it may
  cumulate pessimism (and thus waste of computation
  resources). Indeed, when a job executes for its task's WCET, it may
  spend little time in its critical section, and vice versa (when a
  job enters a critical section's WCET, it may spend little time in
  the task itself).
\item transform the intermediate model into a set of task sets: the
  control flow graph of each task of the intermediate model is
  transformed into an execution tree (going from one suspended state
  of the task to another suspended state). Among the branches without
  accesses to locks, we only keep the one with the highest execution
  time. Branches with accesses to locks are kept as is in the tree.
  Task sets are then built from execution trees by applying a
  cartesian product of the set of execution branches of each
  task. Then, each task set is simple to analyze with tools such as
  Cheddar~\cite{Singhoff:2004:CFR:1032297.1032298}.
\end{enumerate}

\noindent
\textbf{Case study.} We experimented the latest alternative
in~\cite{DBLP:conf/rsp/BordeRCPSD14}. The number of task sets to
analyze grows rapidly with the number of branches in tasks execution
trees. The number of task set configurations to analyze mainly depends
on the characteristics of the input model: the number of configuration
to analyze grows with the number of conditional branches in which
shared data are acquired and released.

\begin{figure}[ht!]
  \centering
  \includegraphics[trim = 12mm 5mm 5mm 10mm, clip, width=\linewidth]{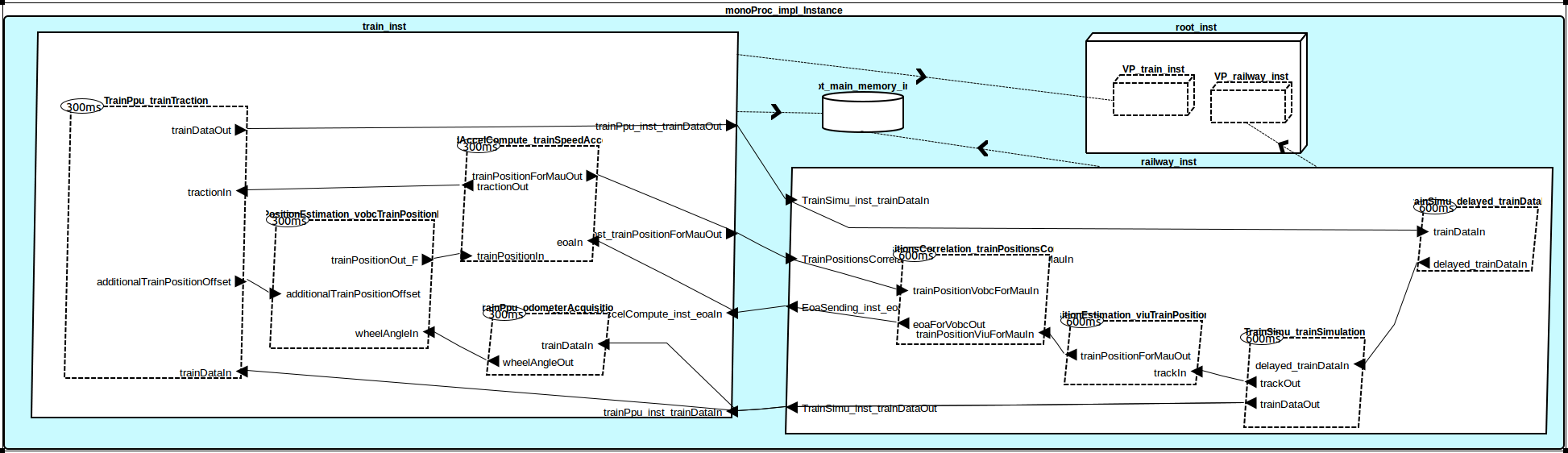}
  \caption{AADL Model of our Case Study}
  \label{fig:case_study}
\end{figure}

\noindent
Figure~\ref{fig:case_study} illustrates the AADL architecture of a
case study from the train industry. In this domain, the main business
objective is to reduce the time interval separating two consecutive
trains while guaranteeing passengers safety. To reach this objective,
the adaptation of powerful CU gives the opportunity to embed more
computation power on-board trains. Functions traditionally deployed on
the wayside infrastructure can then be embedded on-board in order to
reduce response- time of functions. Trains may then be closer to one
another by depending less on the wayside infrastructure. However,
grouping functions on-board the train should not lead to hardware
resources over-consumption otherwise the safety of the system may be
put at risk.

\noindent
Because train application are also made up of components of different
criticality levels, partitioned operating systems are also studied
for future architectures of these
applications. Figure~\ref{fig:case_study} illustrates the AADL
architecture of a simplified application called Communications-Based
Train Control (CBTC). This application is decomposed in two processes:
the Automatic Train Operation (ATO) process, represented on the left
of figure~\ref{fig:case_study}, is responsible for controlling the
position, speed, and acceleration of the train.  The other process,
called Automatic Train Protection (ATP), is represented on the right
of figure~\ref{fig:case_study}: it communicates with the ATO in order
to check the validity of data computed by the ATO.

\noindent
Our objective is to ensure CUs provide enough computation power to
host both processes. The software architecture represented
configure~\ref{fig:case_study} is made up of two AADL processes, eight
AADL threads (four threads in each process), and twelve connections
among ports of these threads.

\noindent
For the case study presented in figure~\ref{fig:case_study}, the
timing analysis of our case study required the analysis of 64 tasks
configurations (different values for the WCET of tasks and the WCET in
critical sections). These configurations were analyzed using the
Cheddar tool
suite~\cite{Singhoff:2004:CFR:1032297.1032298}. Experimentations were
conducted on a 2.7 GHz Intel processor (Intel Core i7-3740QM; 4 cores)
with 3.9 GiB memory and a SSD hard drive disk. The complete process,
from the beginning of the model refinement, until the compilation of
generated code, passing by the analysis of 64 tasks configuration took
2 minutes and 17 seconds. Considering the complexity of the input
architecture, this result seems to be very satisfactory: of course,
the number of configuration to analyze can grow very fast by
increasing the complexity of the input model, but the analysis of
every single configuration is the price to pay for an exhaustive
analysis.

However, to limit the complexity of timing analysis, another strategy
is to consider more deterministic MoCCs and to propose lock-fee
implementations of these MoCCs. We present our work related to such
techniques in the remainder of this chapter.

\section{Periodic delayed communications}
\label{sec:section_3.3}

\textbf{MoCC presentation.} The MoCC we consider in this
section is a variant of message passing communications among periodic
tasks:
\begin{itemize}
\item Communication channels are modeled by directed ports and
  connections to enable various configurations regarding the number of
  sender and receiver.
\item A task $\tau_i$ can receive a set of messages on its input
  ports.
\item A task $\tau_j$ can send a message on its output ports to
  connected input ports.
\item A message sent on an output port p, is eventually received on
  input ports connected to p.
\end{itemize}
We refined this model to ensure deterministic communications among
tasks:
\begin{itemize}
\item During each job $J_j$ of a task $\tau_j$, \textbf{exactly one
  message is sent} on each output port of a task.
\item A message sent by a job is delivered to the receiving task at
  the \textbf{recipient release time following the sender job
    deadline}. More formally, a message sent to $\tau_i$, by the
  $k^{th}$ job of $\tau_j$ is considered delivered at $\lceil \frac{k
    \cdot T_i+D_j}{T_i} \rceil \cdot T_i$ (remember that $T_i$ and
  $D_i$ are respectively the period and deadline of task $\tau_i$, as
  defined in notations used in chapter~\ref{chap:chapter2}).
\item Any message delivered to the $k^{th}$ job of $\tau_i$ should be
  removed from the receiving port at time $k \cdot T_i + D_i$. After
  this time, delivered messages to the $k^{th}$ job of $\tau_i$ are
  considered outdated.
\item Messages delivered to a task are received in the order of sender
  jobs deadlines. When sender jobs deadlines are simultaneous, a
  predefined order noted $\prec$, e.g. task priorities, is used.
\end{itemize}
The model is said ``periodic-delayed'' as messages are periodically
sent and their delivery is delayed until sender job deadlines. Such a
communication model can be modeled in AADL with the following
properties :
\begin{itemize}
\item The \emph{Dispatch\_Protocol} property is set to \emph{Periodic}
  for each thread component: tasks are periodic,
\item The \emph{Period}, and \emph{Deadline} properties are set for
  each thread component (with $Deadline \leq Period$),
\item The \emph{Timing} property is set to \emph{Delayed} for output
  ports of tasks: messages are sent at deadline.
\end{itemize}
Note that the default value of the AADL \emph{Output\_Rate} property
already states that one message is produced per activation of the
producer tasks. Similarly, we use \emph{AllItems} as the default value
for the property \emph{Dequeue\_Protocol}, which means that all the
messages available at release time of the recipient will be considered
as consumed at the end of its job.

\textbf{Lock-free implementation.} In next paragraphs, we show how to
compute message indexes for sent received messages order to implement
these action without locks. The number of received messages at time
$t$ on a queue $q$ of size $Q$ can be computed as follows:
$$Received(q, t) = \sum_{j \in ST_q} \lfloor \frac{t - D_j}{T_j} \rfloor + 1$$
where $ST_q$ is the set of tasks sending messages to $q$.The indexes
of sent values can be computed as follows for queue $q$ and the
$k^{th}$ job of a sender task $\tau_j$:
$$SendIndex(q,j,k) = Redeived(q, k\cdot T_j + D_j) - Followers (q,j,k)$$
where $Followers$ is the number of successors of $\tau_j$ (according to
$\prec$) in $ST_q$ having their deadline at $k \cdot T_j + D_j$. More formally:
$$Follower(q,j,k) = \sum_{s \in ST_q, j \prec s} Collide(s,k \cdot T_j + D_j)$$
where Collide is defined as follows
\[   
Collide(s,t) = 
     \begin{cases}
       \text{1} &\quad\text{if } \frac{t-D_s}{T_s} \in \mathbb{N}\\
       \text{0} &\quad\text{otherwise} \\
     \end{cases}
     \]
Hence, the message sent in $q$ by the $k^{th}$ job of $\tau_j$
($\tau_j$ is the task sending messages in $q$) is stored in slot
$SendIndex(q, i, k)\ modulo\ Q$ ($Q$ is the size of $q$). Besides,
received messages range from $(ReadIndex(q_r, k-1)+1)\ modulo\ Q$ to
$ReadIndex(q, k)\ modulo\ Q$ where $ReadIndex(q, k) = Received(q, k
\cdot T_i)$ for the task $\tau_i$ receiving messages from $q$.
Note that $SendIndex(q,j,k)$ and $ReadIndex(q, k)$ can be computed
independently without any internal state, reason why \textbf{lock free
  implementations} of these functions are possible.
Finally, the size of $q$ can be bound as follows (the proof of this
result is available in~\cite{DBLP:conf/isorc/CadoretRBPS13}):
$$Q \leq \sum_{j \in ST_q} (\lfloor \frac{2 \cdot T_q + D_{max}}{T_j}\rfloor + 1)$$

\begin{figure}[h!]
  \centering
  \includegraphics[trim = 0mm 0mm 0mm 0mm, clip, width=\linewidth]{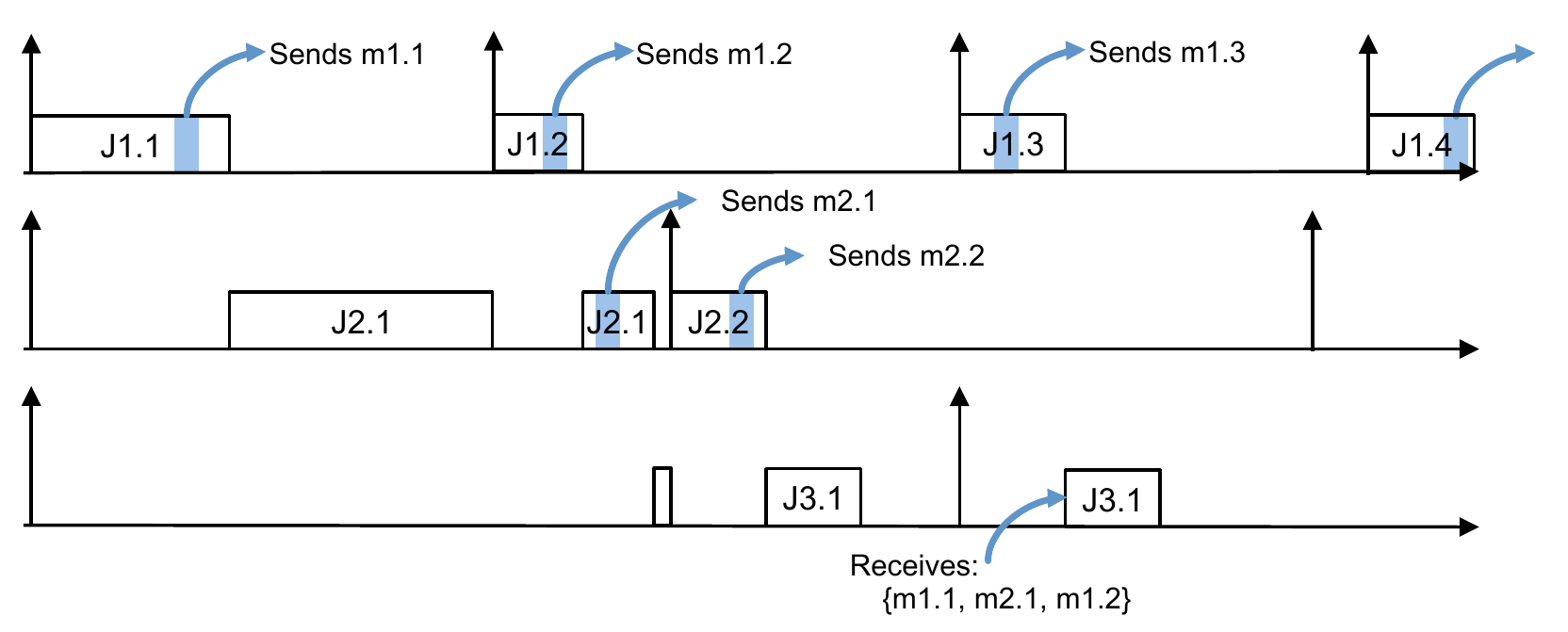}
  \caption{Illustrative Example of periodic-delayed communications}
  \label{fig:pdp_illustration}
\end{figure}

\noindent
\textbf{Illustrative example} To illustrate this task and
communication model, we consider the time-line depicted in
Figure~\ref{fig:pdp_illustration}.
This figure shows communications between three tasks: $\tau_i$ with $i
= 1..3$, $T_i = D_i$, and $T_1 = 5$, $T_2 = 7$, $T_3 = 10$. $\tau_1$
and $\tau_2$ send periodically messages to $\tau_3$ according to the
communication model described above. As illustrated on this figure,
exactly one message is sent during each task job for $\tau_1$ and
$\tau_2$. Note that messages m1.1, m2.1, and m1.2 are only delivered
at time 10. Message m2.2 will not be delivered before 20 time units,
even though job J2.2 (that produces m2.2) already finished its
execution when J3.2 starts in this scenario. This model allows
ensuring deterministic time for message reception independently of
task interleaving and actual execution time. Notice also, that
messages are ordered with respect to sender deadlines, reason why m2.1
is put before m1.2 in the queue even though J1.2 finishes before J2.1
sends message m2.2. This is done to enforce a deterministic order on
message from the point of view of the receiver. Finally, m1.1, m2.1,
and m2.2 are discarded at completion time of J3.2 even if these
message were not used during this job.

\noindent
\textbf{Discussion.} With respect to timing analysis of architecture
models, periodic delayed communications bring the advantage of being
deterministic and can therefore be implemented without locks. As a
consequence, tasks with periodic delayed communications can be
considered as independent tasks, which greatly simplifies timing
analysis. For instance, it reduces the complexity of timing analysis
induced by multiple critical sections (as presented in
section~\ref{sec:section_3.2}). However, these lock free
implementations require to store data structures and execute functions
to retrieve and/or compute the indexes of sent or received
messages. This is why the refined model produced by RAMSES is
important: it allows to check platform resources are still sufficient
even when taking into account the overhead due to the implementation
of communication mechanisms. Last but not least, delayed
communications tend to increase data flow latency, \emph{i.e.}  the
time range separating the reception of inputs from sensors to the
production of commands to actuators in a CPS. This is one of the
reasons why we decided to consider task models made up of DAGs of
tasks. Another reason was the necessity to provide solutions for
scheduling mixed-criticality task sets on multi-core
architectures. This work is presented in next section. Note however,
that these two MoCCs (\emph{i.e.} DAGs and periodic-delayed) are
complementary: periodic delayed communications are often used to break
cycles in tasks dependencies while preserving deterministic MoCC and
lock free implementations. Last but not least, scheduling real-time
DAGs is known to be a difficult problem and its adaptation to mixed
criticality scheduling required a PhD thesis on its own.

\section{Mixed-criticality DAGs analysis}
\label{sec:section_3.4}

The contributions presented in section~\ref{sec:section_3.2} aim at
automating the analysis of tasks sets with critical sections. On
hypothesis of this work was that tasks are executed on mono-core
architectures. The results presented in section~\ref{sec:section_3.3}
show how to implement lock free communications among periodic
tasks. Given the communication model we considered, these results can
be used when scheduling tasks on multi-core architecture. Tasks would
then be considered as independent, which greatly ease the application
of scheduling techniques. However, as stated at the end of previous
section, this MoCC induces important latency on data flows.

\noindent
\textbf{Context.}
A MoCC frequently used to model critical embedded systems consists of
data flow graphs. This model defines \emph{actors} that communicate
with each other in order to make the system run: the system is said to
be \emph{data-driven}. The actors defined by this model can be tasks,
jobs or pieces of code. An actor can only execute if all its
predecessors have produced the required amount of data. Therefore,
actors have \emph{data-dependencies} in their execution. Theory behind
this model and its semantics provide interesting results in terms of
logical correctness: deterministic execution, starvation freedom,
bounded latency, are some examples of properties that can be formally
proven thanks to data-flow graphs.

\noindent
In this work we have considered a simple subclass of data-flow graphs
in which data dependencies are directly captured into Directed Acyclic
Graphs of tasks. In this MoCC, a software architecture is made up of
DAGs in which vertices represents tasks, and edges represent
precendence constraint, \emph{i.e.} a task can only start executing
when all its predecessors have finished their execution. In parallel,
the adoption of multi-core architectures in the real-time scheduling
theory led to the adaptation and development of new scheduling
policies~\cite{davis2009priority}. Processing capabilities offered by
multi-core architectures are quite appealing for safety-critical
systems since there are important constraints in terms of power
consumption and weight.  Nonetheless, this type of architecture was
designed to optimize the average performance and not the worst
case. Therefore, ensuring time correctness becomes harder when
multi-core architectures are considered: in hard real-time systems the
Worst Case Execution Time is used to determine if a system is
schedulable.

\noindent
This observation is one of the main reason for the popularity of the
mixed-criticality scheduling (MCS), intensively studied these last
years~\cite{Burns:2017:SRM:3161158.3131347}. With MCS, tasks can be
executed in different execution modes: in the nominal mode, high and
low criticality tasks are both executed with an optimistic timing
budget. When the system detects a timing failure event (TFE),
\emph{i.e.} a task did not complete its execution within its
optimistic timing budget, the system switches to a degraded mode. In
this mode, high criticality tasks are executed with their pessimistic
timing budget, discarding~\cite{Vestal:2007:PSM:1338441.1338659} low
criticality tasks or degrading them~\cite{Su:2016:EMT:3029795.2984633}
(\emph{i.e.}  reducing their execution frequency).

\textbf{Overview of the work.}
The schedulability problem of real-time tasks in multi-core
architectures is known to be NP-hard. When considering
mixed-criticality multi-core systems, the problem holds its
complexity. Thus, in our contributions we have designed a
meta-heuristic capable of computing scheduling tables for the
execution of Mixed-Criticality DAGs (MC-DAGs). The reason for choosing
scheduling tables is simple: it is known to ease the certification of
critical systems and the ARINC653 scheduling of partitions (using
schedule table to enforce temporal isolation of partitions) is a good
example of this statement.

In addition to this meta-heuristic, we have proposed a method to
evaluate the availability of lower criticality tasks. Indeed, the
initial objective of MCS is to improve computation resources usage by
allowing to configure the system with lower timing budgets than tasks
WCET. In practice, this is only possible if low criticality tasks are
degraded whenever high criticality tasks need more computation
resources. This impacts the quality of low criticality services. We
proposed to measure this impact in terms of availability.

\begin{figure}
  \centering
  \includegraphics[width=0.5\textwidth]{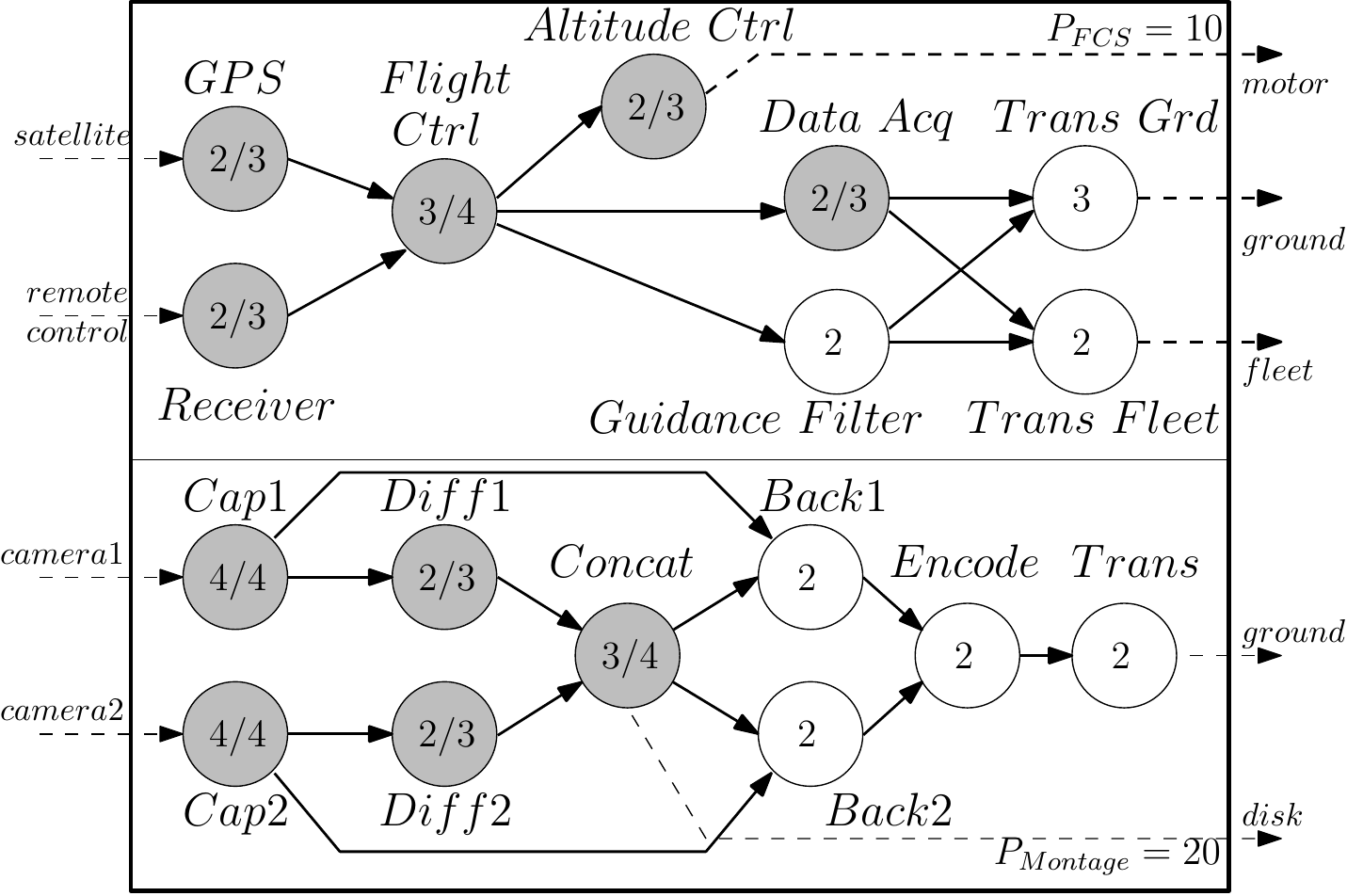}
  \caption{UAV software architecture with two MC-DAGs}
  \label{fig:mcdags}
\end{figure}

\textbf{Illustration.} Figure~\ref{fig:mcdags} illustrate the
structure of a MC-DAG on a motivating example: an UAV for field
exploration. The UAV is composed of two MC-DAGs: the first one takes
care of the Flight Control System (FCS), noted
$G_{FCS}$~\cite{siebert2014mobile}. The second MC-DAG represents a
scientific workflow used for image
processing~\cite{bharathi2008characterization}, noted
$G_{Montage}$. Vertices in gray represent high criticality tasks,
while white vertices are low criticality tasks. Vertices are annotated
with their timing budgets: a single value is given for low criticality
tasks since they are not executed in the high criticality mode. Full
edges represent precedence constraints between tasks, while dashed
edges represent the interface with the system's boundaries: where data
is initially coming from or finally sent to. The idea behind this
motivating example is to demonstrate that the FCS could be executed
next to an image processing workflow on a tri-core architecture.

In the remainder of this section, we present our two main
contributions on this task model: we first present the scheduling
method we proposed, before to explain how we compute the availability
of low criticality tasks. For the sake of simplicity, we present these
results on a dual criticality system with a LO and HI modes
(\emph{i.e.} respectively low and high criticality modes) but we
provided more general results (for any number of criticality levels)
in Roberto Medina's PhD thesis.

\textbf{ALAP implementations of MC-Correct schedules.} To begin with,
we shall recall the definition of a MC-correct scheduling for DAGs, as
defined in~\cite{baruah2016federated}:

\begin{mydef}
  A \textbf{MC-correct} schedule is one which guarantees
  \begin{enumerate}
  \item \textbf{\condlo}: If no vertex of any MC-DAG in
    $\mathcal{G}$ executes beyond its $C_i(LO)$ then all the vertices
    complete execution by the deadlines; and
  \item \textbf{\condhi}: If no vertex of any MC-DAG in
    $\mathcal{G}$ executes beyond its $C_i(HI)$ then all the vertices
    that are designated as being of HI-criticality complete execution by their 
    deadlines.
  \end{enumerate}
  \label{def:mccorrect}
\end{mydef}

In the Real-Time Systems Symposium, 2018, we proposed a sufficient
conditions to guarantee \condhi{} of MC-correct
scheduling~\cite{medina2018scheduling}:

First, for each task $\tau_i$ executing in mode $\chi$, we define the
function $\psi_i^{\chi}$ as follows:
\begin{equation}
  \psi_i^{\chi}(t_1, t_2) = \sum_{s = t_1}^{t_2} \delta_i^{\chi}(s).
  \label{eq:psi}
\end{equation}
\noindent
where
\begin{equation}\nonumber
  \delta_i^{\chi}(s) = \begin{cases}
    1 & \text{if } \tau_i \text{ is running at time } s \text{ in mode 
    } \chi,\\
    0 & \text{otherwise}
  \end{cases}.
\end{equation}

\noindent
This function defines the execution time allocated to task $\tau_i$ in
mode $\chi$ from time $t_1$ to time $t_2$.

\begin{mydef}
  \textbf{Safe Transition Property}
  \begin{equation}
    \psi_i^{LO}(r_{i,k}, t) < C_i(LO) \Rightarrow 
    \psi_i^{LO}(r_{i,k}, t) \geq \psi_i^{HI}(r_{i,k}, t).
    \label{eq:condSwitch}
  \end{equation}
\end{mydef}

\noindent
As one cans see in equation~(\ref{eq:condSwitch}), \condmc{} states
that, while the $k$-th activation of HI task $\tau_i$ has not been
fully allocated in LO mode, the budget allocated to this job in LO
mode must be greater than the one allocated to it in HI
mode. Intuitively, this guarantees that whenever a TFE occurs, the
final budget allocated to the job of $\tau_i$ is at least equal to its
WCET in HI mode.

\begin{figure}[ht!]

  \subfloat[Scheduling table in LO mode (ASAP)] {
    \centering
    \includegraphics[width=\linewidth]{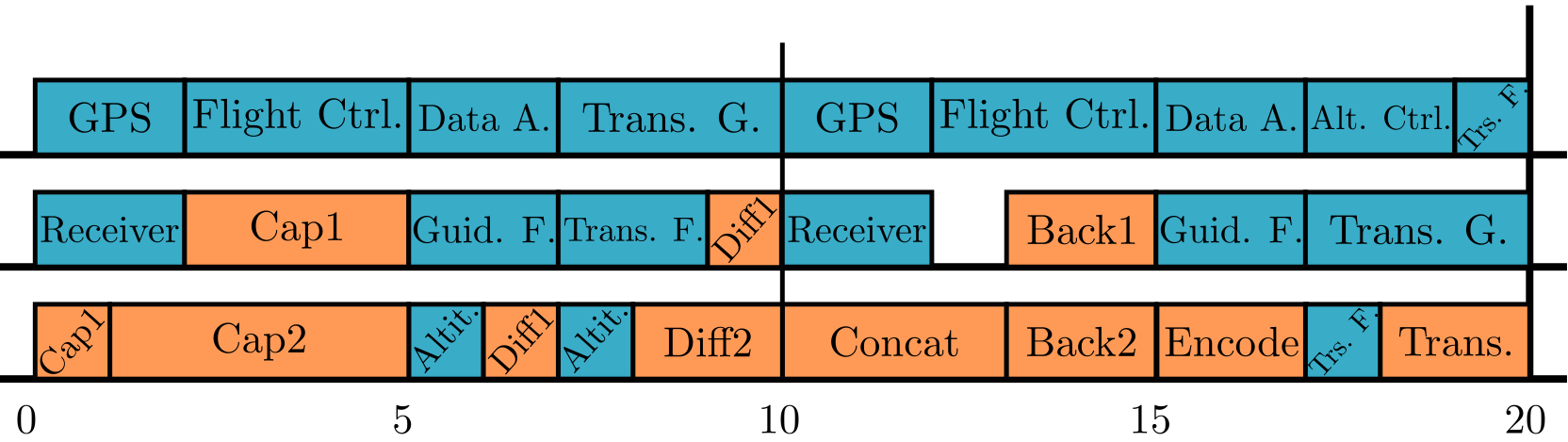}
    \label{fig:sHIf0}
  }\\
  \subfloat[Scheduling table in HI mode (ALAP)] {
    \centering
    \includegraphics[width=\linewidth]{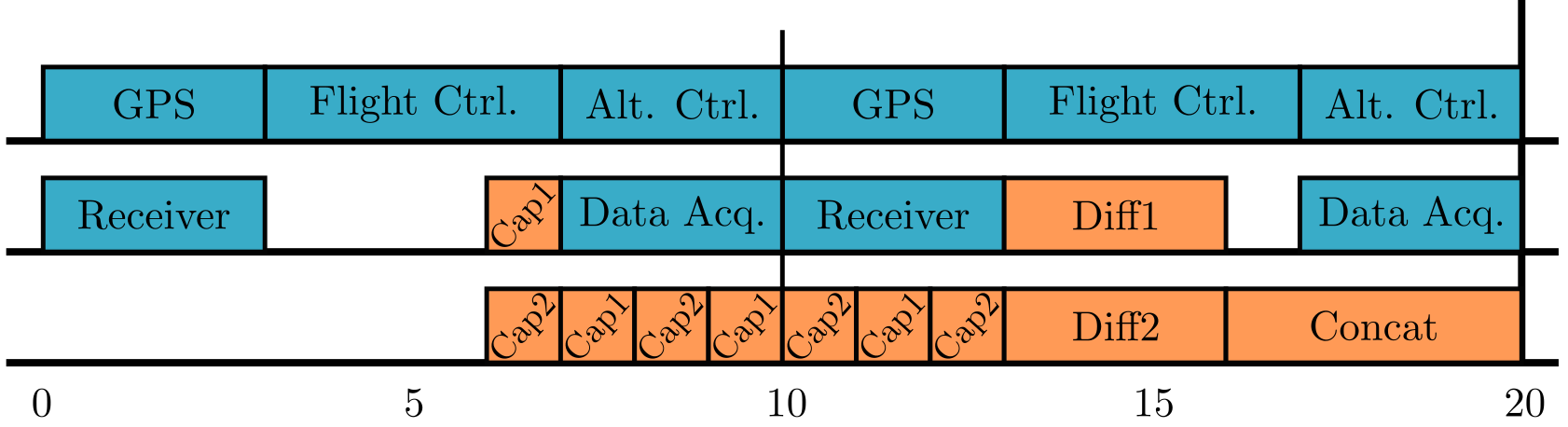}
    \label{fig:sHIf1}
  }
  \caption{Scheduling tables for the UAV, using G-LLF and \condmc{}}
  \label{fig:UAV_scheduling}
\end{figure}

\noindent
Building on the definition \condmc{}, we proposed a meta-heuristic to
build MC-correct schedules. We also proposed several implementations
of this meta-heuristic, based on G-EDF\footnote{Global Earliest
  Deadline First} or G-LLF (Global Least Laxity First): two well known
global schedulers appreciated for their performances. In our
implementations of schedulers for MC-DAGs, G-EDF or G-LLF are used to
assign priorities to tasks. In addition, we proposed to improve the
performance of MC-DAG schedulers by executing tasks As Late As
Possible (ALAP) in HI mode, and As Soon As Possible (ASAP) in LO
mode. Executing tasks ALAP in HI mode, we expect to free execution
slots close to tasks activation, usable by tasks executed ASAP in LO
mode. Figure~\ref{fig:UAV_scheduling} illustrates the scheduling
tables obtained with our MC-DAG scheduler, using G-LLF and enforcing
the respect of \condmc{}.

\noindent
\textbf{Evaluation framework.} A systematic evaluation of the
scheduling methods we proposed was performed during the PhD of Roberto
Medina.
We implemented the \galap{} and \textsc{G-alap-EDF} algorithm in an
open-sourced framework\footnote{MC-DAG Framework -
  https://github.com/robertoxmed/MC-DAG}. In addition, since works
in~\cite{baruah2016federated} have only presented theoretical results,
we also implemented the federated approach. Last but not least, we
developed a MCS generator in order to produce many MCS with random
properties. Thanks to these tools, we generated a set of MCSs and
measured the ratio for which each scheduling method finds a MC-correct
schedule.

The random generation needs to be \textbf{unbiased} and
\textbf{uniformly cover} the possible timing configurations of MCS. To
design this random generation, we first integrated existing methods to
generate DAGs with unbiased topologies~\cite{cordeiro2010random}.
This is an important aspect, since certain DAG shapes tend to be more
schedulable than others. The distribution of execution time for tasks
is not controlled by existing DAG generation approaches.
Yet, the utilization of the 
system is the most important factor used to perform benchmarks on real-time
scheduling techniques. To overcome this limitation, we have integrated
existing methods achieving a uniform distribution of utilizations for
tasks~\cite{bini2005measuring,davis2009priority}.

Parameters for the generation of MCS are:
\begin{itemize}
\item $U$: Utilization of the system in both criticality modes.
\item $|\mathcal{G}|$: Fixed number 
  of MC-DAGs per system.
\item $|V_j|$: Fixed number of vertices per MC-DAG, 
  \emph{i.e.} all MC-DAGs have the same number of vertices.
\item $\rho$: Ratio of HI criticality tasks.
\item $f$: Reduction factor for the utilization of HI tasks in LO
  mode.
\item $e$: Probability to have an edge between two vertices.
\end{itemize}

Once these parameters are set, we first distribute uniformly the
utilization of the system to each MC-DAG. We use the uniform
distribution described in~\cite{bini2005measuring} to assign a
utilization for each MC-DAG. The period/deadline for each MC-DAG is
then assigned randomly: this period is chosen from a predefined list
of numbers in order to avoid prime numbers\footnote{Possible periods:
  $\{100, 120, 150, 180, 200, 220, 250, 300, 400, 500\}$.}  (which are
also avoided in the industrial context). With the assignment of the
period and the utilization of the MC-DAG, we can distribute the
utilization to tasks of the DAG. We use 
UUnifast-discard~\cite{davis2009priority} in this
case. As opposed to the utilization that can be given to DAGs, a
vertex cannot have a utilization greater than 1 since it is a
\emph{sequential} task (parallel execution for a vertex is not
possible). UUnifast-discard is therefore an appropriate method.
The utilization available for LO-criticality tasks is 
given by the difference between the utilization of HI tasks in HI mode and
the utilization of HI tasks in LO mode, the difference being controlled by
parameter $f$.

Once the utilization of the system is distributed among MC-DAGs and
the utilization of MC-DAGs is distributed among tasks, we start the
generation of the topology for the MC-DAGs. We start by creating the
HI-criticality vertices. These vertices are connected following the
probability $e$ given by the user and without creating cycles among
vertices. After the HI-criticality tasks have been created, we create the
LO-criticality tasks. Again vertices are connected following the
probability $e$ chosen by the user and without creating cycles. The
higher the probability $e$, the more dense is the resulting graph:
vertices have more precedence constraints to satisfy, making the
scheduling of the system more difficult.


\textbf{Experimentation setup}: We control the parameters of the MCS 
generator so as to measure their influence on the performance of our method. We 
expect the following parameters to make the scheduling problem 
more difficult: (i) the density of the graphs, (ii) the utilization of the 
system, (iii) the utilization per task of the system, (iv) the number of 
MC-DAGs. Our experiments aim at measuring the effect of these parameters on 
\galap{}'s performance.

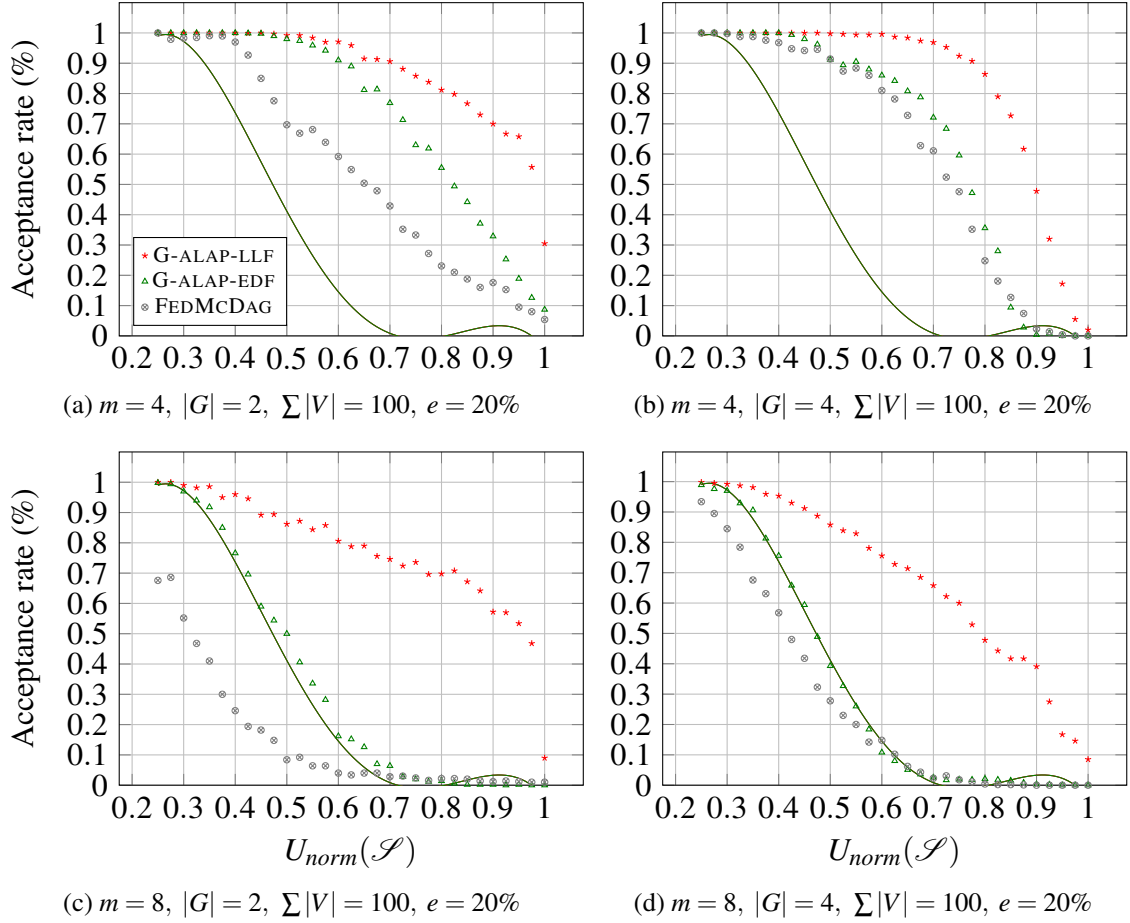
\begin{figure*}
	\centering
	\subfloat[$m=4,~|G|= 2,~ \sum|V|=100, ~ e= 20\%$]{
		\begin{tikzpicture}[every mark/.append style={mark size=1.2pt}]
		\begin{axis}[ylabel=Acceptance rate (\%),
		legend style={nodes={scale=0.7, transform shape}},
		height=6cm, width=0.5\textwidth,
		legend pos=south west,grid,
		ytick={0, 0.1, ..., 1.1},
		xtick={0.2, 0.3, ..., 1.1},
		ymin=0]

		\addplot+[only marks,mark=star,red] table [x=Unorm, 
		y=Lax50Avg] {Results/l2/sched/c4/e20-d2.data};
		\addplot+[only marks,mark=triangle,green!50!black] table [x=Unorm, 
		y=Edf50Avg] 
		{Results/l2/sched/c4/e20-d2.data};
		\addplot+[only marks,mark=otimes,gray] table [x=Unorm, 
		y=Fed50Avg] 
		{Results/l2/sched/c4/e20-d2.data};
		
		\addplot [gray] 
		gnuplot [raw gnuplot] { 
			f(x)=a*x^4+b*x^3+c*x^2+d*x+e;  
			a=0.0001;
			b=0.0001;
			c=0.1;
			d=0.1;
			e=1;
			fit f(x) 'Results/l2/sched/c4/e20-d2.data' using 1:4 via a,b,c,d,e; 
			plot [x=0.25:1] f(x); 
		};
		\addplot [red] 
		gnuplot [raw gnuplot] { 
			f(x)=a*x^4+b*x^3+c*x^2+d*x+e;  
			a=0.0001;
			b=0.0001;
			c=0.1;
			d=0.1;
			e=1;
			fit f(x) 'Results/l2/sched/c4/e20-d2.data' using 1:7 via a,b,c,d,e; 
			plot [x=0.25:1] f(x); 
		};
		\addplot [green!50!black] 
		gnuplot [raw gnuplot] { 
			f(x)=a*x^4+b*x^3+c*x^2+d*x+e;  
			a=0.0001;
			b=0.0001;
			c=0.1;
			d=0.1;
			e=1;
			fit f(x) 'Results/l2/sched/c4/e20-d2.data' using 1:10 via 
			a,b,c,d,e; 
			plot [x=0.25:1] f(x); 
		};
	
		\addlegendentry{\textsc{G-alap-llf}}
		\addlegendentry{\textsc{G-alap-edf}}
		\addlegendentry{\textsc{FedMcDag}}
		
		\end{axis}
		\end{tikzpicture}
		\label{fig:sched-c4-e20-d2-t50}
	}
	\subfloat[$m=4,~|G|= 4,~ \sum|V|=100, ~ e= 20\%$]{
		\begin{tikzpicture}[every mark/.append style={mark size=1.2pt}]
		\begin{axis}[
		legend style={nodes={scale=0.7, transform shape}},
		height=6cm, width=0.5\textwidth,
		legend pos=south west,grid,
		ytick={0, 0.1, ..., 1.1},
		xtick={0.2, 0.3, ..., 1.1},
		ymin=0]
		
		\addplot+[only marks,mark=star,red] table [x=Unorm, 
		y=Lax25] {Results/l2/sched/c4/e20-d4.data};
		\addplot+[only marks,mark=triangle,green!50!black] table [x=Unorm, 
		y=Edf25] 
		{Results/l2/sched/c4/e20-d4.data};
		\addplot+[only marks,mark=otimes,gray] table [x=Unorm, 
		y=Baruah25] 
		{Results/l2/sched/c4/e20-d4.data};
		
		\addplot [gray] 
		gnuplot [raw gnuplot] { 
			f(x)=a*x^4+b*x^3+c*x^2+d*x+e;  
			a=0.0001;
			b=0.0001;
			c=0.1;
			d=0.1;
			e=1;
			fit f(x) 'Results/l2/sched/c4/e20-d4.data' using 1:2 via a,b,c,d,e; 
			plot [x=0.25:1] f(x); 
		};
		\addplot [red] 
		gnuplot [raw gnuplot] { 
			f(x)=a*x^4+b*x^3+c*x^2+d*x+e;  
			a=0.0001;
			b=0.0001;
			c=0.1;
			d=0.1;
			e=1;
			fit f(x) 'Results/l2/sched/c4/e20-d4.data' using 1:3 via a,b,c,d,e; 
			plot [x=0.25:1] f(x); 
		};
		\addplot [green!50!black] 
		gnuplot [raw gnuplot] { 
			f(x)=a*x^4+b*x^3+c*x^2+d*x+e;  
			a=0.0001;
			b=0.0001;
			c=0.1;
			d=0.1;
			e=1;
			fit f(x) 'Results/l2/sched/c4/e20-d4.data' using 1:4 via 
			a,b,c,d,e; 
			plot [x=0.25:1] f(x); 
		};
		
		\end{axis}
		\end{tikzpicture}
		\label{fig:sched-c4-e20-d4-t50}
	}

	\subfloat[$m=8, ~|G|= 2,~ \sum|V|=100,~ e= 20\%$]{
		\begin{tikzpicture}[every mark/.append style={mark size=1.2pt}]
		\begin{axis}[ylabel=Acceptance rate (\%),
		xlabel=$U_{norm}(\mathcal{S})$,
		legend style={nodes={scale=0.7, transform shape}},
		height=6cm, width=0.5\textwidth,
		legend pos=south west,grid,
		ytick={0, 0.1, ..., 1.1},
		xtick={0.2, 0.3, ..., 1.1},
		ymin=0]
		
		\addplot+[only marks,mark=star,red] table [x=Unorm, 
		y=Lax50] {Results/l2/sched/c8/e20-d2.data};
		\addplot+[only marks,mark=triangle,green!50!black] table [x=Unorm, 
		y=Edf50] 
		{Results/l2/sched/c8/e20-d2.data};
		\addplot+[only marks,mark=otimes,gray] table [x=Unorm, 
		y=Fed50] 
		{Results/l2/sched/c8/e20-d2.data};
		
		\addplot [gray] 
		gnuplot [raw gnuplot] { 
			f(x)=a*x^4+b*x^3+c*x^2+d*x+e;  
			a=0.0001;
			b=0.0001;
			c=0.1;
			d=0.1;
			e=1;
			fit f(x) 'Results/l2/sched/c8/e20-d2.data' using 1:2 via a,b,c,d,e; 
			plot [x=0.25:1] f(x); 
		};
		\addplot [red] 
		gnuplot [raw gnuplot] { 
			f(x)=a*x^4+b*x^3+c*x^2+d*x+e;  
			a=0.0001;
			b=0.0001;
			c=0.1;
			d=0.1;
			e=1;
			fit f(x) 'Results/l2/sched/c8/e20-d2.data' using 1:3 via a,b,c,d,e; 
			plot [x=0.25:1] f(x); 
		};
		\addplot [green!50!black] 
		gnuplot [raw gnuplot] { 
			f(x)=a*x^4+b*x^3+c*x^2+d*x+e;  
			a=0.0001;
			b=0.0001;
			c=0.1;
			d=0.1;
			e=1;
			fit f(x) 'Results/l2/sched/c8/e20-d2.data' using 1:4 via 
			a,b,c,d,e; 
			plot [x=0.25:1] f(x); 
		};
		
		\end{axis}
		\end{tikzpicture}
		\label{fig:sched-c8-e20-d2-t50}
	}
	\subfloat[$m=8, ~|G|= 4,~ \sum|V|=100,~ e= 20\%$]{
		\begin{tikzpicture}[every mark/.append style={mark size=1.2pt}]
		\begin{axis}[xlabel=$U_{norm}(\mathcal{S})$,
		legend style={nodes={scale=0.7, transform shape}},
		height=6cm, width=0.5\textwidth,
		legend pos=south west,grid,
		ytick={0, 0.1, ..., 1.1},
		xtick={0.2, 0.3, ..., 1.1},
		ymin=0]
		
		\addplot+[only marks,mark=star,red] table [x=Unorm, 
		y=Lax50] {Results/l2/sched/c8/e20-d4.data};
		\addplot+[only marks,mark=triangle,green!50!black] table [x=Unorm, 
		y=Edf50] 
		{Results/l2/sched/c8/e20-d4.data};
		\addplot+[only marks,mark=otimes,gray] table [x=Unorm, 
		y=Fed50] 
		{Results/l2/sched/c8/e20-d4.data};
		
		\addplot [gray] 
		gnuplot [raw gnuplot] { 
			f(x)=a*x^4+b*x^3+c*x^2+d*x+e;  
			a=0.0001;
			b=0.0001;
			c=0.1;
			d=0.1;
			e=0.1;
			fit f(x) 'Results/l2/sched/c8/e20-d4.data' using 1:2 via a,b,c,d,e; 
			plot [x=0.25:1] f(x); 
		};
		\addplot [red] 
		gnuplot [raw gnuplot] { 
			f(x)=a*x^5+b*x^4+c*x^3+d*x^2+e*x+f;  
			a=0.0001;
			b=0.0001;
			c=0.1;
			d=0.1;
			e=0.1;
			f=0.1;
			fit f(x) 'Results/l2/sched/c8/e20-d4.data' using 1:3 via 
			a,b,c,d,e,f; 
			plot [x=0.25:1] f(x); 
		};
		\addplot [green!50!black] 
		gnuplot [raw gnuplot] { 
			f(x)=a*x^4+b*x^3+c*x^2+d*x+e;  
			a=0.0001;
			b=0.0001;
			c=0.1;
			d=0.1;
			e=1;
			fit f(x) 'Results/l2/sched/c8/e20-d4.data' using 1:4 via 
			a,b,c,d,e; 
			plot [x=0.25:1] f(x); 
		};
		
		\end{axis}
		\end{tikzpicture}
		\label{fig:sched-c8-e20-d4-t25}
	}
	\caption{Comparison to existing multiple MC-DAG scheduling approach}
	\label{fig:results_multiple_dag}
\end{figure*}

\noindent
\textbf{Experimentation results.} Figure~\ref{fig:results_multiple_dag} provides our experimental
results in terms of acceptance rate obtained with different schedulers
and various setups of the MC-DAG generator. For each point in the
figure, the acceptance rate was obtained by generating 500 MC systems
(i.e. sets of MC-DAGs) and measuring the percentage of these systems
for which a MC-correct schedule (see definition~\ref{def:mccorrect})
was found. Each subfigure shows the evolution of the acceptance rate
when the CPU usage increases. From one subfigure to another, the
configuration of the MC-DAG generator was changed. For instance,
subfigure~\ref{fig:sched-c4-e20-d2-t50} shows results obtained with 2
MC-DAGs of 100 tasks each, with an edge probability of 20\% and a
processor with 4 cores. Results shown on
subfigure~\ref{fig:sched-c4-e20-d4-t50} were obtained using the same
configuration except of the number of MC-DAG: 4 MC-DAGs were generated
in this case.

\noindent
On each subfigure, the acceptance rate obtained with different
scheduling strategies are displayed: the red curve corresponds to the
\textsc{G-alap-LLF} scheduler we proposed. It uses G-LLF to set tasks
priority, enforces the respect of~\condmc{}, and executes tasks ALAP
in HI mode and ASAP in LO mode. the green curve corresponds to
\textsc{G-alap-EDF}, following the same principles as
\textsc{G-alap-LLF} but assigning tasks priorities according to
G-EDF. Finally, the grey curve corresponds to the federated approach,
proposed by Baruah in~\cite{baruah2016federated}.

\noindent
Without entering in details into the comparison of these results,
readers can easily observe that \textsc{G-alap-LLF} provides much
better results than the two other methods. When it comes to the
comparison of \textsc{G-alap-EDF} and the federated approach, the
performance gain obtained with G-EDF depends on the system's
configuration. To better understand the results, we shall explain how
the difficulty of the scheduling problems evolve across
subfigures. Scheduling problems obtained with configuration of
subfigure~\ref{fig:sched-c4-e20-d4-t50} are easier than problems
obtained with the configuration of
subfigure~\ref{fig:sched-c4-e20-d2-t50}: when increasing the number of
MC-DAGs on the same number of cores and the same CPU utilization, we
tend to produce smaller tasks which is easier to schedule than more
monolithic task sets. For the same reason, scheduling problems
obtained with configuration of subfigure~\ref{fig:sched-c8-e20-d2-t50}
are easier than problems obtained with the configuration of
subfigure~\ref{fig:sched-c8-e20-d4-t25}, and problems corresponding to
subfigure~\ref{fig:sched-c8-e20-d2-t50} are easier than problems
corresponding to subfigure~\ref{fig:sched-c4-e20-d2-t50}.

This classification of the difficulty of scheduling problems is
confirmed by the experimental results show on the
subfigures. Therefore, the overall comparison of obtained results tend
to show that
\begin{enumerate}
\item \textsc{G-alap-EDF} is better than the federated approach
for systems of intermediate difficulty but the difference between
these methods tend to be reduced for very easy or very difficult
scheduling problems.
\item \textsc{G-alap-LLF} is far better than other approaches in terms
  of acceptance rate.
\end{enumerate}

\noindent
However, solutions based on LLF are known to produce an
important number of preemptions among tasks. In Roberto Medina's PhD,
we have measured the number of preemptions obtained with the different
MC-DAG schedulers mentioned above. Without knowing the cost of a
preemption, it is not possible to know how taking preemptions cost
into account impacts the acceptance rate. However, since we produce
scheduling tables, the number of preemptions is known at design time
and if we were given a value for preemption cost, we could easily
adapt our scheduling method and decide whether the system remains
schedulable.

\textbf{Availability analysis of low criticality services.} In
addition to our contributions on MC-DAGs scheduling, we studied the
impact of mixed criticality scheduling on the quality of services of
low criticality tasks. The objective of mixed criticality scheduling
is to improve computation resources usage, to the price of a
degradation of low criticality services. Indeed, when high criticality
tasks risk to miss their deadlines, low criticality tasks are
discarded (or slowed down).

\noindent
In our work, published at the international conference on Design
Automation and Test in Europe 2018, we proposed a method to evaluate
the impact of mixed criticality scheduling on the quality of low
criticality services. We formalized this quality of service as the
availability of outputs produced by low criticality tasks, and
proposed enhancements of this quality of service for task sets modeled
as MC-DAGs with a discard model (low tasks are temporarily discarded
in case of TFE). Taking advantage of the DAG, we proposed the
following modifications of the MC scheduling: when a TFE occurs in a
low criticality task, we only discard the induced subgraph by the
discard of the faulty task. In addition, we proposed to consider fault
tolerance mechanisms such as the Triple Modular Redundancy of MC-DAGs,
and/or the weakly hard nature of some tasks, which is usually
formalized as follows: tasks able to continue their execution as long
as less than M errors out for K consecutive executions.

\noindent
In order to evaluate the availability of outputs, two types of
precedence relationships among tasks had to be considered: structural
precendence (captured by edges in DAGs) and temporal precedence
(captured by the scheduling table obtained with techniques presented
in previous sections). From these models (\emph{i.e.} MC-DAGs and
scheduling tables) we proposed a model transformation to probabilistic
automata. The probabilistic nature of TFE was captures using the
recent notion of probabilistic
WCET~\cite{Maxim:2017:PAM:3139258.3139276}. Besides, a recovery
towards the LO mode as well as the fault tolerance mechanisms were
also captured with state and transitions of probabilistic automata.

\noindent
To the best of our knowledge, this contribution was the first method
to compute the availability of low criticality tasks of MC systems. In
addition, our experimental evaluation, made with the PRISM
framework~\cite{KNP11}, showed that our enhancements of this
availability were very significant.

\begin{figure}
  \centering
  \includegraphics[width=\textwidth, trim={0cm 7cm 0cm 7cm}, clip]{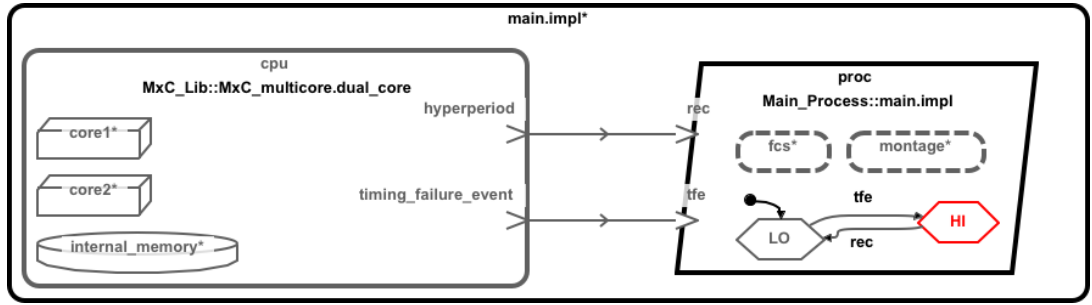}
  \caption{UAV AADL architecture: system level}
  \label{fig:aadl_mcdag_system}
\end{figure}

\noindent
\textbf{Integration in RAMSES.} The work presented in this section has
also been integrated in the RAMSES framework. Interestingly, the
computation of low criticality tasks availability relies on the
computation of scheduling tables. We thus adapted the RAMSES framework
to refine input AADL models by including a representation of the
schedule tables. This model is then used to produce formulas or
automata from which availability of low criticality functions is
computed. We also elaborated a library of AADL models to model
MC-DAGs. Figure~\ref{fig:aadl_mcdag_system} shows the model of the
hardware multi-core platform, \emph{i.e.} two cores modeled as AADL
processors on the left part of the figure. For the software
architecture, which is illustrated on the right part of
figure~\ref{fig:aadl_mcdag_system}, two AADL thread groups represent
the MC-DAGs of the UAV case study. In the software part, LO and HI
modes are modeled with AADL modes, and event ports connect the
hardware part to the software part to represent potential mode
switches (LO to HI in case of TFE in a high criticality task, and
recovery to switch back to LO mode).

\begin{figure}
  \centering
  \includegraphics[width=\textwidth, trim={0cm 6cm 0cm 6cm}, clip]{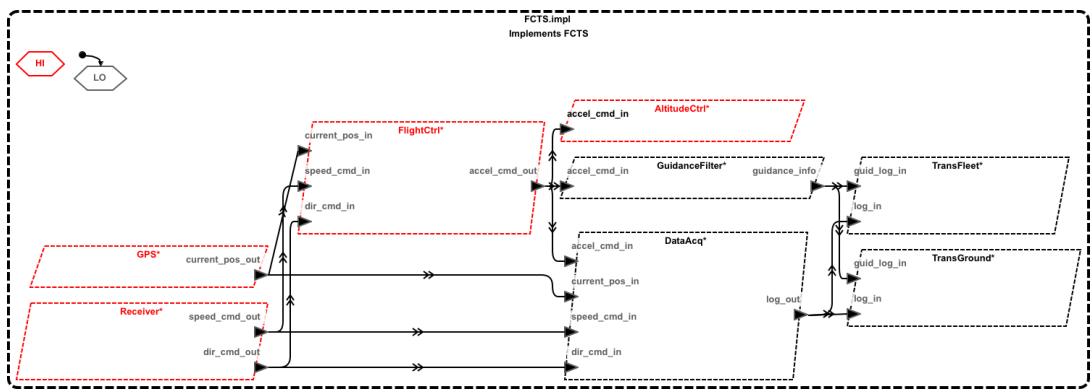}
  \caption{UAV AADL architecture: FCS MC-DAG}
  \label{fig:aadl_mcdag_fcs}
\end{figure}

Figure~\ref{fig:aadl_mcdag_fcs} represents the content of the FCS
MC-DAG modeled as a set of AADL thread with data port
connections. These connections are associated with a \emph{Timing}
property of value \emph{Immediate} to represent in AADL the
precedence constraints of DAGs.

\newpage

\section{Concluding remarks}


In this chapter, we have presented our contributions to answer the
following questions:
\begin{enumerate}
\item how to improve the consistency between models used for analysis
  and code generation purposes?
\item how to improve resource usage induced by pessimistic hypothesis
  in the design of real-time CPS?
\end{enumerate}

\noindent
To answer these questions, we have proposed a model refinement method
and its prototyping in the RAMSES framework. This framework has been
experimented on different MoCCs, showing the added value of model
transformations in the context of CPSs design. By providing answers to
some of the research questions presented here above, our work provides
original methods to improve the reliability of software development in
CPSs.

\noindent
To the best of our knowledge, this work provides a unique AADL
framework for fine grain timing analysis of real-time embedded
systems, as well as source code generation for the most common subsets
of AADL MoCCs used in safety critical real-time systems (partitioned
systems, periodic delayed communications, periodic MC-DAGs). These
contributions have been produced by several PhD students who presented
their results in well established international conferences.

\noindent
This work was inspired by technical discussions with industrial
partners, in particular from the railway domain. Our new results bring
answers to significant concerns in this industry in particular when it
comes to better understand how to improve resource usage while
ensuring safety. These results have also been applied in a project
with the Department of Defence (USA) and the Software Engineering
Institute (SEI), aiming at generating source code for a commercial
operating system implementing the ARINC653 standard.

\noindent
As a follow up on this work, we started new research activities in
order to adapt security techniques to the specificity of critical
real-time embedded systems. In the scope of Maxime Ayrault's PhD, we
aim at studying the integration of resilience mechanisms in connected
cars. More generally, we aim at improving the autonomy of critical
systems, which is the objective of a european collaborative project
proposals we are involved in. These two perspectives are further
described in chapter~\ref{chap:chapter5}.

\noindent
We have focused in this chapter on model transformations for source
code generation, but several other types of model transformations were
implemented in RAMSES. In particular, model transformations for remote
communications, modes, and error management are of interest for timing
analysis since all these transformations require to add task to the
initial model in order to react on incoming messages, mode change
requests, and error occurrences. Building on this experience, we
extended our research activities towards the composition of model
transformations for CPSs. We present our contributions on this topic
in next chapter.


\addtocontents{toc}{\protect\newpage}
\chapter{Composition and formalization of model transformations}
\label{chap:chapter4}

\minitoc

\vspace{1.5cm}

In the previous chapter, we have shown how we use model
transformations to improve software design process in the context of
real-time CPSs. As stated in the introduction of this document,
however, model transformations are difficult to write: they are,
essentially, graphs transformation applications. This is one of the
reasons why dedicated model transformation languages have been
defined. In addition, in order to ease their maintenance and reuse,
model transformations are usually written as small units of
transformation which can be composed into model transformation chains:
the output model of a transformation becomes the input model of the
next transformation of the chain.

However, several problems come from the decomposition of model
transformations into chains of smaller transformations:

\begin{enumerate}
\item transformation convergence: when model transformations are
  chained, each transformation produces an output model that becomes
  the input model of another transformation. This process could be
  repeated infinitely, failing to produce the output model of the
  chain. Ensuring existence of models produced by model transformation
  chains is a difficult problem, mentioned on the upper and left part
  of figure~\ref{fig:approach_overview}.
\item output model correctness: because model transformations are
  complex software applications, model transformation chains rapidly
  become difficult to master. However, it is important to ensure
  output models correctness. This notion can be decomposed into
  qualitative and quantitative correctness where qualitative
  correctness boils to ensure the output model satisfies predefined
  structural constraints whereas quantitative correctness boils to
  ensure the output model exhibits satisfactory NFPs. Ensuring
  correctness of models produced by model transformation chains is a
  difficult problem mentioned on the upper and central part of
  figure~\ref{fig:approach_overview}.
\item variability management: in a model transformation chain, each
  model transformation is subject to variability. Indeed, abstraction
  embodied by a model implies there will exist several implementations
  or refinement variants of this model. In practice, it is very common
  that such variants would have different impacts on NFPs exhibited by
  the resulting model. As these impacts are often in conflict, finding
  the best transformation chain boils to solve a multi-objective
  optimization problem. Solving such problems is a very difficult task
  mentioned on the upper and right part of
  figure~\ref{fig:approach_overview}.

\item transformations validation and verification: model
  transformation chains have to be validated with very rigorous
  methods when they are involved in the implementation process of
  critical software applications. This becomes a very challenging
  problem when model transformations are organized into chains of
  model transformations since transformations have to be tested
  individually and as an integrated chain of transformations. As a
  consequence, validation and verification of model transformation
  chains may become very costly. This problem is not depicted on
  figure~\ref{fig:approach_overview} but it is obviously very relevant
  in the application domains of our work.

\end{enumerate}



In this chapter, we present the work we have conducted to increase the
confidence one can have in model transformation chains. Reading the
contributions presented in this chapter, one will notice their common
focus on structural constraints applied to models produced by model
transformations. Indeed, the notion of structural constraints has been
used in this work to represent: (i) validity constraints for the
applicability of analysis and verification techniques on output
models, (ii) test requirements for the validation of model
transformations, and (iii) validity constraints for the model
transformation variants selection and composition.

\begin{figure}[h!]
\centering
\includegraphics[width=0.9\textwidth, trim={0cm 6cm 0cm 0cm}, clip]{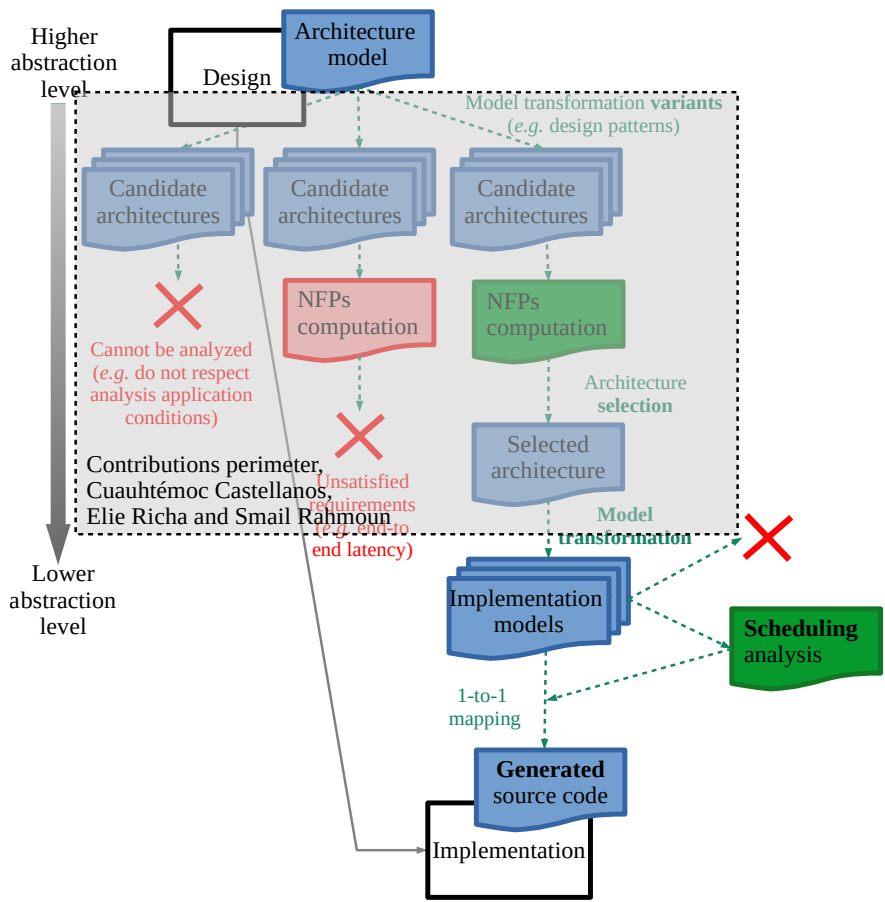}
\caption{Perimeter of research activities presented in this chapter}
\label{fig:approach_overview_part2}
\end{figure}

\noindent
This chapter is organized as follows: section~\ref{sec:section_4.1}
introduces the context of this work with a presentation of model
transformation chains implemented in the RAMSES framework. Building
on this presentation, we illustrate the notion of structural
constraint which enable the application of analysis presented in
previous chapter. In section~\ref{sec:section_4.2}, we describe our
method to build model transformation chains ensuring produced models
respect predefined structural constraints (results of Cuauhtémoc
Castellanos PhD). Section~\ref{sec:section_4.3} presents our
contributions dedicated to model transformation chains validation
(results obtained during Elie Richa PhD). Finally, we present in
section~\ref{sec:section_4.4} our work on model transformation
variants selection and composition to implement model-driven and
multi-objective design space exploration techniques (result obtained
during Smail Rahmoun PhD). Figure~\ref{fig:approach_overview_part2}
show how these contributions are positioned with respect to the
approach we described in chapter~\ref{chap:chapter2}.

\section{Model transformation chains in RAMSES}
\label{sec:section_4.1}

\begin{figure}[h]
  \caption{RAMSES refinements: chain of model transformations}
  \label{fig:ramses_transformation_chains}
\centering \includegraphics[width=\textwidth, trim={0cm 0cm 0cm 0cm},
  clip]{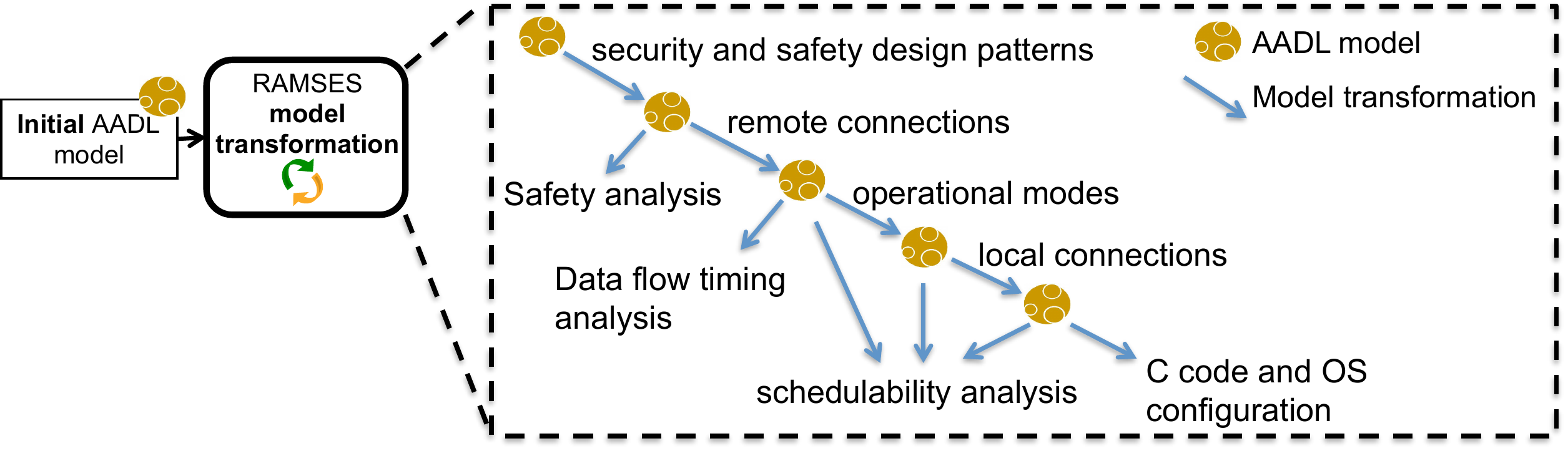}
\end{figure}

In order to produce software applications of CPS from architecture
models, we proposed to implement AADL to AADL model
transformations. One such model transformations have been described in
chapter~\ref{chap:chapter2}. More generally, a step-wise model
transformation process is illustrated on
figure~\ref{fig:ramses_transformation_chains}. In this process,
several model transformations are chained:
\begin{itemize}
\item security and safety design patterns are first applied to
  integrate safety and/or security components such as firewalls,
  encryption/decryption components, software/hardware redundancy,
  etc.
\item remote connections are then transformed in order to incorporate
  communication tasks in the software architecture of the
  application.
\item operational modes are treated in a similar way: dedicated mode
  management tasks are added to the software architecture to handle
  mode change requests.
\item connections among ports of tasks deployed on the same processors
  are then mapped into global variables and runtime services calls, as
  described in chapter~\ref{chap:chapter2}.
\end{itemize}

In the remainder of this section, we give more details about the
implementation of these transformations in RAMSES. The objective of
this presentation is to provide enough information about the technical
context in which the research work presented in this chapter have been
conducted. We start with an illustration of a model transformation,
and use this example to explain the model transformation language used
in RAMSES.

\noindent
Several model transformation methods, languages, and tools, have been
studied to help MDE experts develop their frameworks. A classification
of model transformation approaches was proposed by Czarnecki and
al. in ~\cite{Czarnecki:2006:FSM:1165093.1165106}. From this
classification, we decided to use \textit{\textbf{ATL}} for the
implementation of the RAMSES framework. We chose this language for the
simplicity of its semantics, as well as for the quality of the
associated model transformation tools.

In terms of semantics, ATL is a rule-based transformation language
which execution relies mainly on a pattern matching
semantics~\cite{MTIP05}: in ATL, a transformation consists of a set of
declarative \emph{matched rules}, each specifying a \emph{source
  pattern} and a \emph{target pattern}. The source pattern is made up
of (i) a set of objects identifiers, typed with meta-classes from the
source meta-model and (ii) an optional OCL~\cite{std:ocl} constraint
acting as a \emph{guard} of the rule. The target pattern is a set of
objects of the target meta-model and a set of \emph{bindings} that
assign values to the attributes and references of the target objects.

\begin{figure}[h]
  \centering
  \caption{Overview of the TMR (also called 2oo3) transformation}
  \label{fig:tmr_transformation}
\includegraphics[width=\textwidth, trim={0cm 0cm 0cm 0cm},
  clip]{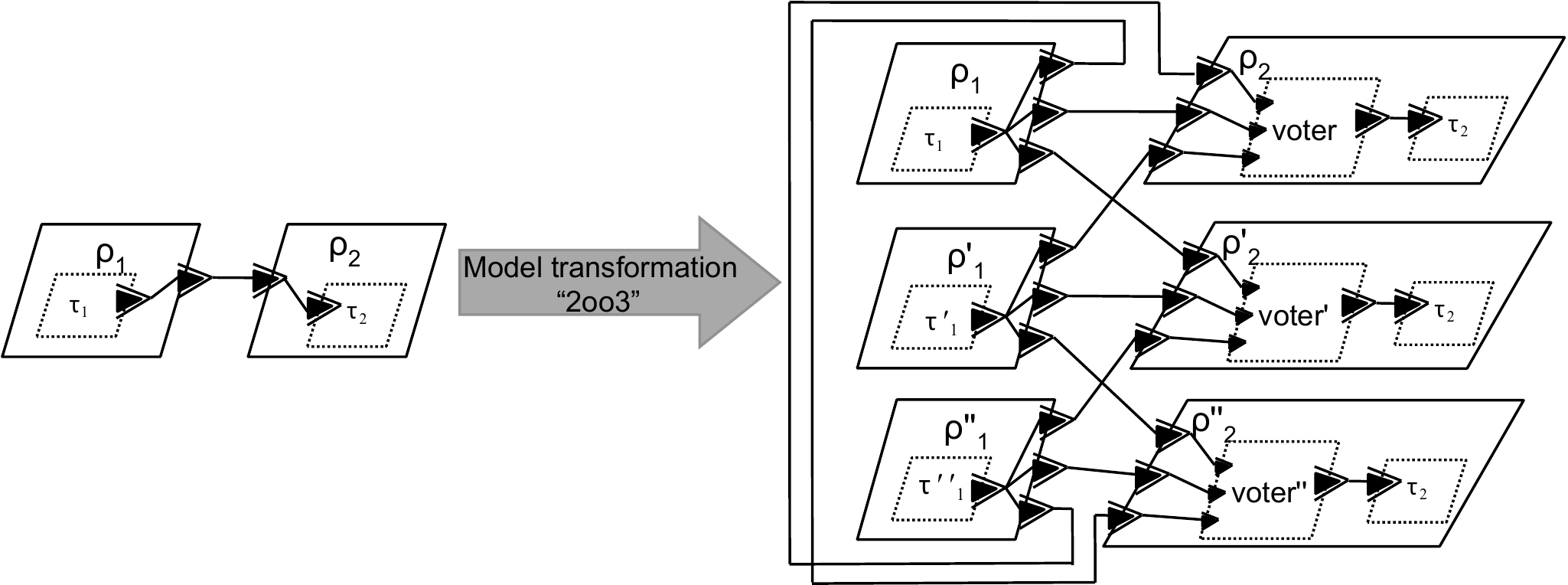}
\end{figure}

Figure~\ref{fig:tmr_transformation} provides an illustration of the
application of a model transformation to implement a safety design
pattern called Triple Modular Redundancy
(TMR)~\cite{Lyons:1962:UTR:1661979.1661984}, also called two out of
three (2oo3). Listings~\ref{lst:atl_process_2oo3}
and~\ref{lst:atl_connection_2oo3} provide snippets of the ATL code
used to implement the 2oo3 model transformation for components
replication. This transformation will be used as an illustrative
example in the remainder of this chapter. In
listing~\ref{lst:atl_process_2oo3}, ATL rule \texttt{m\_Process\_2oo3}
transforms every AADL process component into three process components
identified with the following \emph{target object} identifiers:
\texttt{proc1\_2oo3}, \texttt{proc2\_2oo3}, and
\texttt{proc3\_2oo2}. For the sake of concision, this listing does not
develop the creation of these processes. Given the execution semantics
of ATL, this rule will match any AADL component instance of the
process category.

\begin{figure}
\begin{lstlisting}[basicstyle=\scriptsize,captionpos=t,caption=ATL rule for 2oo3: processes replication,label=lst:atl_process_2oo3,frame=single,xleftmargin=\parindent,language=ATL,numbers=left,escapeinside={(*}{*)},frame=single]
rule m_Process_2oo3
{
  from
    c : AADLI!ComponentInstance (c.category = #process)
  to
    proc1_2oo3: AADLBA!ProcessSubcomponent (...),
    proc2_2oo3: AADLBA!ProcessSubcomponent (...),
    proc3_2oo3: AADLBA!ProcessSubcomponent (...),
}
\end{lstlisting}
\end{figure}

In listing~\ref{lst:atl_connection_2oo3}, ATL rule
\texttt{m\_PortConnection\_2oo3} transforms connections among
process components in the source model, into connections among their
replicas in the target model.
This is represented in rule \texttt{m\_PortConnection\_2oo3} with
the creation of \texttt{cnx1\_1\_2oo3}, \texttt{cnx1\_2\_2oo3},
etc. For the sake of concision, the creation of only one of the
connections in the target model is fully developed in this
listing. This rule will match, in the source model of the
transformation, any connection $cnx_c$ between two process components
of the input model.

\begin{figure}
\begin{lstlisting}[basicstyle=\scriptsize,captionpos=t,caption=ATL rule for 2oo3: connections replication,label=lst:atl_connection_2oo3,frame=single,xleftmargin=\parindent,language=ATL,numbers=left,escapeinside={(*}{*)},frame=single]
rule m_PortConnection_2oo3
{
 from
   cnx: AADLI!ConnectionReference(cnx.isProcessPortsConnection())
 using
 {
   cSrc : AADLI!ComponentInstance = cnx.getSrcCptInstance();
   cDst : AADLI!ComponentInstance = cnx.getDstCptInstance();
 }
 to
   --  feature f_1: PROC_1_src -> PROC_1_dst --
   cnx1_2oo3: AADLBA!PortConnection (
     name <- cnx.getName()+'_1',
     source <- sourceCE1_1,
     destination <- destinationCE1_1 
   ),
   sourceCE1_1: AADLBA!ConnectedElement (
     connectionEnd <- cnx.source,
     context <- thisModule.resolveTemp(cSrc, 'proc1_2oo3') 
   ),
   destinationCE1_1: AADLBA!ConnectedElement (
     connectionEnd <- cnx.destination,
     context <- thisModule.resolveTemp(cDst, 'proc1_2oo3') 
   ),
   --  feature f_1: PROC_1_src -> PROC_2_dst --
   cnx2_2oo3: AADLBA!PortConnection
   ...
   -- other connections ommitted for the sake of concision
}
\end{lstlisting}
\end{figure}

\noindent
One of the reasons for the simplicity of the ATL language is the
definition of its resolve mechanisms: when a source object identifier
is referenced in the right hand side of a binding, a resolve operation
is automatically performed to find the rule that matched the source
objects, and the first output pattern object created by that rule is
used for the assignment to the target reference. This is referred to
as the default resolve mechanism. Another non-default resolve
mechanism allows resolving a (set of) source object(s) to an arbitrary
target pattern object instead of the first one as in the default
mechanism. It is invoked via the following ATL standard operation:
\texttt{thisModule.resolveTemp(obj, tgtPatternName)} as shown in
previous listing. Last but not least, the semantics of ATL ensures
transformations convergence: each rules is applied at most once per
pattern it matches in the input model. This property is not verified
by all model transformation languages: for some of them, rules are
executed as long as they match on the input model and the target
elements produced by previously executed rules. As a consequence, we
do not consider in our work the problem of model transformation chains
convergence presented in the introduction of this chapter.

\noindent
This brief presentation of ATL will allow us to present our
contributions on model driven engineering in the context CPSs. In the
next section, we present the framework we proposed in order to chain
model transformations in a way that guarantees the output model of the
chain can be analyzed. Contributions presented in
chapter~\ref{chap:chapter2} provide examples of such analysis.

\section{Automatic construction of transformation chains}
\label{sec:section_4.2}

Chaining model transformations properly may become a difficult
task. In particular, one of the objective of models in the domain of
CPS is to enable analysis of Non-Functional Properties. However, this
requires to conform to a set of validity constraints: for instance,
models used to ensure real-time systems schedulability (presented in
chapter~\ref{chap:chapter2}) assume that tasks are periodic (or
sporadic in some cases). However, error management tasks are often
aperiodic and have a high priority. As a consequence, hypothesis for a
timing analysis does not hold in this case. In practice, this is not
an issue from a timing analysis perspective since designers proceed to
timing analysis in nominal conditions (i.e. in the absence of
errors). However, this becomes an issue from a model transformation
perspective since it requires to ensure that timing analysis is
performed before model transformations dedicated to errors management
but after all other transformations leading to an analyzable model in
terms of timing analysis. Solving this problem boils to find model
transformation chains producing output models enforcing the respect of
application conditions on output models.

\noindent
In the introduction of this chapter, we have introduced the problem of
ensuring model transformation chains produce correct output
model. This notion of correctness was decomposed into qualitative and
quantitative correctness. In his PhD, Cuauhtémoc Castellanos studied
this problem considering qualitative correctness defined as a set of
structural constraints on models produced by model transformation
chains. Such constraints would also enable the verification of
qualitative correctness by enforcing the respect of application
conditions for analysis of NFPs on the output model. Cuauhtémoc
Castellanos also proposed to formalize design patterns as model
transformations. Design patterns composition was thus implemented by
chaining model transformations. Structural constraints on intermediate
models were formalized as a set of pre-conditions and post-conditions
of transformations. Additional structural constraints were also
defined in order to enforce the applicability of analysis techniques
on output models of a chain. OCL was first used to formalize these
constraints, while ATL was used to define model transformations.
The main problem we addressed in this PhD was: given a set of model
transformations with their preconditions and post conditions, and a
set of structural constraints on the output models of transformation
chains, in which order should transformations be chained in order to
produce an output model which satisfies structural constraints on the
output model?
When transformations are commutative, which was the object of previous
works~\cite{DBLP:conf/models/EtienABP12}, the order has no
importance. However, model transformations we consider in our work are
mostly non-commutative since they are refinement transformations.
In order to find a correct sequence of non-commutative model
transformations, we proposed to formalize model transformations in
Alloy~\cite{Jackson02}. Alloy is a modeling language to defined
constraint satisfaction problems with a \emph{relational
  algebra}. With Alloy, we specify a set of constraints a solution to
the problem must satisfy. These constraints are expressed in
first-order logic, which matches a subset of OCL. Once these
constraints are solved, a model instance satisfying all the
constraints expressed in Alloy is generated (if it exists).

\begin{figure}[h]
  \centering
  \caption{Model transformation chains production with Alloy}
  \label{fig:transformation_chains_alloy}
\includegraphics[width=\textwidth, trim={0cm 13cm 7cm 0cm},
  clip]{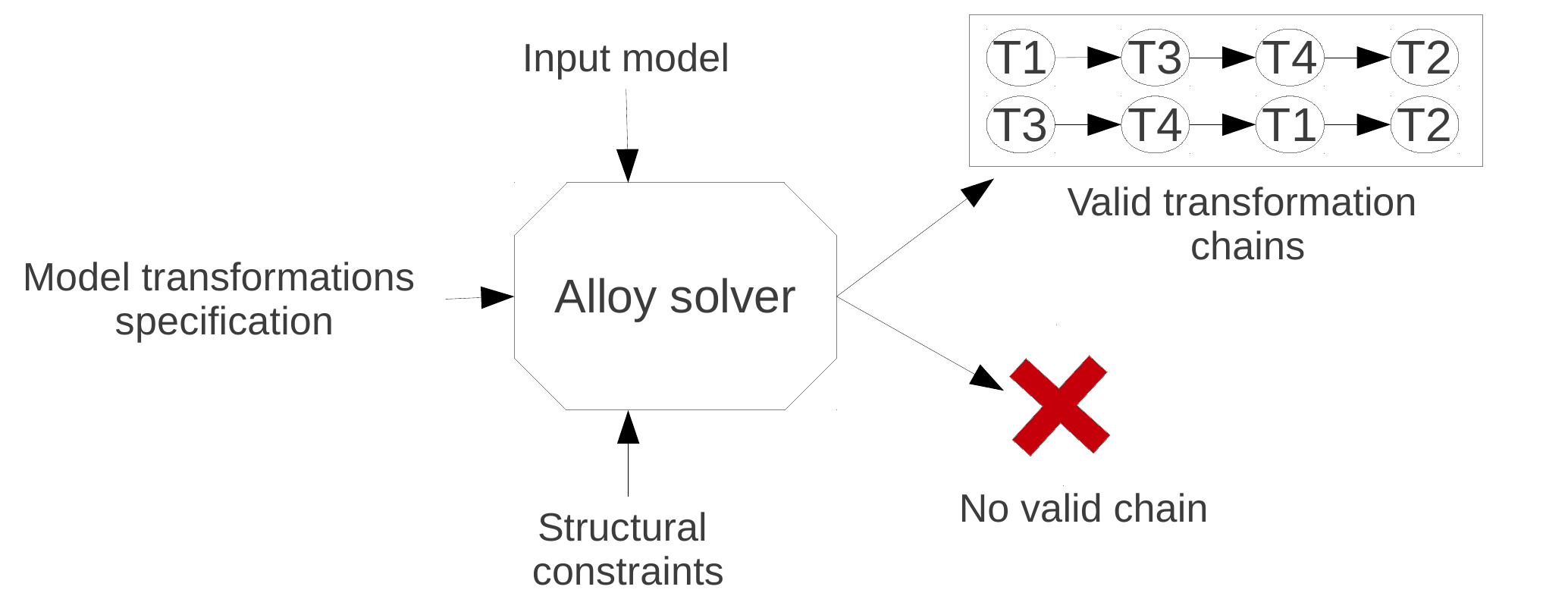}
\end{figure}

Figure~\ref{fig:transformation_chains_alloy} illustrates the approach
we proposed in this research work: from an input model, a set of
transformations along with their applications preconditions and
post-conditions, we asked the Alloy solver to produce a model
transformation chains and an output model which respects predefined
post-conditions.

In this work, we specified how to formalize ATL model
transformations and the chaining problem in
Alloy~\cite{Jackson02}. However, we did not implement a higher order
transformation (HOT, a transformation that takes as input and/or
produce as output a model transformation) from ATL to Alloy. Instead,
we focused on solving important scalability issues we faced when using
Alloy solvers to find correct chains of non-commutative
transformations. We thus decided to study and improve the scalability
of the method proposed
method~\cite{DBLP:conf/euromicro/CastellanosBPSV15}.

As a case-study, we presented in this PhD the formalization of a
safety design pattern called Triple Modular Redundancy (TMR), also
called two out of three (2oo3). We also defined a security pattern
used in cyber-security called red-black separation. This research work
helped us to formalize such model transformations, highlight their non
commutativity, and define constraint satisfying model transformation
chains.

The formalization of these transformations in Alloy led us to realize
that the approach defined in
figure~\ref{fig:transformation_chains_alloy} suffered from important
scalability issues. More precisely, these scalability issues came from
the chaining process itself and not from the execution of each
transformation in isolation. In order to improve the scalability of
this approach, we needed to define more transformations than just the
safety and security patterns mentioned above. The formalization in
Alloy of all the transformations implemented in RAMSES would have
taken too much time, and the implementation of a HOT from ATL to Alloy
would have been risky without improving first the scalability of the
method.

We thus focused on simpler model transformations (\emph{i.e.} easier
to formalize in Alloy) but more complex chains. We decided to
formalize design patterns from the Gang of Four~\cite{GOF} in
Alloy. This contribution, as well as the general approach illustrated
on figure~\ref{fig:transformation_chains_alloy}, were published in the
Euromicro conference on Software Engineering and Advanced Applications
(SEAA) 2014~\cite{DBLP:conf/euromicro/CastellanosBPVD14}. Studying
more precisely the reasons for the scalability limitation of our
approach, we defined an improved version in 2015 and published our
results at SEAA~2015~\cite{DBLP:conf/euromicro/CastellanosBPSV15}.

\begin{figure}[h]
  \centering
  \subfloat[Performance of our initial approach] {
    \centering
    \includegraphics[width=0.5\textwidth, trim={0cm 0cm 0cm 0cm},
      clip]{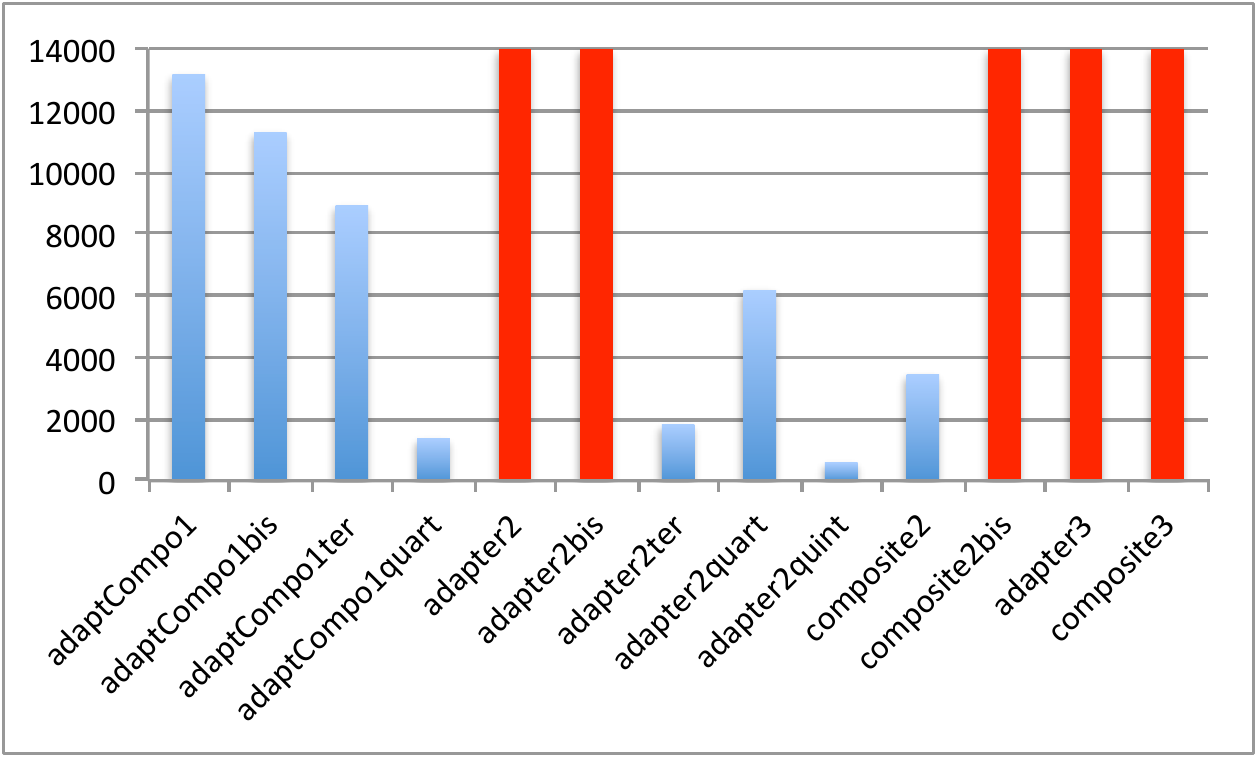}
    \label{fig:solving_time_alloy}
  }  
  \subfloat[Performance of the improved the chaining process] {
    \centering
    \includegraphics[width=0.5\textwidth, trim={0cm 0cm 0cm 0cm},
      clip]{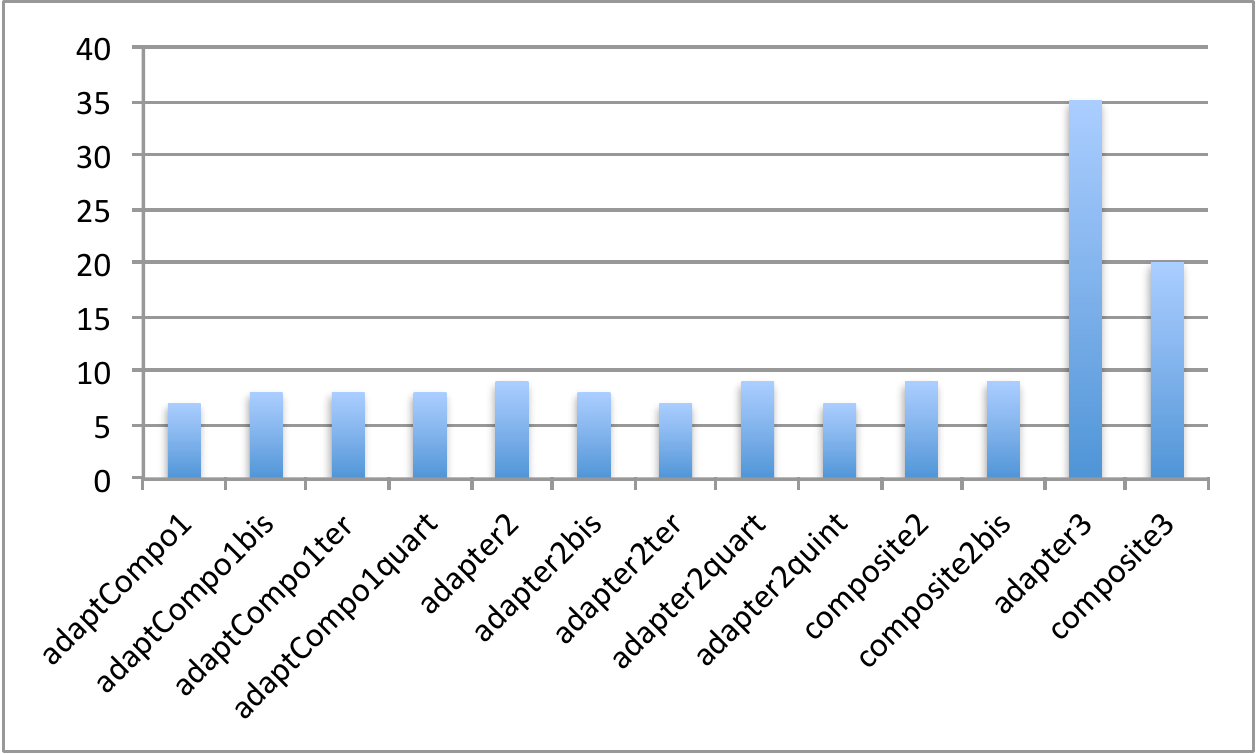}
    \label{fig:solving_time_alloy_improved}
  }
  \caption{Mean time to find transformation chains}
  \label{fig:solving_time_alloy_improvement}
\end{figure}

Figure~\ref{fig:solving_time_alloy_improvement} shows the performance
improvements on a set of model transformation chains:
figure~\ref{fig:solving_time_alloy} shows the solving time (in
seconds) obtained on different scenarios with our initial approach
while figure~\ref{fig:solving_time_alloy_improved} shows the solving
time (in seconds) obtained on the same scenarios but with our improved
approach. Each bar thus corresponds to the solving time for finding a
transformation chain on a given input model and a set of design
patterns of the GoF to apply. For instance, the case ``adapter3''
(which appears to be the most difficult case) consists in applying
three time the adapter pattern on an input model made up of 8 UML
classes. These experiments were conducted on a bi-processor
Intel\texttrademark Xeon\texttrademark CPU E7-4870 at 2.40 GHz with 52
GB RAM and 3 exploration threads.

The reasons for these improvements are further explained
in~\cite{DBLP:conf/euromicro/CastellanosBPSV15} and can be summarized
as follows:

\begin{itemize}
\item partial solutions and parallelization: instead of submitting to
  the Alloy solver the complete chaining problem, we submit partial
  instances of the solution where a partial solution is the result of
  executing a sub-chain. This strategy significantly reduced the size
  of each problem submitted to the solver, thus leading to better
  performances. In addition, it enabled to parallelize of the
  exploration.
\item when selecting one model transformation to apply, early pruning
  was implemented by checking the following constraint: either the set
  of remaining transformations to apply is empty, or there exists at
  least one model transformation that can be applied on the output
  model.
\end{itemize}

We also assessed the scalability of our method on ``long'' model
transformation chains: with 12 transformations to chain (leading to
12! possible orders) and input models of about 30 elements
impacted by the transformations, it took about 2 hours to find a
correct transformation chain. The same case study would not have been
solvable without the optimizations we proposed in this work.

However, an important limitation of this work came from the focus on
structural constraints only. Indeed, beyond such structural
constraints, resulting non functional properties are of prime
importance. This is the reason why we decided to study model
transformations composition as a multi-objective optimization problem,
as described in section~\ref{sec:section_4.4}. In parallel, we
continued studying the validation of model transformation chains
considering their formalization as algebraic graph
transformations. This work is presented in
section~\ref{sec:section_4.3}.

\section{Precondition construction in algebraic graph transformations}
\label{sec:section_4.3}

As stated in the introduction of this chapter, model transformation
chains are complex applications, hence difficult to validate and
verify. In the PhD of Elie Richa, we studied the validation of model
transformation chains in the context of source code generators
qualification. Indeed, when a code generator is used to produce source
code of a critical system, the generated code needs to be
certified. Using qualified tools, the certification effort can be
reduced. Qualifying a code generator is as rigorous and demanding as
certifying critical embedded software. This is the reason why tool
providers need to adopt efficient methodologies in the development and
verification of code
generators~\cite{DBLP:conf/models/RichaBPBR14,DBLP:conf/icmt/RichaBP15,Richa2017}.

As illustrated on figure~\ref{fig:ramses_transformation_chains}, we
consider code generators made up of a transformation chain. Qualifying
an Automatic Code Generator (ACG) requires extensive testing to show
the compliance of the implementation with its requirements. Both the
testing of components in isolation (\emph{i.e.} unit testing) and
the testing of the tool as a whole (\emph{i.e.}  integration testing)
are required.

Given the ACG is a transformation $T$, a unit is a transformation step
$T_i$. Unit testing then consists of producing test models $M_{i,j}$
in the intermediate meta-model $MM_i$, executing $T_i$ over these test
models, and validating the resulting models $M_{i+1,j}$ with a
suitable test oracle. Conversely, integration testing considers the
complete chain, producing test models $M_0$ in the input meta-model
$MM_0$, executing the complete chain, and validating the final result
$M_{N,j}$ (where N is the number of transformations in the chain) with
a suitable test oracle.

A unit test requirement $tr_{i,j}$ of a transformation step $T_i$ is a
constraint over its input meta-model which must be satisfied at least
once during the testing campaign.

Taking inspiration from the work
in~\cite{inc:bauer2011test-suite-quality-for}, we notice the
following: an integration test exercises the complete tool, i.e. all
intermediate transformation steps $T_i$. During the execution of an
integration test, the intermediate models $M_i$ manipulated along the
way can cover unit test cases of the intermediate
transformations. This interesting property of transformation chains
would allow us to use only integration testing to cover unit test
cases.

However, we now need a way to produce new models to cover these unit
test requirements. Given a non-satisfied test requirement $tr_{i,j}$
how can we produce a test model $M_0$ in the input meta-model of the
chain such that upon execution of the integration test, $tr_{i,j}$ is
satisfied?

\begin{figure}[h]
  \centering
  \caption{Backward translation of test requirements}
  \label{fig:post2pre_principles}
\includegraphics[width=\textwidth, trim={0cm 0cm 0cm 0cm},
  clip]{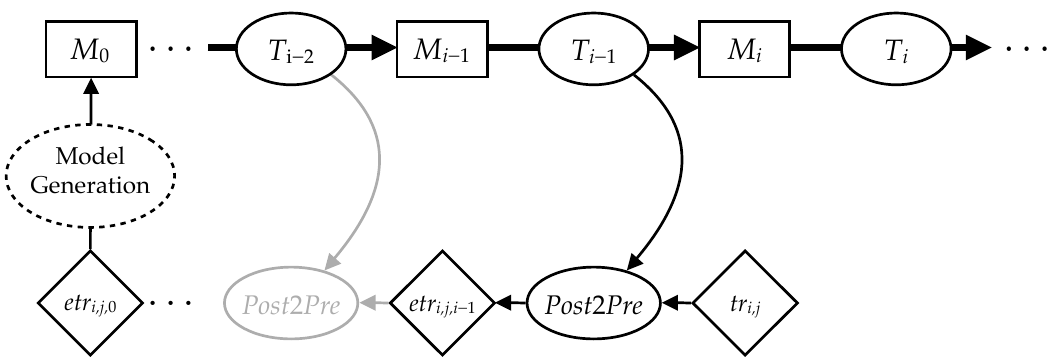}
\end{figure}

In order to answer this question, we proposed a new approach in the
PhD of Elie Richa, as illustrated in
figure~\ref{fig:post2pre_principles}: given a non satisfied test
requirement $tr_{i,j}$ , we consider $tr_{i,j}$ as a post-condition of
the previous transformation step $T_{i-1}$, and design a construction
\emph{Post2Pre} that transforms the post-condition into an equivalent
precondition that ensures the satisfaction of the post-condition. We
call this pre-condition the \emph{equivalent test requirement}
$etr_{i,j,i-1}$ of $tr_{i,j}$ at step $T_{i-1}$.

Recently, the construction of \emph{Post2Pre} has been studied in in
the theory of Algebraic Graph Transformation (AGT)
in~\cite{tec:habel2006weakest-preconditions-for-high-level}
and~\cite{inc:poskitt2013verifying-total-correctness-graph}.

The implementation of this approach in the context of ATL model
transformations required to propose two main contributions. These
contributions were all implemented as components of our Java and
EMF-based tool called ATLAnalyser\footnote{ATLAnalyser,
  https://github.com/eliericha/atlanalyser}:
\begin{enumerate}
\item a translation of OCL (used to express test requirements) into
  \emph{Nested Graph Constraints} and ATL model transformations into
  AGTs. This contribution was presented at the International
  Conference on Model Transformations 2015 and received the \emph{best
    paper award} of the conference. Using ATL instead of AGT, the
  objective was to reduce the complexity of \emph{Post2Pre} by using
  model transformations with a simpler semantics than AGT.  One of the
  challenges of this translation was the translation of the ATL
  \textit{resolveTemp} mechanism, for which dedicated model
  transformation traces had to be defined in the AGT. In addition, an
  ATL rule had to be decomposed in several AGT rules in order to
  implement the different phases of an ATL transformation engine:
  mainly (i) input patterns matching and output objects instantiation,
  and (ii) bindings and objects resolution.
\item an automatic construction of the \emph{weakest liberal
  precondition (wlp)}: a liberal precondition of a graph
  transformation is a precondition for which the existence of a graph
  resulting from the transformation is not guaranteed, and the
  termination of the program is not guaranteed either. In the context
  of ATL transformations, the termination is however guaranteed, and
  the existence of a graph resulting from the transformation should
  also be guaranteed by the definition of the application
  preconditions of the transformation. A liberal precondition is thus
  sufficient in our context. Besides, a precondition c is the weakest
  precondition if it is implied by all other preconditions. However,
  $wlp$ is theoretically infinite for the kind of transformations that
  we analyze. We thus proposed to implement a bounded version of the
  $wlp$, called $scopedWlp$ which proceeds to the construction of
  $wlp$ on a bounded number of transformation rules iteration. In this
  work, we proved $scopedWlp$ provides results applicable to the
  original unbounded transformations.
\end{enumerate}

Both contributions were tested on different model transformations. The
translation from ATL to AGT was validated on several model
transformations available online. The validation method consisted in
transforming an ATL transformation into an AGT transformation.Then
both were executed on the same input model and output models were
compared to ensure they are identical. The ATL to AGT translation was
also applied to model transformation from MATLAB/Simulink to C source
code in CodeGen, a qualifiable source code generator developed by
AdaCore. The corresponding results have been published in the
international journal on Software and System
Modeling~2018~\cite{Richa2017}.

The automated \emph{Post2Pre} construction was tested on simple
transformations but the computational complexity of the algorithms
made the results difficult to produce on realistic examples.We
proposed simplification strategies, but the resulting prototype was
still unable to scale. The theoretical results, however, as well as
the first implementation of this complex technique, is a step forward
towards the formal proof of model transformations. To go beyond these
limitations, significant improvements in the algorithms of the
Post2Pre constructions are necessary but such contributions were out
of the initial scope of Elie Richa's PhD.

Last but not least, we present in next section our contributions on model
transformation variants selection and composition, aiming at producing
models answering at best the trade-off between conflicting NFPs.

\section{Design space exploration by composition of model transformations}
\label{sec:section_4.4}

The last problem we defined in the introduction of this chapter is
variability management in model refinements. Indeed, from our
experience in architecture models refinement, we realized that the
quality of a model transformation chains depends not only on its
structural correctness, but more importantly on the quality attributes
of the resulting model. The correctness of transformation chains was
the object of PhDs described in sections~\ref{sec:section_4.2}
and~\ref{sec:section_4.3}. In the PhD of Smail Rahmoun, we decided to
study the quality of model transformation chains with respect to the
quality attributes of the resulting
models~\cite{Rahmoun2015,DBLP:conf/iceccs/RahmounBP15,Rahmoun2017}. This
research work was also aiming at facilitating the transition from
requirements specification to early architecturel
design~\cite{LoniewskiBBI13b,LoniewskiBBI13a}.

We thus defined model transformations composition as a multi-objective
optimization problem. Indeed, as stated in the introduction of this
document, design alternatives often come into conflict with respect to
their impacts on NFP: a design alternative improves a NFP at the cost
of degrading another NFP of a CPS. In this work, we proposed to define
design alternatives from model transformation variants. We then used
genetic algorithms to compose these variants and have them evolve
towards satisfactory architectures. In order to apply genetic
algorithms to the selection and composition of model transformation
variants, we proposed a generic encoding on which genetic operators
(\emph{i.e.} mutations, crossover) can be applied. Last but not least,
we proposed to express constraints on the output model as boolean
constraints on the application of transformations. To do so, we defined
a dedicated language and the notion of Transformation Rules Catalog
(TRC), as well as a translation of these constraints into a
satisfiability (SAT) problem. Once the SAT problem is solved, model
transformation rules are structured in a way that guarantees that the
application of genetic operators would only produce valid
transformations, \emph{i.e.} transformations producing models
respecting structural validity constraints. Again, structural
constraints take an important role in this work but using genetic
algorithms, we also proposed a framework aiming at improving
non-functional properties.

\begin{figure}[!ht]
\centering
\includegraphics[trim= 0 80 0 0,clip,width=1.0\linewidth]{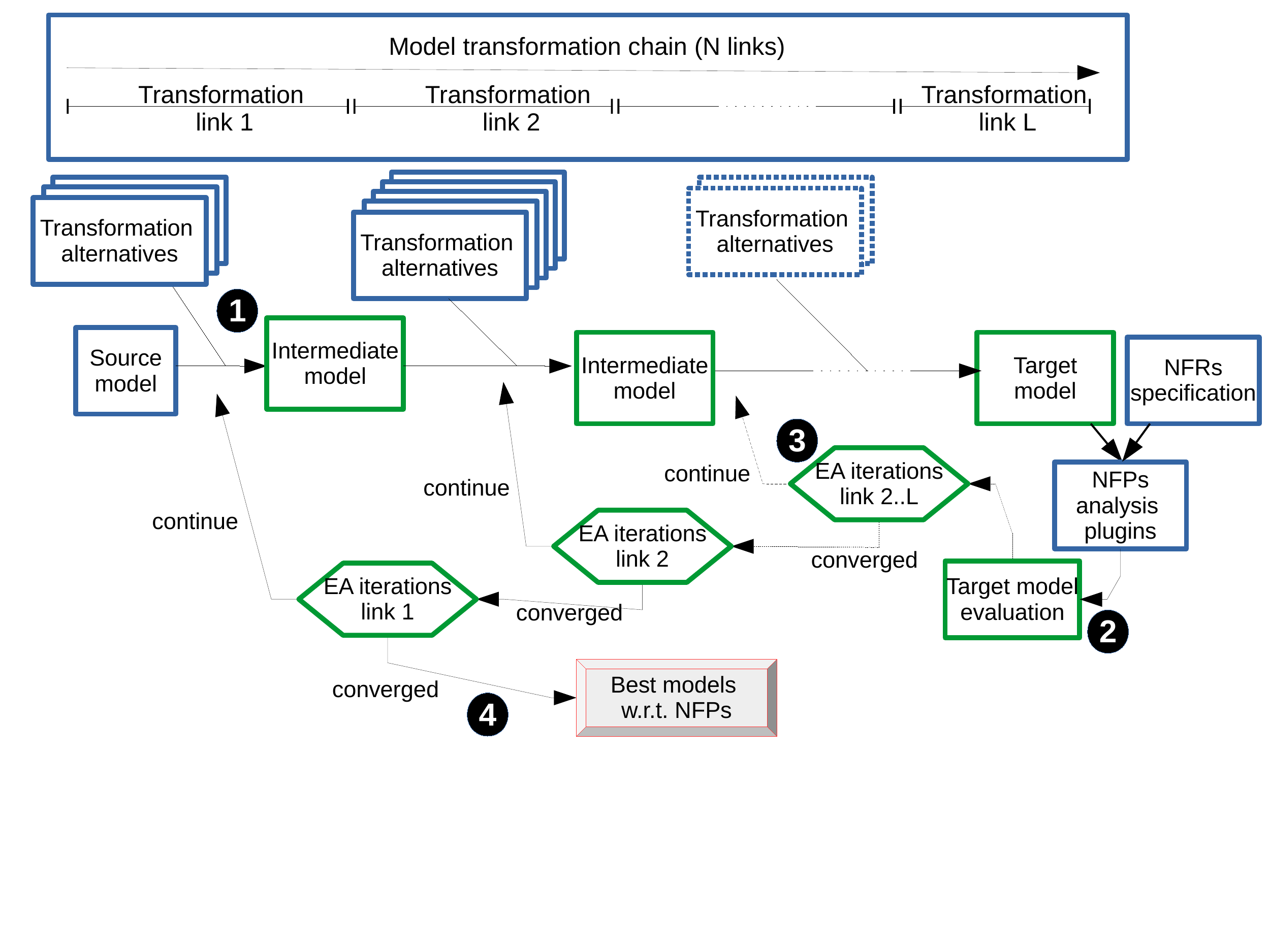}
\caption{Approach Overview}
\label{fig:mu_ramses_approach}
\end{figure}

Figure~\ref{fig:mu_ramses_approach} gives an overview of the process
we proposed in Rahmoun's PhD: from the definition of a model
transformation chain, having for each link a set of transformation
alternatives, we first produce an intermediate model that results from
the composition of these alternatives. This first step is highlighted
with bullet 1 on the figure, and repeated for each link of the
transformation chain until the target model is produced. The
composition mechanism, used in this step of the process, is explained
at the end of this section. Once produced, the target model is
analyzed with respect to NFPs (bullet 2 in
figure~\ref{fig:mu_ramses_approach}), and the analysis results are
used to evaluate composite transformations.

The process we propose is iterative: each iteration produces,
executes, and evaluates a (sub)chain of composite transformations. In
addition, because of the combinatorial complexity of the design space
exploration, it is not possible to enumerate, execute, and evaluate
all the composite transformations. As a consequence, we rely on
evolutionary algorithms (EAs) to implement this iterative exploration
(see bullet 3 in figure~\ref{fig:mu_ramses_approach}). In addition, we can see
in figure~\ref{fig:mu_ramses_approach} that the proposed process is made up of
embedded loops, each loop being dedicated to explore composite
transformations of a given link in the transformation chain. When an
inner loop has converged, other transformation candidates may be
evaluated for the outer loop, thus producing a new intermediate model
for the inner loop. The convergence criteria for each loop relies on
convergence criteria of EAs and is parameterized by an end-user of our
approach.

As far as structural constraints are concerned, we aim at validating
them \textbf{a priori}, i.e. before executing the transformation. As
far as NFPs are concerned, we aim at validating them \textbf{a
  posteriori}, i.e. after executing the transformation. To reach the
objective of \textbf{a priori} validation, we defined application
constraints on model transformation rules in transformation rules
catalog.

Listing~\ref{lst:trc_ramses} provides a subset of the Transformation
Rule Catalog (TRC) we used to describe transformation alternatives for
components replication: \textit{2oo3} and \textit{2*2oo2}. This TRC is
made up of two main parts:
\begin{enumerate}
\item a description of model transformation alternatives, from
  line~\ref{part1_beginning} to line~\ref{part1_end}, lists the set of ATL
  modules and rules being part of each alternative. The
  \textit{2oo3} alternative is made up of transformation rules
  described in section~\ref{sec:section_4.1}. The \textit{2*2oo2} alternative is
  made up of very similar transformation rules.
\item a specification of validity constraints, from
  line~\ref{part2_beginning} to line~\ref{part2_end}. The first one,
  from line~\ref{ct1_beginning} to line~\ref{ct1_end}, specifies that
  when \texttt{m\_PortConnection\_2\_2oo2} is applied on a connection
  (identified as \texttt{cnx} in the constraint), it is necessary
  to apply rule \texttt{m\_Process\_2\_2oo2} on both ends of the
  connection (retrieved executing a OCL helpers called
  \texttt{getDestinationProcess} and \texttt{getSourceProcess} on
  \texttt{cnx}). The second constraint, from
  line~\ref{ct2_beginning} to line~\ref{ct2_end}, specifies that when
  applying \texttt{m\_Process\_2\_2oo2} on a process component
  \texttt{processInstance}, \texttt{m\_PortConnection\_2\_2oo2} should
  be applied on all the connections having \texttt{processInstance} as
  a source (retrieved by applying the OCL helper
  \texttt{getSourceConnectionReference} on
  \texttt{processInstance}). Gathering these two constraints lead to
  ensure that the 2*2oo2 alternative is applied to sets of
  interconnected process components. Very similar constraints are
  expressed for the application of the 2oo3 alternative in the
  remaining of the TRC.
\end{enumerate}

\begin{figure}[ht!]
\begin{lstlisting}[basicstyle=\scriptsize,morekeywords={requires,excludes,and,or,Alternatives,Modules,Constraints,Apply,modules,rules,Helpers,helper},numbers=left,captionpos=b,caption=TRC for the AADL refinement alternatives,label=lst:trc_ramses,frame=single,xleftmargin=\parindent,escapeinside={(*}{*)}]
Modules(*\label{part1_beginning}*)
{
  2_2oo2.atl: m_Process_2_2oo2, m_PortConnection_2_2oo2;
  2oo3.atl: m_Process_2oo3, m_PortConnection_2oo3;
}

Alternatives {
  replication_2_2oo2 { modules: 2_2oo2.atl },
  replication_2oo3 { modules: 2oo3.atl }
}(*\label{part1_end}*)

Constraints {(*\label{part2_beginning}*)
  // 2*2oo2
  Apply(replication_2_2oo2.m_PortConnection_2_2oo2, {cnx}) (*\label{ct1_beginning}*)
  [
    requires (replication_2_2oo2.m_Process_2_2oo2, 
      {getSourceProcess(cnx)}
    ) and requires (replication_2_2oo2.m_Process_2_2oo2, 
      {getDestinationProcess(cnx)}
    )
  ];(*\label{ct1_end}*)
  Apply(replication_2_2oo2.m_Process_2_2oo2, {processInstance}) (*\label{ct2_beginning}*)
  [
    requires (replication_2_2oo2.m_PortConnection_2_2oo2, 
      {getSourceConnectionReference(processInstance)}
    )
  ];(*\label{ct2_end}*)
  // similar constraints for 2oo3
  ...
}(*\label{part2_end}*)
\end{lstlisting}
\end{figure}

When applying ATL model transformation variants, such as the 2oo3 and
2*2oo2 replication patterns, we consider as optimization variables the
choice of each variant applied to elements of the input model. To
define such alternatives more formally, we first provide a definition
of transformation rules instantiation:

\begin{mydef}
\label{TRI_definition}
A \textbf{transformation rule instantiation} $TRI_{i}$ is the
application of a transformation rule on an ordered set of elements
from the source model. In the remainder of this section, we say such
TRIs are non-confluent. It can be represented as a tuple
$<R,E_{i},A_{i}>$, where:
\begin{enumerate}
 \item $R$ represents the applied transformation rule;
 \item $E_{i}$ is $i^{th}$ tuple of elements in the source model;
 \item $A_{i}$ is the set of actions that $R$ executes when it is
   applied to $E_{i}$.
\end{enumerate}
\end{mydef}

Given this definition, alternative TRIs exist when more than one rule
can be applied to the same tuple of elements in the source
model. Formally, this means :

\begin{equation}
\begin{split}
\exists (R,R')\ s.t.\ R \neq R' and\ TRI_i=<R,E_i,A_{i}>\\ and\ TRI_j=<R',E_j,A_{j}>\ and\ E_i=E_j
\end{split}
\end{equation}

According to the semantics of ATL, such situation has to be solve by
selecting, among all the possible TRIs, a subset where non-confluence
has been eliminated.

To do so, we rely on a simple selection function defined as follows:

\begin{pushright}$Sel:$\hspace*{0.5cm}$ T $\hspace*{0.6cm}$\rightarrow$\hspace*{0.2cm}$ \mathbb{B}=\{True, False\}$\\
  \hspace*{0.8cm} $TRI $\hspace*{0.4cm}$\rightarrow$\hspace*{0.2cm}$ b$, where $b$ is $True$ if
  $TRI$ should be included, and\\
  \hspace*{4.8cm} $False$ if $TRI$ should be
  excluded from $T$.
\end{pushright}

The selection of TRIs may be decomposed into the following formulas:\\
\textit{1)} $AtLeastOne$, dedicated to ensure that at least one of the
non-confluent TRIs, gathered in a set $S$, is selected:
\begin{small}
\begin{equation}
\label{eq:selection_one}
AtLeastOne(S) = \underset{i=1}{\overset{P}{\bigvee}} Sel(TRI_i)
\end{equation}
\end{small}
\textit{2)} $AtMostOne$, dedicated to ensure that at most one of the
non-confluent TRIs is selected:
\begin{small}
\begin{equation}
\label{eq:only_one}
AtMostOne(S) = \underset{i=1,j=1,i \neq j}{\overset{i=P,j=P}{\bigwedge}} \neg(Sel(TRI_i) \wedge Sel(TRI_j))
\end{equation}
\end{small}
Combining equations~(\ref{eq:selection_one}) and~(\ref{eq:only_one}),
we obtain $SelectOne$, dedicated to select exactly one TRI from $S$
(set of non-confluent TRIs):
\begin{small}
\begin{equation}
\label{eq:exactly_one}
SelectOne(S) = AtLeastOne(S) \wedge AtMostOne(S)
\end{equation}
\end{small}

Alternative TRIs can be detected by applying the pattern matching part
of alternative model transformations on an input model. Enforcing the
respect of equation (\ref{eq:exactly_one}) boils to select exactly
one alternative per element of the input model.

In addition, validity constraints expressed in the TRC are transformed
into boolean validity constraints on TRIs selection of the form:

\begin{small}
\begin{equation}
\label{eq:dependencies_definition}
ValidityConstraints =
\underset{i=1}{\overset{N}{\bigwedge}}(Sel(TRI_i) \Rightarrow BoolExpr(T_i))
\end{equation}
\end{small}

where $T_i$ is a subset of TRIs, and $BoolExpr$ is a boolean
expression over $TRIs$ in $T_i$, using (i) the $Sel$ function, (ii)
simple boolean operators $and$, $or$, and $not$ ($\wedge$, $\vee$, and
$\neg$).

Finally, for all the sets of alternative rule instantiations
${S_i}_{i\in[0..Q]}$, the selection of a valid set of TRIs boils to
evaluate the satisfiability of the boolean formula:
\begin{small}
\begin{equation}
\label{eq:selection}
\underset{i=1}{\overset{Q}{\bigwedge}} (SelectOne(S_i)) \wedge ValidityConstraints
\end{equation}
\end{small}

However, our objective is not to define a set of valid model
transformations, but to define the genome encoding of a genetic
algorithm. To do so, instead of solving the SAT problem induced by
equation~(\ref{eq:selection}) at once, we aim at grouping TRIs
involved in the same validity constraints into partitions called
Atomic Transformation Instantiations (ATI). Doing so, each ATI become
a potential gene in the genome of individuals in the genetic
algorithm population.

Grouping TRIs into ATIs is done as follows: we first reorganize
equation~(\ref{eq:selection}) under a conjunctive normal form.  We
call $B$ the set of boolean expressions in the conjunction, and build
a partition of $B$: we group such expressions into smallest non-empty
subsets of $B$ in such a way that every $TRI$ is used in expressions
of one and only one of the subsets. These subsets are called the parts
of the partition, and we note $\beta_q$ the boolean formula
corresponding to the $q^{th}$ part of the partition. We note $ATI_{q,
  i}$ the $i^{th}$ solution of $\beta_q$. We then structure a
composite transformation (CT) by choosing, for each possible value of
q, one of the solutions of $\beta_q$. An individual of the genetic
algorithm population is then a CT structured as an array of
$ATI_{i,j}$ where $i$ is one of the partitions of~(\ref{eq:selection})
and $j$ is the identifier of a solution of this partition.

\begin{figure}[!ht]
\centering
\includegraphics[clip,width=\linewidth]{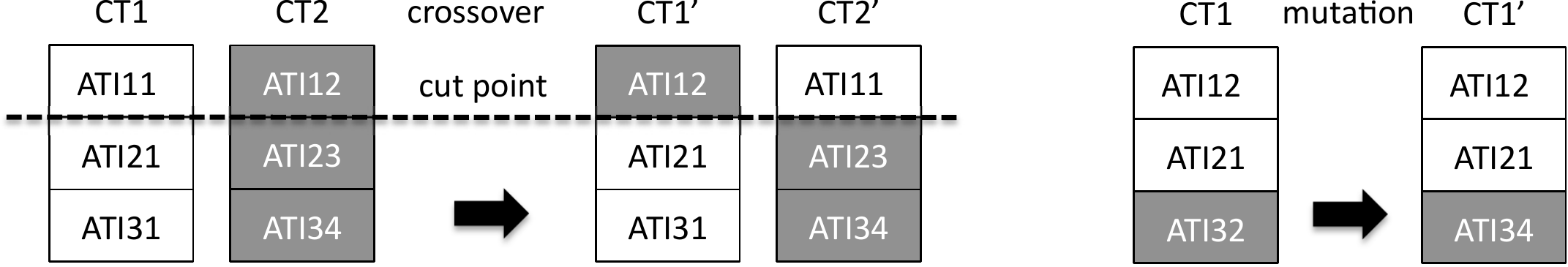}
\caption{Crossover and Mutation Operators}
\label{fig:crossover_mutation}
\end{figure}

Figure~\ref{fig:crossover_mutation} illustrates the structure of a
genome as an array of atomic transformation instantiations
$ATI_{i,j}$, as well as the application of the crossover operator on
two genomes $CT_1$ and $CT_2$ and the application of the mutation
operator $CT_1$ to obtain $CT_1'$.

Note that the construction of the genome by partitioning solutions of
equation~(\ref{eq:selection}) ensures that the result of the genetic
operations (\emph{i.e.} crossover, mutation) leads to the production
of valid transformations.

Resulting composite transformations are then applied to input models
in order to produce intermediate models on which NFPs can be analyzed.

This method has been successfully applied on two case-studies inspired
from the railway domain:
\begin{enumerate}
\item a selection of implementations of the periodic delayed
  communications introduced in section~\ref{sec:section_3.3}. Indeed,
  three implementations of this communication patterns were proposed,
  with conflicting impact on three NFPs: timing performance, memory
  footprint, and maintainability. The corresponding results were
  published in the ICECCS conference~\cite{Rahmoun2017}.
\item a chain of model transformations made up of (i) the selection of
  the replication pattern (2oo3 or 2*2oo2), (ii) the binding of
  replicated processes and connections onto processors and busses with
  the following validity constraint: replicated components should not
  use the same hardware resources. The corresponding results were
  published in the international journal on Software and Systems
  Modeling~2018~\cite{Rahmoun2017}.
\end{enumerate}

\section{Concluding remarks}

In this chapter, we have presented our contributions to answer the
following questions:
\begin{enumerate}
\item how to verify the correctness of model transformations
  structured as model transformation chains?
\item how to compose model transformations to define optimal (or near
  optimal) refinement steps?
\end{enumerate}

\noindent
To answer these questions, we have proposed to formalize model
transformations along with structural constraints on the output models
of these transformations. This formalization was then used either to
search for model transformation chains enforcing the structural
constraints, or to enhance a model transformation validation framework
by automating the generation of test requirements. Reusing this notion
of structural constraints, we also defined a framework combining model
transformations, SAT solving techniques and genetic algorithms to
automate the search for near-optimal architecture refinements.

By providing answers to some of the research questions presented here
above, our work provides original methods to improve the reliability
of software development in CPSs: it enables to check as early as
possible the correctness of model transformations, and to automate the
search for near-optimal refinements.

The work we have done on model transformation of AADL models gave us
several inputs to start new research activities in the domain of Model
Driven Engineering for CPSs. In particular, we plan to extend our
framework to deal with security in CPSs.

Indeed, security is becoming an important concern in the design and
development of CPS. The way to deal with interactions among security
design patterns and other design techniques of CPS is still an open
question. In the PhD of Jean Oudot, started with IRT SystemX and the
Nanyang Technological University (NTU) Singapore, we aim at proposing
new techniques to (i) evaluate architectures security, and (ii)
optimize them with respect to this criteria as well as traditional
concerns of CPS (timing performance, safety, cost, and/or energy
consumption). This work aims at extending Smail Rahmoun's results with
a dedicated focus on cyber security, which would require to
drastically revise the design space exploration method since the
nature of threats in cyber security is very different from the nature
of threats in safety.



\chapter{Conclusion and Perspectives}
\label{chap:chapter5}



\minitoc

\vspace{1.5cm}

In this chapter, we conclude this document before to present of our
future research directions.

\section{Concluding remarks}

Our concluding remarks aim at summarizing our cintributions, as well
as comparing our contributions to the state of the art. This
comparison will focus on a few existing frameworks, and the comparison
will be based on the set of expecte capabilities for model-based
engineering frameworks dedicated to critical, real-time and embbeded
software.

\subsection{Overview of contributions}

Our initial objective was to propose methods to improve the
reliability of CPSs design process. We decided to pursue this
objective in the framework of MDE techniques. Our first intent was to
use models as a mean to bridge the gap between requirements definition
and source code production. Indeed, models help bridging this gap by
representing systems under design with different abstraction levels,
ranging from high level specifications to implementation models. We
proposed to use architecture models as design artifacts, and model
transformations to automate their refinement steps up to the
production of source code.

\noindent
Building on this first idea, we defined a set of challenging problems,
digged into existing solutions, and proposed improvements over the
state-of-art. We highlight hereafter some noticeable aspects of our
work:

\begin{enumerate}
\item as often as possible, we prototyped our contributions on top of
  AADL in order to improve their applicability to realistic case
  studies: AADL is a standardized language built from collaborations
  with companies in the domain of CPS, and AADL is quite a complex
  language (its meta-model is made up of more than 260 meta classes).
\item we defined the first implementation of a deterministic model of
  computation and communications based on mixed criticality DAGs and
  periodic delayed communications. We defined the corresponding subset
  of AADL and prototyped the compilation of this AADL subset in a
  source code generator.
\item we proposed to formalize model transformations to ease their
  validation and verification. We have shown how this formalization
  can help to find correct model transformation chains and validate
  model transformations using integration tests.
\item we invented a design space exploration framework based on model
  transformations composition, combining model transformations,
  constraint solving and multi-objective optimization techniques.
\end{enumerate}

\noindent
These contributions have been integrated in the RAMSES in order to
experiment them on industrial case studies. This effort was necessary
to perpetuate the works of PhD students I had the opportunity to
co-supervise. In the remainder of this section, we compare RAMSES with
similar frameworks from the state of the art.

\subsection{Comparison of RAMSES with existing frameworks}

In order to compare RAMSES with existing frameworks from the state of
the art, we propose to use the following citeria. The selection of
these citeria relies on our expertise in the domain of criticcal and
real-time embedded systems. It is incomplete, and aims primarily at
showing the wide spectrum of contributions prototyped in the RAMSES
framework.

\noindent
The criteria we have selected are the capabilities of existing frameworks to:
\begin{itemize}
\item[C1] \textbf{refine} software models; necessary to bridge the gap
  between requirements specification and software implementation while
  progressively producing and analysing models.
\item[C2] \textbf{source code generation}; here, the objective is to
  fasten the production of software while easing its maintenance by
  relying on automate code generation techniques.
\item[C3] explore and select the best design alternatives; this is all
  the more important in the context of critical real-time embedded
  systems where non-functional properties play a very important
  role. In addition, this capability requires to deal with
  \textbf{multi-objective optimization} problems.
\item[C4] \textbf{interoperate} with other frameworks, in order to
  avoid re-defining existing analysis or verification techniques, but
  re-use them instead.
\item[C5] produce \textbf{scheduling configuration}; this criteria is
  obviously specific to the domain of real-time system, but a very
  fundamental one since software integration issues in this domain are
  often related to scheduling and timing performance issues.
\end{itemize}

\begin{figure}[h]
\centering
\includegraphics[width=\textwidth, trim={0cm 0cm 0cm 0cm}, clip]{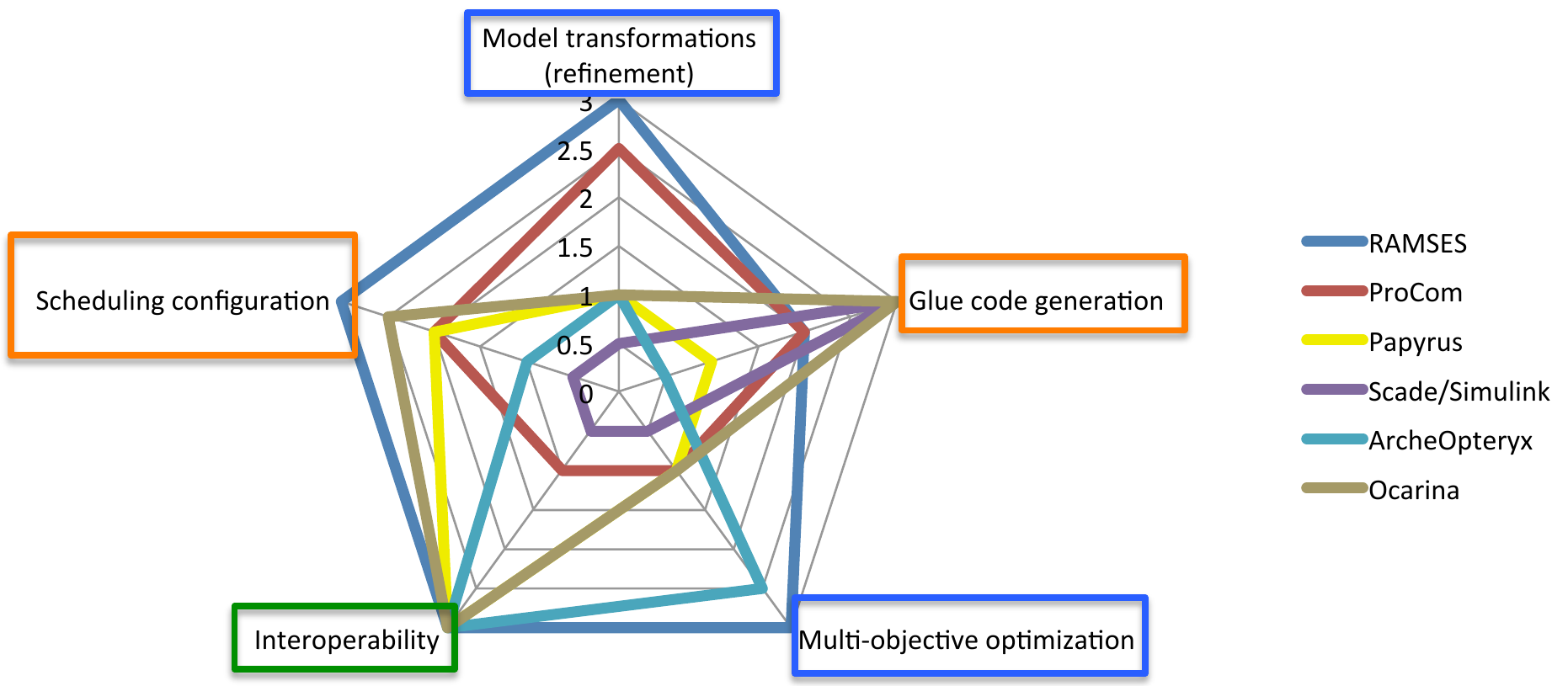}
\caption{Comparison between RAMSES and similar frameworks}
\label{fig:RAMSES_vs_SoA}
\end{figure}

\noindent
In order to compare RAMSES with existing frameworks from the
state-of-the art, we only selected existing frameworks that, to our
best knowledge, cover all these criteria. Without this selection,
potentially incomplete, the body of literature covering these criteria
is far too prolific to provide a meaningful comarison. The frameworks
we have selected are: ProCom~\cite{ProCom},
Papyrus/MARTE~\cite{Gerard15},
Scade~\footnote{\url{https://www.ansys.com/products/embedded-software/ansys-scade-suite}},
Simulink~\footnote{\url{https://www.mathworks.com/products/simulink.html}},
ArcheOpteryx~\cite{DBLP:conf/mompes/AletiBGM09} and
Ocarina~\cite{LasnierZPH09}. Figure~\ref{fig:RAMSES_vs_SoA} provides
an overview of the comparison between RAMSES and these selected
frameworks. In the reminder of this section, we provide a brief
argumentation for this comparison.

\subsubsection{C1: Model transformations (refinements)}
Models refinement is the main focus of the method introduced in this
document. Among the selected frameworks, the ProCom component model
has been used to define model
refinements~\cite{ProCom-Refinement,Yin2071}.

\noindent
Other framework, based on rich modeling languages (e.g. MARTE and
AADL), could also implement similar refinements as those proposed in
RAMSES. This is typically the case for Papyrus, ArcheOpteryx, and
Ocarina. However, to our best knowledge, model refinements have not
been prototyped in the considered frameworks.

\noindent
Scade and Simulink could also be used to proceed to models refinement,
but their scope and utility would be less significant as these
languages mainly focus on modeling the funcional aspects of critical
real-time and embedded software.

\subsubsection{C2: Source/glue code generation}
In terms of source code generation for the functional parts of a
software architecture, Scade and Simulink are, by far, the
references. When it comes to code generation for the technical
architecture (\emph{i.e.} operationg system, middleware), Ocarina is
the most mature of the frameworks we considered. Developped in the
context of the ASSERT project, it has been maintained for the needs of
the European Space Agency in the scope of the TASTE project. ProCom
also provides code generation capabilities, though the focus of this
work was more on the preservation of the semnatics than on the support
of various programming languages or operating
systems~\cite{ProCom-Borde}. Papyrus has integrated code generation
techniques as well, mainly focusing on code generation for component
diagrams and state-charts. Finally, code generation was not the main
focus of ArcheOpteryx.

\subsubsection{C3: Multi-objective optimization}
With respect to design space exploration, ArcheOpteryx is, among the
frameworks we considered, the richest in terms variety of prototyped
optimization methods. However, by combining model transformations and
optimisation techniques, RAMSES allows to (i) reuse a generic design
exploration framework, i.e. the design alternatives being expressed
using transformation rules catalogs, (ii) focus only on valid design
alternatives in terms of structural constraints satisfaction, and
(iii) reuse existing analysis method to evaluate the quality
attributes of these alternatives.

\noindent
Other frameworks have been used to implement preliminary works on
models optimization, \emph{e.g.} is Papyrus/MARTE~\cite{MuraMP08}, or
in Ocarina~\cite{GillesH11}.

\noindent
Finally, Scade and Simulink are less usable for automated design space
exploration as they mainly focus on modeling the funcional aspects of
critical real-time and embedded software.

\subsubsection{C4: Interoperability}

This criteria is mainly achieved by relying on standardized modeling
languages, which is the case for RAMSES, Papyrus, Ocarina, and
ArcheOpteryx. The ProCom component model is not standardized but the
language specification is publicly
available~\cite{Bures2008ProComT}. Finally, Scade and Simulink use
proprietary modeling languages.

\subsubsection{C5: Scheduling configuration}
Ocarina can be used to anlyse various types of real-time schedulers,
and to configure the scheduler of several operating systems. However,
compared to RAMSES, this framework supports a smaller subset of models
of computation. For instance, our recent works on DAGs and
Mixed-criticality scheduling is not part of the subset of AADL
supported by Ocarina. This statement is also true for periodic delayed
communications, even though it has less impact on the scheduler
configuration.

\noindent
Papyrus/MARTE and ProCom can also be used to analyse and configure the
scheduler of real-time operating system. However, only a few research
works report new contributions in this area.

\noindent
Finally, scheduling configuration has not been the main focus of
research works in ArcheOpteryx, Scade, or Simulink.

\noindent As on can easily understand, our work is at the intersection
of the fields of model-driven engineering, operational research, and
scheduling of real-time systems. The perspectives we present in the
remainder of this chapter aim at extending this work in three main
directions: (i) security of software architectures for connected CPSs,
(ii) autonomy of complex and/or critical CPSs, and (iii) uncertainty
management in design space exploration activities. We develop these
perspectives in the next sections.

\section{Future Research Directions}

\subsection{Software Security in Cyber-Physical Systems.}

Cyber-physical systems are increasingly connected with their external
environment. In particular, in the transportation domain (rail,
automotive and avionics), vehicles are now connected to their
infrastructure, to mobile devices, and to the internet. This evolution
allows manufacturers to deliver new services, but exposes these
systems to malicious actions from hackers.  In parallel to this, more
and more functionalities are introduced into the transportation
systems in the form of software systems. Some of these functionalities
are also critical, since their failure, or a malicious takeover of
these functions can have catastrophic consequences. The link between
vulnerabilities and safety of these systems was considered negligible
as long as the potentially affected functions were not critical, but
this is less and less true.

\noindent
Indeed, security vulnerabilities nowadays can jeopardize the safety of
critical embedded real-time systems, and thus endanger its users. The
introduction of security counter-measures in the architectures of
these embedded systems is necessary, provided they do not degrade
significantly the safety, performance or cost of the system.

\noindent
We plan to study this problem with two complementary viewpoints:
\begin{enumerate}
\item in the PhD of Jean Oudot (started in 09/2017), we aim at defining
  a design space exploration framework focusing on security metrics
  and counter measures. This will require to quantify a CPS
  architecture security, and the impact of security counter measures
  on security, safety, performance, and cost. The objective is to
  define the best set of counter measures to deploy in the CPS
  architecture. Another objective of this work is to propose a generic
  exploration framework, \emph{i.e.} usable in different fields of
  cyber-physical systems and in particular in the transportation
  domain: avionics, railway, and automotive. This PhD is funded by the
  IRT\footnote{Institut de Recherche Technologique/Technological
    Research Institute} SystemX in the CTI (Cybersecurity of
  Intelligent Transports) and is co-supervised with Arvind Easwaran
  from the Nanyang Technological University (NTU) of Singapore.
\item in the PhD of Maxime Ayrault (started in 10/2018), funded by the
  industrial chaire on Connected Cars and Cyber Security (C3S) we
  adopt a complementary viewpoint: how to delay attacks and/or their
  propagation? The basic idea behind this work is to consider that (i)
  all vulnerabilities and attacks paths cannot be anticipated at
  design time, and (ii) once a system is deployed, attackers have time
  to study the system and discover new vulnerabilities and attacks
  paths. In this work, we will study the integration of moving target
  defense mechanisms in CPSs architectures: the idea of MTD is to
  reconfigure the CPS architecture periodically so that
  vulnerabilities become unstable and attacks difficult to discover
  and deploy. Finally, the objective will be to maximize the attacker
  learning time and minimize the probability of success of an
  attack. This approach is particularly challenging in CPSs since
  computation and communication resources are very limited.
\end{enumerate}

\noindent
The approach outlined in the work of Maxime Ayrault can be generalized
and extended to improve the autonomy of CPS. We present this
perspective of our work in the next section.

\subsection{Autonomy by Reconfiguration of CPS Applications}

Architecture models such as AADL provide means to describe different
software configurations as well as services or techniques to
reconfigure software applications. In AADL, this is done by defining
operational modes along with a particular configuration per
mode. Reconfigurations are described with transitions between
operational modes. This type of reconfigurations is pseudo-dynamic:
they are executed at runtime but all the possible configurations and
transitions among them have been defined at design time.

\noindent
Modern CPS are expected to be used for a very long time: a train, for
instance, is exploited for decades. The algorithms running on these
systems are thus expected to evolve over time, as well as the devices
they are connected to, they interact with. In addition, some of these
systems cannot be stopped for maintenance purpose, and more generally
maintenance should disrupt as little as possible the operation of
CPSs. Eventually, the objective of this perspective is to improve the
robustness of CPSs by having capabilities to reconfigure them even in
case of situations or evolution needs that were not foreseen in their
initial development.

\noindent
In this context, we plan to study how MDE techniques presented in this
document can help to define reconfigurations beyond pseudo-dynamic
reconfigurations. The idea would be to (i) model platform
reconfiguration capabilities, (ii) use design exploration techniques
to find new configurations, and (iii) develop new design space
exploration techniques to find out how to progressively move from the
current configuration to the targeted configuration.

\noindent
This boils down to find ways to update systems as dynamically as
possible, delegating design space exploration and target configuration
selection to an external model based infrastructure. This model-based
infrastructure would be responsible for finding how to deploy the new
configuration, and how to move from the current configuration to the
new one.

\noindent
The objective of using models to reason about adaptation is to shorten
as much as possible the delay between the detection of new needs and
the implementation of a new solution. Another objective to automate
the migration from one solution to another one.

\noindent
This perspective is ambitious, and is actually being developed as a
European project proposal with partners met in the European COST
action MPM4CPS (Multi-Paradigm Modeling for Cyber Physical Systems).

\subsection{Uncertainty Management in Design Space Exploration}

Uncertainty management in design decisions is the third perspective
presented in this chapter. It is strongly linked to the topic of
architectures exploration as this activity, exploration, aims to
provide architecture candidates for a decision process to select one
(or a subset) of them.

\noindent
This perspective is also related to requirements engineering insofar
architecture exploration aims to develop an architecture (or set of
architectures) that satisfy at best a set of non-functional
requirements. However, these requirements are often in conflict:
improving certain non-functional properties of an architecture
requires adopting solutions that degrade the architecture with respect
to other non-functional properties. It is in this context that we have
extended the requirements modeling language called RDAL (Requirements
Analysis and Definition
Language)~\cite{DBLP:conf/euromicro/LoniewskiBBI13,DBLP:conf/re/BlouinBSGDBTN17}
and have developed a method of architectures exploration based on the
composition of model transformations~\cite{Rahmoun2017}.  However,
these studies do not consider an important aspect of model-based
design: uncertainty.

Indeed, in the early stages of the development process, decisions rely
on models containing uncertain data. Those data are uncertain either
because they are difficult to estimate by essence (\emph{e.g.}  errors
probability distribution), or because the implementations they could
be measured from are not available yet (\emph{e.g.} in construction).

As a consequence, decision-makers should be aware of how much
confidence they may have in the data upon which they base their
judgement. Somehow, this problem could be tackled by considering design
activities as an optimization problem aiming at maximizing design
margins (\emph{i.e.} difference between an estimated quality attribute
and the lowest acceptable level of quality for this
attribute). Roughly, increasing margins is expected to improve the
confidence of the decision maker, at least with respect to the risk of
necessary rework during the integration and validation phases of the
development process.

However, as explained in the introduction of this document, increasing
margins may also lead to poor quality of service. This is, for
example, the philosophy behind mixed-criticality scheduling as the
objective is to improve computation resources usage by reducing
margins in low criticality modes while preserving margins for high
criticality tasks in high mode.

\noindent
In essence, a model is an abstraction of an object that focuses on
predominant characteristics of this object. The evaluation of this
abstraction therefore contains a degree of uncertainty since it deals
with a model and not the object itself.  Beyond the activity of
abstraction, the information contained in a model may come from more
or less reliable sources. For instance, this information can be
estimated (by expert judgment for example) or imprecise (using
inaccurate or approximate analysis methods). To some extent, RAMSES
reduces uncertainty by executing successive refinements and
revaluations of real-time embedded system architectures. This reduces
the uncertainty due to abstraction, exhibiting architectures for which
the non-functional properties are evaluated with lower and lower
abstraction level.

\noindent
However, uncertainty must also be considered, evaluated, and
processed, at the decision-making level, in order to guide the
exploration of architectures. This requires considering, in
particular, uncertainty due to estimated or imprecise information
contained in a model. In the future, we aim at studying uncertainty
modeling, and its impact on the related decision-making and
architectural exploration techniques.

\noindent
Existing studies consider a formalization with fuzzy
mathematics~\cite{DBLP:conf/icse/EsfahaniMR13} to guide the
construction of products from a set of feature variants. It would be
interesting to extend such an approach in the scope of design space
exploration by model refinements. In particular, considering
uncertainty metrics would help the robustification of design processes
in a semi-automatic way: uncertainty would evolve by interleaving
automatic exploration phases with studies dedicated to lower the level
of uncertainty and update NFP properties. In addition, the risk
associated to uncertainty could be controlled by feedbacks from
previous experience.

\noindent
More generally, this perspective aims to rationalize the attitude of
decision makers facing uncertainty in an architecture design
activity. We plan to pursue this objective by taking advantage of
model transformation traces, analysis results with uncertainty
modeling, and a description of the multi-criteria decision process.


\begin{spacing}{0.9}


\bibliographystyle{apalike}
\cleardoublepage
\bibliography{References/references} 

\begin{thebibliography}{}

\bibitem[Aleti et~al., 2009]{DBLP:conf/mompes/AletiBGM09}
Aleti, A., Bj{\"{o}}rnander, S., Grunske, L., and Meedeniya, I. (2009).
\newblock Archeopterix: An extendable tool for architecture optimization of
  {AADL} models.
\newblock In {\em {ICSE} 2009 Workshop on Model-Based Methodologies for
  Pervasive and Embedded Software, {MOMPES} 2009, May 16, 2009, Vancouver,
  Canada}, pages 61--71.

\bibitem[Allen, 1997]{Allen97Thesis}
Allen, R. (1997).
\newblock {\em A Formal Approach to Software Architecture}.
\newblock PhD thesis, Carnegie Mellon, School of Computer Science.
\newblock Issued as CMU Technical Report CMU-CS-97-144.

\bibitem[Amnell et~al., 2004]{10.1007/978-3-540-40903-8_6}
Amnell, T., Fersman, E., Mokrushin, L., Pettersson, P., and Yi, W. (2004).
\newblock Times: A tool for schedulability analysis and code generation of
  real-time systems.
\newblock In Larsen, K.~G. and Niebert, P., editors, {\em Formal Modeling and
  Analysis of Timed Systems}, pages 60--72, Berlin, Heidelberg. Springer Berlin
  Heidelberg.

\bibitem[Arendt et~al., 2010]{Arendt:2010:HAC:1926458.1926471}
Arendt, T., Biermann, E., Jurack, S., Krause, C., and Taentzer, G. (2010).
\newblock Henshin: Advanced concepts and tools for in-place emf model
  transformations.
\newblock In {\em Proceedings of the 13th International Conference on Model
  Driven Engineering Languages and Systems: Part I}, MODELS'10, pages 121--135,
  Berlin, Heidelberg. Springer-Verlag.

\bibitem[{As-2 Embedded Computing Systems Committee SAE}, 2004]{aadl04}
{As-2 Embedded Computing Systems Committee SAE} (2004).
\newblock {Architecture Analysis \& Design Language ({AADL})}.
\newblock SAE Standards n$^{o}$ AS5506.

\bibitem[Baruah, 2016]{baruah2016federated}
Baruah, S. (2016).
\newblock The federated scheduling of systems of mixed-criticality sporadic dag
  tasks.
\newblock In {\em Real-Time Systems Symposium}. IEEE.

\bibitem[Bauer et~al., 2011]{inc:bauer2011test-suite-quality-for}
Bauer, E., K{\"u}ster, J., and Engels, G. (2011).
\newblock Test suite quality for model transformation chains.
\newblock In Bishop, J. and Vallecillo, A., editors, {\em Objects, Models,
  Components, Patterns}, volume 6705 of {\em Lecture Notes in Computer
  Science}, pages 3--19. Springer Berlin Heidelberg.

\bibitem[Berthomieu et~al., 2009]{DBLP:conf/adaEurope/BerthomieuBCDFV09}
Berthomieu, B., Bodeveix, J., Chaudet, C., Dal{-}Zilio, S., Filali, M., and
  Vernadat, F. (2009).
\newblock Formal verification of {AADL} specifications in the topcased
  environment.
\newblock In {\em Reliable Software Technologies - Ada-Europe 2009, 14th
  Ada-Europe International Conference, Brest, France, June 8-12, 2009.
  Proceedings}, pages 207--221.

\bibitem[Bharathi et~al., 2008]{bharathi2008characterization}
Bharathi, S., Chervenak, A., Deelman, E., Mehta, G., Su, M.-H., and Vahi, K.
  (2008).
\newblock Characterization of scientific workflows.
\newblock In {\em Workshop on Workflows in Support of Large-Scale Science}.
  IEEE.

\bibitem[Bini and Buttazzo, 2005]{bini2005measuring}
Bini, E. and Buttazzo, G.~C. (2005).
\newblock Measuring the performance of schedulability tests.
\newblock {\em Real-Time Systems Symposium}, 30(1).

\bibitem[Blouin et~al., 2017]{DBLP:conf/re/BlouinBSGDBTN17}
Blouin, D., Barkowski, M., Schneider, M., Giese, H., Dyck, J., Borde, E.,
  Tamzalit, D., and Noppen, J. (2017).
\newblock A semi-automated approach for the co-refinement of requirements and
  architecture models.
\newblock In {\em {IEEE} 25th International Requirements Engineering Conference
  Workshops, {RE} 2017 Workshops, Lisbon, Portugal, September 4-8, 2017}, pages
  36--45.

\bibitem[Borde and Carlson, 2011]{ProCom-Borde}
Borde, E. and Carlson, J. (2011).
\newblock Towards verified synthesis of procom, a component model for real-time
  embedded systems.
\newblock In {\em Proceedings of the 14th International {ACM} Sigsoft Symposium
  on Component Based Software Engineering, {CBSE} 2011, part of Comparch '11
  Federated Events on Component-Based Software Engineering and Software
  Architecture, Boulder, CO, USA, June 20-24, 2011}, pages 129--138.

\bibitem[Borde et~al., 2014]{DBLP:conf/rsp/BordeRCPSD14}
Borde, E., Rahmoun, S., Cadoret, F., Pautet, L., Singhoff, F., and Dissaux, P.
  (2014).
\newblock Architecture models refinement for fine grain timing analysis of
  embedded systems.
\newblock In {\em 25nd {IEEE} International Symposium on Rapid System
  Prototyping, {RSP} 2014, New Delhi, India, October 16-17, 2014}, pages
  44--50.

\bibitem[Bougueroua et~al., 2004]{DBLP:conf/pdcn/BouguerouaGM04}
Bougueroua, L., George, L., and Midonnet, S. (2004).
\newblock Task allowance for failure prevention of fixed priority scheduled
  real-time java systems.
\newblock In {\em Proceedings of the {IASTED} International Conference on
  Parallel and Distributed Computing and Networks, Innsbruck, Austria, February
  17-19, 2004}, pages 375--380.

\bibitem[Bozga et~al., 2009]{Bozga:2009:MSS:1629335.1629347}
Bozga, M.~D., Sfyrla, V., and Sifakis, J. (2009).
\newblock Modeling synchronous systems in bip.
\newblock In {\em Proceedings of the Seventh ACM International Conference on
  Embedded Software}, EMSOFT '09, pages 77--86, New York, NY, USA. ACM.

\bibitem[Bruneton et~al., 2006]{Bruneton:2006:FCM:1152333.1152345}
Bruneton, E., Coupaye, T., Leclercq, M., Qu{\'e}ma, V., and Stefani, J.-B.
  (2006).
\newblock The fractal component model and its support in java: Experiences with
  auto-adaptive and reconfigurable systems.
\newblock {\em Softw. Pract. Exper.}, 36(11-12):1257--1284.

\bibitem[Bures et~al., 2008a]{Bures1279}
Bures, T., Carlson, J., Crnkovic, I., Sentilles, S., and Feljan, A.~V. (2008a).
\newblock Procom - the progress component model reference manual, version 1.0.
\newblock Technical Report ISSN 1404-3041 ISRN MDH-MRTC-230/2008-1-SE.

\bibitem[Bures et~al., 2008b]{Bures2008ProComT}
Bures, T., Carlson, J., Crnkovic, I., Sentilles, S., and Vulgarakis, A.
  (2008b).
\newblock Procom - the progress component model reference manual, version 1.0.

\bibitem[Burns and Davis, 2017]{Burns:2017:SRM:3161158.3131347}
Burns, A. and Davis, R.~I. (2017).
\newblock A survey of research into mixed criticality systems.
\newblock {\em ACM Comput. Surv.}, 50(6):82:1--82:37.

\bibitem[Cadoret et~al., 2012]{DBLP:conf/iceccs/CadoretBGP12}
Cadoret, F., Borde, E., Gardoll, S., and Pautet, L. (2012).
\newblock Design patterns for rule-based refinement of safety critical embedded
  systems models.
\newblock In {\em 17th {IEEE} International Conference on Engineering of
  Complex Computer Systems, {ICECCS} 2012, Paris, France, July 18-20, 2012},
  pages 67--76.

\bibitem[Cadoret et~al., 2013]{DBLP:conf/isorc/CadoretRBPS13}
Cadoret, F., Robert, T., Borde, E., Pautet, L., and Singhoff, F. (2013).
\newblock Deterministic implementation of periodic-delayed communications and
  experimentation in {AADL}.
\newblock In {\em 16th {IEEE} International Symposium on
  Object/Component/Service-Oriented Real-Time Distributed Computing, {ISORC}
  2013, Paderborn, Germany, June 19-21, 2013}, pages 1--8.

\bibitem[Castellanos et~al., 2015]{DBLP:conf/euromicro/CastellanosBPSV15}
Castellanos, C., Borde, E., Pautet, L., Sebastien, G., and Vergnaud, T. (2015).
\newblock Improving reusability of model transformations by automating their
  composition.
\newblock In {\em 41st Euromicro Conference on Software Engineering and
  Advanced Applications, {EUROMICRO-SEAA} 2015, Madeira, Portugal, August
  26-28, 2015}, pages 267--274.

\bibitem[Castellanos et~al., 2014]{DBLP:conf/euromicro/CastellanosBPVD14}
Castellanos, C., Borde, E., Pautet, L., Vergnaud, T., and Derive, T. (2014).
\newblock Automatic production of transformation chains using structural
  constraints on output models.
\newblock In {\em 40th {EUROMICRO} Conference on Software Engineering and
  Advanced Applications, {EUROMICRO-SEAA} 2014, Verona, Italy, August 27-29,
  2014}, pages 158--165.

\bibitem[Cordeiro et~al., 2010]{cordeiro2010random}
Cordeiro, D., Mouni{\'e}, G., Perarnau, S., Trystram, D., Vincent, J.-M., and
  Wagner, F. (2010).
\newblock Random graph generation for scheduling simulations.
\newblock In {\em International ICST conference on simulation tools and
  techniques}. Institute for Computer Sciences, Social-Informatics and
  Telecommunications Engineering.

\bibitem[Czarnecki and Helsen, 2006]{Czarnecki:2006:FSM:1165093.1165106}
Czarnecki, K. and Helsen, S. (2006).
\newblock Feature-based survey of model transformation approaches.
\newblock {\em IBM Syst. J.}, 45(3):621--645.

\bibitem[Davis and Burns, 2009]{davis2009priority}
Davis, R.~I. and Burns, A. (2009).
\newblock Priority assignment for global fixed priority pre-emptive scheduling
  in multiprocessor real-time systems.
\newblock In {\em Real-Time Systems Symposium}, pages 398--409. IEEE.

\bibitem[Ehrig et~al.,
  2006]{boo:ehrigfundamentals-algebraic-graph-transformation}
Ehrig, H., Ehrig, K., Prange, U., and Taentzer, G. (2006).
\newblock {\em Fundamentals of algebraic graph transformation}, volume 373.
\newblock Springer.

\bibitem[Esfahani et~al., 2013]{DBLP:conf/icse/EsfahaniMR13}
Esfahani, N., Malek, S., and Razavi, K. (2013).
\newblock Guidearch: guiding the exploration of architectural solution space
  under uncertainty.
\newblock In {\em {ICSE}}, pages 43--52. {IEEE} Computer Society.

\bibitem[Etien et~al., 2012]{DBLP:conf/models/EtienABP12}
Etien, A., Aranega, V., Blanc, X., and Paige, R.~F. (2012).
\newblock Chaining model transformations.
\newblock In {\em Proceedings of the First Workshop on the Analysis of Model
  Transformations, AMT@MODELS 2012, Innsbruck, Austria, October 2, 2012}, pages
  9--14.

\bibitem[Fleck et~al., 2016]{10.1007/978-3-319-42064-6_6}
Fleck, M., Troya, J., and Wimmer, M. (2016).
\newblock Search-based model transformations with momot.
\newblock In Van~Gorp, P. and Engels, G., editors, {\em Theory and Practice of
  Model Transformations}, pages 79--87, Cham. Springer International
  Publishing.

\bibitem[Gamma, 1994]{GOF}
Gamma, E. e.~a. (1994).
\newblock {\em Design Patterns - Elements of Reusable Object-Oriented
  Software.}
\newblock Addison-Wesley.

\bibitem[G{\'{e}}rard, 2015]{Gerard15}
G{\'{e}}rard, S. (2015).
\newblock Once upon a time, there was papyrus...
\newblock In {\em {MODELSWARD} 2015 - Proceedings of the 3rd International
  Conference on Model-Driven Engineering and Software Development, ESEO,
  Angers, Loire Valley, France, 9-11 February, 2015}, pages IS--7.

\bibitem[Guernic et~al., 2003]{DBLP:journals/jcsc/GuernicTL03}
Guernic, P.~L., Talpin, J., and Lann, J.~L. (2003).
\newblock {POLYCHRONY} for system design.
\newblock {\em Journal of Circuits, Systems, and Computers}, 12(3):261--304.

\bibitem[Habel et~al., 2006]{tec:habel2006weakest-preconditions-for-high-level}
Habel, A., Pennemann, K.-H., and Rensink, A. (2006).
\newblock Weakest preconditions for high-level programs: Long version.
\newblock Technical Report 8/06, University of Oldenburg.

\bibitem[Halbwachs et~al., 1991]{Halbwachs91thesynchronous}
Halbwachs, N., Caspi, P., Raymond, P., and Pilaud, D. (1991).
\newblock The synchronous dataflow programming language lustre.
\newblock In {\em Proceedings of the IEEE}, pages 1305--1320.

\bibitem[Henzinger et~al., 2001]{Henzinger:2001:GTL:646787.703890}
Henzinger, T.~A., Horowitz, B., and Kirsch, C.~M. (2001).
\newblock Giotto: A time-triggered language for embedded programming.
\newblock In {\em Proceedings of the First International Workshop on Embedded
  Software}, EMSOFT '01, pages 166--184, Berlin, Heidelberg. Springer-Verlag.

\bibitem[Jackson, 2002]{Jackson02}
Jackson, D. (2002).
\newblock Alloy: A lightweight object modelling notation.
\newblock {\em ACM Transactions on Software Engineering and Methodology
  (TOSEM)}, 11(2).

\bibitem[Jaou{\"{e}}n et~al., 2014]{DBLP:conf/adaEurope/JaouenBPR14}
Jaou{\"{e}}n, A., Borde, E., Pautet, L., and Robert, T. (2014).
\newblock {PDP} 4ps : Periodic-delayed protocol for partitioned systems.
\newblock In {\em Reliable Software Technologies - Ada-Europe 2014, 19th
  Ada-Europe International Conference on Reliable Software Technologies, Paris,
  France, June 23-27, 2014. Proceedings}, pages 149--165.

\bibitem[Jouault and Kurtev, 2005]{MTIP05}
Jouault, F. and Kurtev, I. (2005).
\newblock Transforming models with atl.
\newblock In {\em Proceedings of the Model Transformations in Practice Workshop
  at MoDELS 2005}, Montego Bay, Jamaica.

\bibitem[Koziolek et~al., 2011]{PerOpteryx}
Koziolek, A., Koziolek, H., and Reussner, R. (2011).
\newblock Peropteryx: Automated application of tactics in multi-objective
  software architecture optimization.
\newblock In {\em the ACM SIGSOFT Conference Quality of Software
  Architectures}, NY, NY, USA. ACM.

\bibitem[Kwiatkowska et~al., 2011]{KNP11}
Kwiatkowska, M., Norman, G., and Parker, D. (2011).
\newblock {PRISM} 4.0: Verification of probabilistic real-time systems.
\newblock In Gopalakrishnan, G. and Qadeer, S., editors, {\em Proc. 23rd
  International Conference on Computer Aided Verification (CAV'11)}, volume
  6806 of {\em LNCS}, pages 585--591. Springer.

\bibitem[Lasnier et~al., 2009]{LasnierZPH09}
Lasnier, G., Zalila, B., Pautet, L., and Hugues, J. (2009).
\newblock Ocarina : An environment for {AADL} models analysis and automatic
  code generation for high integrity applications.
\newblock In {\em Reliable Software Technologies - Ada-Europe 2009, 14th
  Ada-Europe International Conference, Brest, France, June 8-12, 2009.
  Proceedings}, pages 237--250.

\bibitem[Lee and Messerschmitt, 1987]{Lee1987SynchronousDF}
Lee, E.~A. and Messerschmitt, D. (1987).
\newblock Synchronous data flow.
\newblock {\em Proceedings of the IEEE}, 75:1235--1245.

\bibitem[Leveque et~al., 2011]{ProCom-Refinement}
Leveque, T., Carlson, J., Sentilles, S., and Borde, E. (2011).
\newblock Flexible semantic-preserving flattening of hierarchical component
  models.
\newblock In {\em 37th {EUROMICRO} Conference on Software Engineering and
  Advanced Applications, {SEAA} 2011, Oulu, Finland, August 30 - September 2,
  2011}, pages 31--38.

\bibitem[Li et~al., 2011]{AQOSA}
Li, R., Etemaadi, R., Emmerich, M. T.~M., and Chaudron, M. R.~V. (2011).
\newblock An evolutionary multiobjective optimization approach to
  component-based software architecture design.
\newblock In {\em Evolutionary Computation (CEC), 2011 IEEE Congress on}, pages
  432--439.

\bibitem[Liu and Layland, 1973]{journals/jacm/LiuL73}
Liu, C.~L. and Layland, J.~W. (1973).
\newblock Scheduling algorithms for multiprogramming in a hard-real-time
  environment.
\newblock {\em J. ACM}, 20(1):46--61.

\bibitem[Loniewski et~al., 2013a]{LoniewskiBBI13a}
Loniewski, G., Borde, E., Blouin, D., and Insfr{\'{a}}n, E. (2013a).
\newblock An automated approach for architectural model transformations.
\newblock In {\em Information System Development - Improving Enterprise
  Communication, [Proceedings of the 22nd International Conference on
  Information Systems Development, {ISD} 2013, Seville, Spain]}, pages
  295--306.

\bibitem[Loniewski et~al., 2013b]{LoniewskiBBI13b}
Loniewski, G., Borde, E., Blouin, D., and Insfr{\'{a}}n, E. (2013b).
\newblock Model-driven requirements engineering for embedded systems
  development.
\newblock In {\em 39th Euromicro Conference on Software Engineering and
  Advanced Applications, {SEAA} 2013, Santander, Spain, September 4-6, 2013},
  pages 236--243.

\bibitem[Loniewski et~al., 2013c]{DBLP:conf/euromicro/LoniewskiBBI13}
Loniewski, G., Borde, E., Blouin, D., and Insfr{\'{a}}n, E. (2013c).
\newblock Model-driven requirements engineering for embedded systems
  development.
\newblock In {\em 39th Euromicro Conference on Software Engineering and
  Advanced Applications, {SEAA} 2013, Santander, Spain, September 4-6, 2013},
  pages 236--243.

\bibitem[Lyons and Vanderkulk, 1962]{Lyons:1962:UTR:1661979.1661984}
Lyons, R.~E. and Vanderkulk, W. (1962).
\newblock The use of triple-modular redundancy to improve computer reliability.
\newblock {\em IBM J. Res. Dev.}, 6(2):200--209.

\bibitem[Maxim et~al., 2017]{Maxim:2017:PAM:3139258.3139276}
Maxim, D., Davis, R.~I., Cucu-Grosjean, L., and Easwaran, A. (2017).
\newblock Probabilistic analysis for mixed criticality systems using fixed
  priority preemptive scheduling.
\newblock In {\em Proceedings of the 25th International Conference on Real-Time
  Networks and Systems}, RTNS '17, pages 237--246, New York, NY, USA. ACM.

\bibitem[Medina et~al., 2017]{MedinaBP17}
Medina, R., Borde, E., and Pautet, L. (2017).
\newblock Directed acyclic graph scheduling for mixed-criticality systems.
\newblock In {\em Reliable Software Technologies - Ada-Europe 2017 - 22nd
  Ada-Europe International Conference on Reliable Software Technologies,
  Vienna, Austria, June 12-16, 2017, Proceedings}, pages 217--232.

\bibitem[Medina et~al., 2018a]{MedinaBP18}
Medina, R., Borde, E., and Pautet, L. (2018a).
\newblock Availability enhancement and analysis for mixed-criticality systems
  on multi-core.
\newblock In {\em 2018 Design, Automation {\&} Test in Europe Conference {\&}
  Exhibition, {DATE} 2018, Dresden, Germany, March 19-23, 2018}, pages
  1271--1276.

\bibitem[Medina et~al., 2018b]{medina2018scheduling}
Medina, R., Borde, E., and Pautet, L. (2018b).
\newblock Scheduling multi-periodic mixed criticality dags on multi-core
  architectures.
\newblock In {\em 2018 IEEE Real-Time Systems Symposium (RTSS)}. IEEE.

\bibitem[Mens and Van~Gorp, 2006]{Mens:2006:TMT:1706639.1706924}
Mens, T. and Van~Gorp, P. (2006).
\newblock A taxonomy of model transformation.
\newblock {\em Electron. Notes Theor. Comput. Sci.}, 152:125--142.

\bibitem[MG, 2009]{OMG2009}
MG (2009).
\newblock Uml profile for marte: Modeling and analysis of real-time embedded
  systems.

\bibitem[Mura et~al., 2008]{MuraMP08}
Mura, M., Murillo, L.~G., and Prevostini, M. (2008).
\newblock Model-based design space exploration for {RTES} with sysml and
  {MARTE}.
\newblock In {\em Forum on specification and Design Languages, {FDL} 2008,
  September 23-25, 2008, Stuttgart, Germany, Proceedings}, pages 203--208.

\bibitem[OMG, 2005]{QVT-specification}
OMG (2005).
\newblock {\em MOF QVT Final Adopted Specification}.
\newblock Object Modeling Group.

\bibitem[{OMG}, 2014]{std:ocl}
{OMG} (2014).
\newblock {O}bject {C}onstraint {L}anguage ({OCL}).
\newblock \url{http://www.omg.org/spec/OCL}.

\bibitem[Pennemann,
  2008]{inc:pennemann2008development-correct-graph-transformation}
Pennemann, K.-H. (2008).
\newblock Development of correct graph transformation systems.
\newblock In {\em Graph Transformations}, volume 5214 of {\em LNCS}, pages
  508--510. Springer.

\bibitem[Poskitt, 2013]{phd:poskitt2013verification-graph-programs}
Poskitt, C.~M. (2013).
\newblock {\em Verification of Graph Programs}.
\newblock PhD thesis, University of York.

\bibitem[Poskitt and Plump,
  2013]{inc:poskitt2013verifying-total-correctness-graph}
Poskitt, C.~M. and Plump, D. (2013).
\newblock Verifying total correctness of graph programs.
\newblock In {\em Revised Selected Papers, Graph Computation Models (GCM
  2012)}, Electronic Communications of the ECEASST 61.

\bibitem[Rahmoun et~al., 2015a]{Rahmoun2015}
Rahmoun, S., Borde, E., and Pautet, L. (2015a).
\newblock Automatic selection and composition of model transformations
  alternatives using evolutionary algorithms.
\newblock In {\em 2015 European Conference on Software Architecture Workshops},
  pages 25:1--25:7, New York, NY, USA. ACM.

\bibitem[Rahmoun et~al., 2015b]{DBLP:conf/iceccs/RahmounBP15}
Rahmoun, S., Borde, E., and Pautet, L. (2015b).
\newblock Multi-objectives refinement of {AADL} models for the synthesis
  embedded systems (mu-ramses).
\newblock In {\em 20th International Conference on Engineering of Complex
  Computer Systems, {ICECCS} 2015, Gold Coast, Australia, December 9-12, 2015},
  pages 21--30.

\bibitem[Rahmoun et~al., 2017]{Rahmoun2017}
Rahmoun, S., Mehiaoui-Hamitou, A., Borde, E., Pautet, L., and Soubiran, E.
  (2017).
\newblock Multi-objective exploration of architectural designs by composition
  of model transformations.
\newblock {\em Software {\&} Systems Modeling}.

\bibitem[Renault et~al., 2009]{Renault:2009:AMM:1590961.1591431}
Renault, X., Kordon, F., and Hugues, J. (2009).
\newblock Adapting models to model checkers, a case study: Analysing aadl using
  time or colored petri nets.
\newblock In {\em Proceedings of the 2009 IEEE/IFIP International Symposium on
  Rapid System Prototyping}, RSP '09, pages 26--33, Washington, DC, USA. IEEE
  Computer Society.

\bibitem[Reussner et~al., 2011]{reussner2011a}
Reussner, R., Becker, S., Burger, E., Happe, J., Hauck, M., Koziolek, A.,
  Koziolek, H., Krogmann, K., and Kuperberg, M. (2011).
\newblock {The Palladio Component Model}.
\newblock Technical report, KIT, Fakultät für Informatik, Karlsruhe.

\bibitem[Richa et~al., 2015]{DBLP:conf/icmt/RichaBP15}
Richa, E., Borde, E., and Pautet, L. (2015).
\newblock Translating {ATL} model transformations to algebraic graph
  transformations.
\newblock In {\em Theory and Practice of Model Transformations - 8th
  International Conference, {ICMT} 2015, Held as Part of {STAF} 2015, L'Aquila,
  Italy, July 20-21, 2015. Proceedings}, pages 183--198.

\bibitem[Richa et~al., 2017]{Richa2017}
Richa, E., Borde, E., and Pautet, L. (2017).
\newblock Translation of atl to agt and application to a code generator for
  simulink.
\newblock {\em Software {\&} Systems Modeling}.

\bibitem[Richa et~al., 2014]{DBLP:conf/models/RichaBPBR14}
Richa, E., Borde, E., Pautet, L., Bordin, M., and Ruiz, J.~F. (2014).
\newblock Towards testing model transformation chains using precondition
  construction in algebraic graph transformation.
\newblock In {\em Proceedings of the Workshop on Analysis of Model
  Transformations co-located with {ACM/IEEE} 17th International Conference on
  Model Driven Engineering Languages {\&} Systems (MoDELS 2014), Valencia,
  Spain, September 29, 2014}, pages 34--43.

\bibitem[Siebert and Teizer, 2014]{siebert2014mobile}
Siebert, S. and Teizer, J. (2014).
\newblock Mobile 3d mapping for surveying earthwork projects using an unmanned
  aerial vehicle (uav) system.
\newblock {\em Automation in Construction}, 41.

\bibitem[Singhoff et~al., 2004]{Singhoff:2004:CFR:1032297.1032298}
Singhoff, F., Legrand, J., Nana, L., and Marc{\'e}, L. (2004).
\newblock Cheddar: A flexible real time scheduling framework.
\newblock In {\em Proceedings of the 2004 Annual ACM SIGAda International
  Conference on Ada: The Engineering of Correct and Reliable Software for
  Real-time \&Amp; Distributed Systems Using Ada and Related Technologies},
  SIGAda '04, pages 1--8, New York, NY, USA. ACM.

\bibitem[Su et~al., 2016]{Su:2016:EMT:3029795.2984633}
Su, H., Zhu, D., and Brandt, S. (2016).
\newblock An elastic mixed-criticality task model and early-release edf
  scheduling algorithms.
\newblock {\em ACM Trans. Des. Autom. Electron. Syst.}, 22(2):28:1--28:25.

\bibitem[Vestal, 2007]{Vestal:2007:PSM:1338441.1338659}
Vestal, S. (2007).
\newblock Preemptive scheduling of multi-criticality systems with varying
  degrees of execution time assurance.
\newblock In {\em Proceedings of the 28th IEEE International Real-Time Systems
  Symposium}, RTSS '07, pages 239--243, Washington, DC, USA. IEEE Computer
  Society.

\bibitem[Vulgarakis et~al., 2009]{ProCom}
Vulgarakis, A., Suryadevara, J., Carlson, J., Seceleanu, C.~C., and Pettersson,
  P. (2009).
\newblock Formal semantics of the procom real-time component model.
\newblock In {\em 35th Euromicro Conference on Software Engineering and
  Advanced Applications, {SEAA} 2009, Patras, Greece, August 27-29, 2009,
  Proceedings}, pages 478--485.

\bibitem[Yin et~al., 2011]{Yin2071}
Yin, H., Borde, E., and Hansson, H. (2011).
\newblock Composable mode switch for component-based systems.
\newblock In Sebastian~Fischmeister, L. T.~P., editor, {\em 3rd Workshop on
  Adaptive and Reconfigurable Embedded Systems (APRES 2011)}, pages 19--22.

\end{thebibliography}



\end{spacing}





\printthesisindex 

\end{document}